\documentclass[twocolumn,mathlines]{aastex631}
\usepackage[utf8]{inputenc}
\usepackage{multirow}
\usepackage{blindtext}
\usepackage[T1]{fontenc}
\usepackage{textcomp}
\usepackage{amsmath,amssymb}
\usepackage{color}
\usepackage{float}
\usepackage{url}
\usepackage{wrapfig}
\usepackage{graphicx}
\usepackage{subfigure}
\usepackage{rotating}
\usepackage{enumitem}\setlist[description]{font=\textendash\enskip\scshape\bfseries}
\usepackage{lmodern}
\usepackage{makecell} 
\usepackage{caption}
\graphicspath{ {./images/} }

\newcommand{\beq}{\begin{equation}}
\newcommand{\eeq}{\end{equation}}
\newcommand{\bdm}{\begin{displaymath}}
\newcommand{\edm}{\end{displaymath}}
\newcommand{\T}[1]{\tilde{#1}}

\definecolor{Gray}{gray}{0.9}
\definecolor{orange}{rgb}{0.9,0.5,0}

\begin{document}

\title{The Effect of Vera C. Rubin Observatory Cadence Selections on Kilonova Detectability}

\author[0009-0004-9687-3275]{Cristina Andrade}
\affil{School of Physics and Astronomy, University of Minnesota, Minneapolis, Minnesota 55455, USA}

\author{Raiyah Alserkal}
\affil{Department of Physics and Astronomy, University of Sussex, Brighton, BN1 9RH, UK}
\affil{Department of Physics, American University of Sharjah, PO Box 26666, Sharjah, UAE}

\author[0000-0002-6514-318X]{Luis Salazar Manzano}
\affil{Department of Astronomy, University of Michigan, Ann Arbor, Michigan 48109, USA}
\affil{Department of Physics and Astronomy, University of Texas Rio Grande Valley, Brownsville, Texas 78520, USA}

\author{Emma Martin}
\affil{School of Physics and Astronomy, University of Minnesota, Minneapolis, Minnesota 55455, USA}

\author[0000-0002-8977-1498]{Igor Andreoni}
\altaffiliation{Neil Gehrels Fellow}
\affil{Joint Space-Science Institute, University of Maryland, College Park, MD 20742, USA}
\affil{Department of Astronomy, University of Maryland, College Park, MD 20742, USA}
\affil{Astrophysics Science Division, NASA Goddard Space Flight Center, Mail Code 661, Greenbelt, MD 20771, USA}
\affil{University of North Carolina, Chapel Hill, NC 27514, USA}

\author[0000-0002-8262-2924]{Michael W. Coughlin}
\affil{School of Physics and Astronomy, University of Minnesota, Minneapolis, Minnesota 55455, USA}

\author[0000-0003-1585-8205]{Nidhal Guessoum}
\affil{Department of Physics, American University of Sharjah, PO Box 26666, Sharjah, UAE}

\author[0000-0002-9396-7215]{Liliana Rivera Sandoval}
\affil{Department of Physics and Astronomy, University of Texas Rio Grande Valley, Brownsville, Texas 78520, USA}

\begin{abstract}
%Binary neutron-star mergers are notable astrophysical phenomena; they offer us an opportunity to examine important astrophysical processes, including heavy-element nucleosynthesis, merger-driven mass ejections, wide spectrum electromagnetic emissions, and now gravitational waves. 
The discovery of the optical/infra-red counterpart (AT2017gfo) to the binary neutron star gravitational-wave detection (GW170817), which was followed by a short gamma-ray burst (GRB170817), marked a groundbreaking moment in multi-messenger astronomy. To date, it remains the only confirmed joint detection of its kind. However, many experiments are actively searching for similar fast-fading electromagnetic counterparts, known as kilonovae. Fortunately, the Vera C. Rubin Observatory's Legacy Survey of Space and Time (LSST) provides excellent prospects for identifying kilonova candidates either from, or independent of, gravitational-wave and gamma-ray burst triggers. Cadence choices for LSST surveys are especially important for maximising the likelihood of kilonovae detections. In this work, we explore the possibility of optimizing Rubin Observatory's ability to detect kilonovae by implementing a fast transient metric shown to be successful with an existing wide field survey, e.g. the Zwicky Transient Facility (ZTF). We study existing LSST cadences, how detection rates are affected by filter selections, the return timescales for visits of the same area in the sky, and other relevant factors. Through our analysis, we have found that employing \texttt{baseline} cadences and utilizing triplet families like \texttt{presto\_gap} produced the highest likelihood of kilonova detection.

%We assess the benefit of our findings to related scientific interests, including maximising a range of fast transient discoveries.

\end{abstract}

\section{Introduction}
Multi-messenger astronomy is the exploration of the universe through the interpretation of a combination of cosmic signals, including electromagnetic radiation, gravitational waves, neutrinos, and cosmic rays \citep{Bartos&Kowalski_2017}.  Black hole-neutron star and binary neutron star mergers are areas of particular interest in multi-messenger astronomy, as they produce potentially detectable gravitational-wave signals, in addition to electromagnetic ones. These systems and associated phenomena give insight into stellar binary evolution and are favorable sites for the production of heavy elements  \citep{10.3389/fspas.2020.609460}; detections of these phenomena also provide prospects to constrain the neutron star equation of state \citep[e.g.,][]{Metzger_2017} as well as the Hubble Constant \citep[e.g.,][]{Coughlin_2020}. 

When two neutron stars merge, or, potentially, when a neutron star and a black hole merge, the explosion produces an optical event called a kilonova (KN), and it has long been theorized that this event is associated with short gamma-ray bursts and part of the resulting ``afterglow'' electromagnetic radiation at lower energies/frequencies \citep{Li_1998}. Early predictions of optical counterparts to neutron star mergers were supported by a potential KN detected following a short gamma-ray burst (GRB130603B) \citep{Tanvir_2013}. However, a breakthrough in multi-messenger astronomy occurred with the simultaneous detection of a gravitational-wave signal (GW170817) \citep{PhysRevLett.118.221101} and electromagnetic multi-wavelength emission (GRB170817) \citep{Abbott_2017}; this event had all the characteristics expected from a binary neutron star merger, including an optical and near-infrared counterpart (AT2017gfo), known as a kilonova \citep{Coulter_2017}. Optimizing identification methods for these events, particularly their optical parts, offers an opportunity to study the astrophysical heavy-element nucleosynthesis as well as merger-driven mass ejections \citep{Barnes_2020, toivonen2024expectkilonovalightcurve}. 

The most common way to search for kilonovae (KNe), and similar fast-fading transients, is through follow-up observations of GRB and GW triggers. When such a signal is detected, an announcement is sent through the global network of observatories, and observations are then attempted in other bands; from UV to Radio \citep{Abbott_2016,Chaudhary_2024}. Although this was successful in the case of GW170817, wide-field surveys can potentially make detections independent of GRB and GW triggers \citep{Andreoni_2021}; this is potentially a more difficult task because fast fading transients are hard to discover.

The Vera C. Rubin Observatory is an optical observatory being built in north-central Chile to execute its 10-year Legacy Survey of Space and Time (LSST). Its mission is to probe the universe for dark energy and dark matter, take an inventory of the solar system, explore the transient optical sky, and map the Milky Way \citep{Ivezic_2019,Lochner_2022, Hernitschek_2022}. The LSST is a general term used to describe a set of surveys performed over 10 years; they include Wide-Fast-Deep (WFD), which is the main survey, Deep Drilling Fields (DDFs), as well as additional mini-surveys with varied sky regions or survey parameters \citep{Bianco_2021, Li_2022, Gris_2023}. It will image approximately 20,000 square degrees of the sky in 6 photometric filters (u, g, r, i, z, y) \citep{jones_r_lynne_2020_4048838} with its wide field view of 9.6 deg$^2$ and a 3.2 Gigapixel camera \citep{Ivezic_2019}.

Amongst other science cases, LSST provides excellent opportunities to detect KNe. This can take a few forms. One possibility is the follow-up of gravitational-wave candidates. \citet{Andreoni_2022} detailed potential target-of-opportunity strategies. Another  possibility is serendipitous detections of kilonovae. \citet{Andreoni_2021} introduced a variety of metrics to support this study, including one based on an algorithm in production for the Zwicky Transient Facility \citep{Andreoni_Coughlin_2021, Clarke_2024} to find fast transients, which relies on finding objects fading faster than 0.3\,mag per day in at least one passband. We note that this is a very generic filter designed to find all kinds of fast transients, beyond kilonovae, including fast blue optical transients, gamma-ray burst afterglows, shock breakout, etc.

For KNe or other objects/phenomena, detection is not the only goal of observational studies. It is important that kilonovae (or other) candidates found by surveys (Rubin Observatory or others) can be characterized in real time so that follow-up observations can confirm their nature \citep{Bonito_2023, Prisinzano_2023, Di_Criscienzo_2023}. In this paper, we expand on previous studies to examine how filter selection and cadence strategies influence kilonova detection while also considering the impact of population variability on characterization. This is especially important as KNe are expected to appear mostly red and, importantly, fade slower in red/near-IR bands than in bluer bands \citep{Smartt_2017}. All of this builds upon infrastructure developed by the LSST project and the broader community to cadences around various optimized science goals \citep{Bianco_2021, Raiteri_2022}.

We use simulated surveys to analyze how KNe detection rates vary with different filter selections, and whether they improve when red/near-infrared filters are used more often. We employ metrics designed to flag transient detection as well as ones that identify fast-fading transient based on their flux evolution (see \S\ref{sec:metrics}).  We also use the existing \texttt{TgapsMetric} from \citet{Bellm_2022} to explore the effect of time gaps between observations of the same area in the sky on transient discovery. Additionally, we study the variation of population on detection \citep{Ragosta_2024}. We perform these studies on modified LSST baseline cadences, as well as surveys from existing LSST cadence families, particularly \texttt{baseline}, \texttt{rolling}, \texttt{triplets}, \texttt{long gaps no pairs} and \texttt{suppress repeats} cadences from the v2.0, v2.1, and v3.0 survey strategy releases, and observe their effectiveness at detecting KNe. These cadence families and their function will be further explained in Section \ref{sec:cadences}. We include in some of our simulations the draft of the v3.0 survey strategies (v2.99) and the v3.3 through v3.6 that had been released at the time of writing. 

\begin{table*}[t]
\centering
\begin{tabular}{l c c c c c c}%{lllllll}
\hline
\hline
Kilonova(e) & Injected & Luminosity & $M_{dyn}$ & $M_{wind}$ & $\theta$ & $\phi$\\
model(s) & KNe & Distance [Mpc] & [$M_{\odot}$] & [$M_{\odot}$] & [deg] & [deg] \\
\Xhline{3\arrayrulewidth}
& & & & & & \\
\textbf{Single model} & $5\times10^{5}$ & 10 - 600 & GW170817 & GW170817 & GW170817 & GW170817 \\
 & & & & & & \\
\Xhline{3\arrayrulewidth}
 & & & GW170817 & GW170817 & & \\
\textbf{Single $\phi$ population} & $5\times10^{5}$ & 10 - 1400 & Optimistic & Optimistic & Full variation & GW170817 \\
 & & & Pessimistic & Pessimistic & & \\
\Xhline{3\arrayrulewidth}
 & & & GW170817 & GW170817 & 0, 25.8, 36.9, 45.6, & \\
\textbf{$\theta$ space population} & $5\times10^{5}$ & 10 - 1400 & Optimistic & Optimistic & 53.1, 60.0, 66.4, & Full variation  \\
 & & & Pessimistic & Pessimistic & 72.5, 78.5, 84.3, 90 & \\
\Xhline{3\arrayrulewidth}
 & & & GW170817 & GW170817 & & 0, 15, 30, \\
\textbf{$\phi$ space population} & $5\times10^{5}$ & 10 - 1400 & Optimistic & Optimistic & Full variation & 45, 60, \\
 & & & Pessimistic & Pessimistic & & 75, 90 \\
\Xhline{3\arrayrulewidth}
\end{tabular}
\caption{Injection and model parameters used in each of our simulation scenarios. The grid values for $\theta$ are 0, 25.8, 36.9, 45.6, 53.1, 60.0, 66.4, 72.5, 78.5, 84.3, and 90 degrees, and for $\phi$ are 0, 15, 30, 45, 60, 75, and 90 degrees.}
\label{tab:table1}
\end{table*}

\section{Methods}

The Rubin Observatory is currently designing a scheduling system that can respond swiftly to unexpected events and can be optimized throughout the survey. The Operations Simulator (Opsim) was developed by the Rubin team to generate simulations of the 10-year survey and its field selection and image acquisition process \citep{Delgado_2014}. They routinely provide new sets of \texttt{Opsims} runs, organized in ``families,'' with modifications made as a response to recommendations from various optimization studies \citep{jones_r_lynne_2020_4048838}. Each \texttt{Opsims} survey strategy differs in the parameters specified; the \texttt{footprint} family, for example, modifies the survey footprint \citep{Bianco_2021}. The Rubin Observatory team has also released the Metric Analysis Framework (MAF), an open-access software package that allows the creation of metric calculations associated with specific scientific goals, and each metric is used to evaluate existing survey simulations \citep{Jones_2014}.

\begin{figure}[hbt!]
\centering
\includegraphics[scale=0.4]{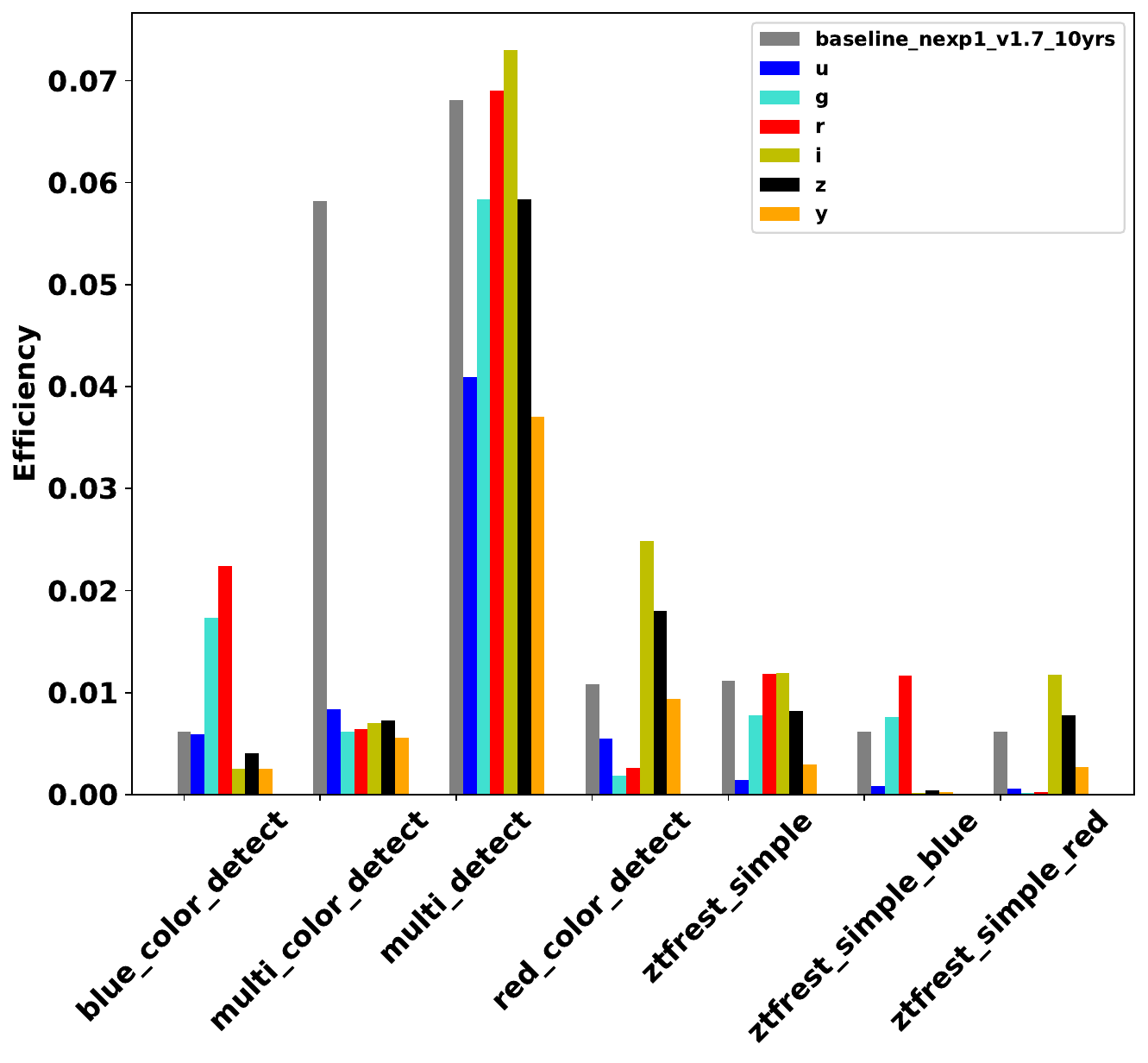}
\centering
\caption{This histogram presents a preliminary comparison of LSST’s six photometric filters using an initial simulated cadence. The goal of this test was to determine which bandpass, when paired with a detection metric, yields the highest kilonovae recovery efficiency.}
\label{fig:cadencecomparison}
\end{figure}

The results in this project’s simulations were obtained using the LSST LINCC JupyterHub platform \footnote{\url{https://lsst.dirac.dev/}}, where we were granted access to LSST simulation strategies and metrics. We use the most recent \texttt{Opsim} cadence simulations to study how effective various LSST cadence options are at KNe discovery. 

The process for assessing LSST sensitivity to fast transients begins by injecting numerous light curves of merger events, from a grid of models described in \cite{bulla2019possis} and \cite{Dietrich_2020} -- see \S\ref{sec: light curves}. Then, with the \texttt{kneMetrics} function,  we explore the extent to which survey parameters, including the adoption of red/near-IR filters, of varied return times, and of other factors defined by recent LSST cadences, allow cadence simulations to better detect the associated KNe. The cadence used as the fiducial strategy for our preliminary simulations is the \texttt{baseline\_nexp2\_v1.7.1} cadence. For the following simulations, we used the \texttt{baseline\_v2.1} cadence. The existing \texttt{TgapMetric} \citep{Bellm_2022}, from \texttt{lsst.sims.maf.metrics.TgapsMetric}, was utilized to calculate the time gap lengths between observations of the same area in the sky in each cadence simulation.

\subsection{Kilonova Light Curves}
\label{sec: light curves}

In our analysis, light curves were injected into simulated \texttt{Opsim} cadences (see Table \ref{tab:table1}), and we evaluated the efficiency of the Rubin Observatory for KNe detections in the surveys. The theoretical light curve model that re-constructs the kilonova was taken from \cite{Dietrich_2020}, which uses the POSSIS \citep{bulla2019possis} Monte Carlo radiative transfer code. This is an axially symmetric model characterized by two ejecta mass components:  $M_{dyn}$ and $M_{wind}$.  The dynamical ejecta mass $M_{dyn}$ is lanthanide-rich in composition and distributed within an angle $\pm\phi$ around the merger plane. The disk wind ejecta mass $M_{wind}$ is released after the merger and is lanthanide-free in composition and distributed at higher latitudes. Dynamical ejecta, which is rich in lanthanides, heavier r-process elements, has high opacity that traps radiation, leading to a redder and longer-lasting kilonova signal. In contrast, disk wind ejecta, which is lanthanide-free, consists of lighter r-process elements, resulting in lower opacity and a bluer, faster-fading optical counterpart \citep{Tanvir_2017}. In addition to these three parameters ($M_{dyn}$, $M_{wind}$ and $\phi$), a fourth one on which this model is varied is the observer viewing angle $\theta$. We suggest referring to Figure 1 of \citep{bulla2019possis} for a graphical diagram of the two-component KN model.

For the first set of simulations (first row of Table \ref{tab:table1}), we injected $5\times10^{5}$ light curves distributed uniformly in the volume defined by the luminosity distance range 10--600\,Mpc and using a model with parameters $M_{dyn}=0.005$ $M_{\odot}$, $M_{wind}=0.050$ $M_{\odot}$, $\phi=30^\circ$ and $\theta=25.8^\circ$ that approximate AT2017gfo \citep{Andreoni_2021}. The simulations run with these parameters will be referred to as Single Model simulations.

For the study on how varying KNe parameters affects detection (see \ref{sec:knepop}), we applied the second to fourth row of Table \ref{tab:table1}, injecting $5\times10^{5}$ light curves out to a luminosity distance of 1400\,Mpc. In this case, we considered three population scenarious using a variety of model parameter values with the aim of representing a KNe population. In each of them, we included KNe with optimistic, pessimistic, and GW170817-like ejecta masses. The optimistic and pessimistic masses chosen are physically realistic according to numerical relativity simulations. In the aforementioned scenario, ``pessimistic'' and ``optimistic'' are defined as favorable or unfavorable to detection. The optimistic masses, both $M_{dyn}$ (0.02 $M_{\odot}$) and $M_{wind}$ (0.070, 0.080, 0.090 $M_{\odot}$), are larger in value which makes the KNe particularly bright; while the pessimistic masses $M_{dyn}$ (0.005 $M_{\odot}$) and $M_{wind}$ (0.010, 0.020, 0.030 $M_{\odot}$) are smaller making fainter light curves. The first scenario of the second set of simulations includes a kilonova viewed from 11 angles uniformly distributed in cos$(\theta)$, all cases set with a lanthanide-rich composition opening angle $\phi$ of 30$^\circ$. In the second and third scenarios, we explore the parameter space of $\theta$ or $\phi$, respectively, while fully varying the other angle ($\phi$ or $\theta$) as allowed by the model grid (see the last two columns of Table \ref{tab:table1}).   
%The parameter values, which were determined in a physically realistic model by numerical relativity simulations, are:
%\begin{itemize}
%    \item M(dyn) = 0.005M$\odot$
%    \item M(wind) = 0.050M$\odot$
%    \item Viewing angle ($\theta$) = 25.8°
%    \item Lanthanide-rich composition opening angle ($\Phi$) = 30°
%\end{itemize}

\subsection{Metrics}
\label{sec:metrics}

Providing appropriate metric criteria is necessary for defining/determining when a KNe will be (or said to be) discovered in a sky survey or simulation. In the first part of this study, a set of metrics is taken from \texttt{kneMetrics} and its \texttt{KNePopMetric} functions, which have 7 individual metrics that were developed for transient detection criteria (see below).
A detailed description of the detection criteria for each metric is given below (from \cite{Andreoni_2021}):
\begin{itemize}
    \item \texttt{multi\_detect}: 2 or more transient detections. 
    \begin{itemize}
        \item \texttt{multi\_color\_detect}: 2 or more detections with at least 2 filters.
        %\item multi-detect red:
        \item \texttt{red\_color\_detect}: 2 or more detections with at least 2 red filters (i,z,y).
        %\item multi-detect blue: 
        \item \texttt{blue\_color\_detect}: 2 or more detections with at least 2 blue filters (u,g,r).
    \end{itemize}
    \item \texttt{ztfrest\_simple}: detections are made when sources are found to be rising faster than 1 mag day$^{-1}$ and fading faster than 0.3 mag day$^{-1}$.
    \begin{itemize}
        \item \texttt{ztfrest\_simple\_red}: applied only to red (i,z,y) bands
        \item \texttt{ztfrest\_simple\_blue}: applied only to blue (u,g,r) bands.
    \end{itemize}
\end{itemize}

The \texttt{multi\_detect} and related metrics follow standard transient detection criteria, 2 or more transient detections, which provides information on the celestial coordinates of a source, whereas the \texttt{ztfrest\_simple} and related metrics detection criteria allow for proper source characterization, i.e., KNe discovery. \texttt{ZTFReST} metrics are taken from studies made by the ZTF Realtime Search and Triggering (\texttt{ZTFReST}) project, which is dedicated to optimizing fast-fading transient discovery \citep{Andreoni_Coughlin_2021}. In particular, \texttt{ztfrest\_simple} is designed for KNe discovery. Simple detection may be appropriate for gravitational-wave follow-ups but is not effective in regular fast transient surveys \citep{Andreoni_2021}, thus additional photometric criteria to separate KNe from other transients may be required.

We also employed the \texttt{TgapMetric} from \texttt{lsst.sims.maf.metrics.TgapsMetric} to understand how KNe detections are affected by the return times between observations in the same area of the sky. Most Rubin Observatory cadences consist of pairs of visits that occur in the same night and are separated by approximately 30 minutes, with a return visit after a few days. Current visit separations provide little constraint on temporal evolution within a night, and ideal cadences for transient discovery and similar variability characterization would be uniform in log(time) and sensitive to variations on all timescales \citep{Bellm_2022}. 

\subsection{LSST Cadences}
\label{sec:cadences}

In this paper, we study how effective the most recent LSST strategies, or cadences, are at detecting KNe. They consist of a set of strategy variations, all of which evolve from the previous releases (v1.5 to v1.7.1). The survey simulations v1.5 and v1.7.1 heavily focused on the metrics discussed in the previous section. The versions that followed, versions 2.0, 2.1, 2.99, 3.0 and v3.2 survey simulations, placed emphasis on other variables as well as the metrics. At the time of writing, versions 3.3, 3.4, 3.5 and 3.6 baseline cadences were released and studied. We study all the available families in the v2.0 and v2.1 survey strategies, with special focus on the families designed for transient science; the \texttt{baseline}, \texttt{rolling}, \texttt{triplets} and \texttt{long gaps no pairs} families. Given some of our preliminary findings, we paid particular attention to the \texttt{suppress repeats} family for our populations study.

The \texttt{retro} simulations are intended to provide a bridge from v1.X to v2.X baseline simulation editions, introducing modifications to the footprint and scheduler code separately and in stages. This provides a gradual change in factors in the cadence families. The \texttt{retro} cadence within the \texttt{baseline} family allows the reader to understand the potential changes in their metric results from the v1.0 series of runs to the v2.0 by incorporating priority variables from the baseline. \texttt{Baseline\_retrofoot} is very similar to \texttt{baseline\_v2.0}; however, it uses the v1.7 footprint. Additionally, \texttt{retro\_baseline} is similar to v1.7, but it removes the rolling cadence. 

The idea behind \texttt{rolling} cadences is to divide the footprint area into regions, and to enhance the sampling rate one region at a time, at the expense of decreasing the sampling rate in the other regions. This cadence family concentrates observations on subsets of the sky for specific periods, allowing for improved temporal resolution in those regions. Thus, it provides the best means of allocating additional observations into the 2 hours to 1 night return time window critical for rapid discovery of fast-evolving transients. Cadences with a rapid timescale is a significant parameter for fast-transients \citep{Feigelson_2023}. This family covers many cadence variations, arranged primarily by the level of impact of the rolling cadence, i.e. least to most. The \texttt{no\_roll} cadence, as its name indicates, has no rolling cadence. Moreover, there are variations on the number of stripes (2, 3, or 6), the areas of the sky (WFD only or the additional bulge WFD-area, etc.), and the strength of the rolling (50, 80, or 90\%). In the \texttt{rolling} cadence runs, ``rw'' refers to ``rolling weight,'' or how much emphasis is put on the rolling cadence. The \texttt{baseline\_v2.0} has a 2-band rolling cadences, and for this season performs similarly to the \texttt{rolling\_ns2\_rw0.9}. 

The presto cadences fall under the \texttt{triplets} family; they add a third visit within each night, repeating one of the filters used in the initial pair. This family was proposed by \citealt{Bianco:2019} with the aim of adding brightness change in the same filter to the color information given by the initial pair. The main variation within this family is the time interval between the first pair of visits and the third visit, which could increase detection \citep{Alves_2023, Schwamb_2023}.\texttt{Presto\_gap} cadence family names indicate the time between visits in the header. For instance, \texttt{presto\_gap1.5\_v2.0\_10yrs} specifies triples spaced 1.5 hours apart (g+r, r+i, i+z), while \texttt{presto\_gap2.0\_v2.0\_10yrs} uses a 2.0-hour spacing. The \texttt{presto\_gap} cadences always take \texttt{triplets} while \texttt{presto\_half\_gap} only half of the time. The standard runs of this family take the initial pairs in successive filters (g+r, r+i, i+z) while in the \texttt{gap\_mix} runs the pair of filters are more spaced (g+i, r+z, i+y). There are also \texttt{long\_gaps} runs within this family that use even longer separations.

For our population study, we tested the effect of the \texttt{long gaps no pairs} family. This family extends the revisit time between the pair of visits, originally separated by 33 minutes, from 2 to 7 hours. In some cadences, the long gaps are applied throughout the whole survey while in other cadences they start on year 5. In the \texttt{baseline} cadence, some fields are observed more than 2 times within the same night; the \texttt{suppress repeats} family was proposed to redistribute these additional visits into different nights. For this family, a basis function is added to the Rubin scheduler algorithm, and is characterized by a suppression factor that indicates how strong the influence on the scheduler is.

\subsubsection{Basline Cadence Family}

The difference between the previous strategies and the v2.0 run is that the WFD survey now includes a low-dust-extinction area and a Galactic Plane extension, which increases the WFD footprint by 15\% \citep{lsst_pstn053} . Versions 2.0 and v2.1 feature a survey footprint with expanded dust-free area. The WFD area-level visits in the Galactic Bulge and Magellanic Clouds, coverage of the Northern Ecliptic Spur, South Celestial Pole, and remainder of the Galactic Plane is maintained, at lower levels \citep{lsst_pstn055}. 
Filter balance is modified in different areas of the sky. 

Following, we focus on the evolution of the \texttt{baseline} cadence family. The v2.99 cadences are drafts of the v3.0 survey strategy which includes recommendations from the Survey Cadence Optimization Committee (SCOC) Phase 1 and Phase 2 (\citep{lsst_pstn053}, \citep{lsst_pstn055}). The v3.0 survey strategy expended the WFD survey footprint to include the Virgo Cluster and enhanced coverage of the Galactic Plane. The time spent in deep drilling fields increased to 6.5\% of survey time and the $u$-band visits improved through a change to a single 30 second exposure. This version also introduced the \texttt{triplet} survey mode on top of a rolling weight which added a third visit every 6 nights for science with short timescales. %Version 3.1 was unreleased for internal testing.

As we transition to v3.2, we see changes to the \texttt{triplet} survey mode which is now triggered every 3 nights instead of 6. The evolution of the baseline survey cadence from v3.2 to v3.3 involved a significant update with the transition to a new set of throughput curves. This change, which included a new set of mirror coatings, had a notable impact on the survey's observational capabilities. While it led to improved throughput in $g$, $r$, $i$, $z$ and $y$ filters, it also resulted in a decrease in throughput in the $u$-band. This shift in throughput affected the depth of observations, particularly in the $u$-band, with a noticeable drop of about 0.2\,mag. 

Subsequent investigations of other cadences, notably in v3.4, focused on further understanding the effects of changes in $u$-band exposure time and filter balance \citep{Bucar_Bricman_2023}. Version 3.4 saw improvements in masking unavailable sky areas and scheduling scripted surveys like DDFs further from twilight. Through simulations varying $u$-band exposure time, the project aimed to assess the sensitivity of metrics to $u$-band depth and the overall impact of altering u-band depth on survey performance. A series of cadence families from this version were simulated: \texttt{good\_seeing}, \texttt{long\_u}, and \texttt{bluer\_indx}. Although kilonova fade most slowly in the near-IR, running cadence families like \texttt{long\_u} contributed to our understanding of the Rubin Observatory's potential to detect kilonova sooner, as early photometry of these events often occurs in the $u$-band \citep{Smartt_2017}. 

The most recent editions of the v3.5 and v3.6 survey strategies for \texttt{baseline} implemented changes to balance uniformity such as reducing the uniform rolling strategy to 3 cycles from 4. Minor adjustments were made to extend time spent in DDFs, but stayed within previous SCOC recommendations. We observed that changes were made to increase exposure time in the $u$-band to 38 seconds per visit while limiting the number of those visits. Version 3.6 saw a dip in survey efficiency, not particular to KNe, by increasing downtime in Year 1 to simulate more realistic transition into full operation and the effect of ``jerk'' on slew time \citep{lsst_pstn056}. The application of each iteration of these survey strategies informs the optimization and implementation of observational cadences by the LSST by improving transient detection, classification, and photometric accuracy to maximize scientific returns from its time-domain survey \citep{Dal_Tio_2022,Street_2023,Gizis_2022}.
%However, since they were released at the time of writing of this paper, the four v2.99 runs were only included in the first population scenario (second row of Table 1).

\subsection{Simulating Kilonovae Detection Efficiency}
\label{sec:simulation}
 We used metrics specifically designed for analyzing kilonovae such as those represented by the \texttt{KNe\_lc} and \texttt{KNePopMetric} classes, slicers, metric bundles, and various utility functions specific to the MAF. 

The simulation procedure begins by initializing parameters for simulating kilonova events that are user configurable. These include the minimum and maximum distances for kilonova injection ($d_{\text{min}}$ and $d_{\text{max}}$), the number of light curves to generate ($n\_events$), and the chosen cadence name ($runName$). The kilonova parameters for injection ($inj\_params\_list$) include properties such as the mass of the dynamic and wind-driven masses ($M_{dyn}$ and $M_{wind}$) and angles of interest ($\phi$ and $\theta$) that influence the characteristics of the generated light curves. The angles of interest are expanded on in Section \ref{sec:Effects}. 
We take $M_{dyn}$ to be 0.005 $M_{\odot}$ and $M_{wind}$ 0.020 for the simulations in this paper.

Next, the script employs the \texttt{generateKNPopSlicer} function to create a slicer, facilitating the placement of a user specified number of kilonova events ($n\_events$), represented by light curves, using a seed to generate random coordinates across the celestial sphere based on the distances previously specified. 

From here, the script establishes a connection to the Rubin Observatory's operational database (opsdb) that contains the simulation data of the LSST survey cadences. 

A metric is then defined using the \texttt{KNePopMetric} class. This metric calculates the detection efficiency of KNe events in the simulated survey data. This metric object is configured with the kilonova parameters for injection ($inj\_params\_list$) and the list of KNe model files obtained from the \texttt{get\_filename} function.

A metric bundle is defined using the \texttt{metricBundles} module, specifying the metric, slicer, SQL constraints (if any), and options for plotting and summarizing results. This bundle encapsulates the evaluation of the KNe detection metric for each slicer point in the survey database.

The metric bundle group is created and executed, which evaluates the metric for each slicer point. Summary metrics, specifically the mean metric, are defined using the \texttt{MeanMetric} class. The script creates a \texttt{MetricBundleGroup} and runs the metric calculation and analysis for the selected cadence. For our simulation, the \texttt{outputLc} parameter is set to \texttt{False}, indicating that the light curves themselves will not be output as part of the metric calculation. The code specifies an SQL query (as an empty string SQL) to filter the data from the database. This can be used to select specific subsets of the data for analysis, but in this case, no filtering is applied. 

\begin{figure}[hbt!]
\centering
\includegraphics[scale=0.5]{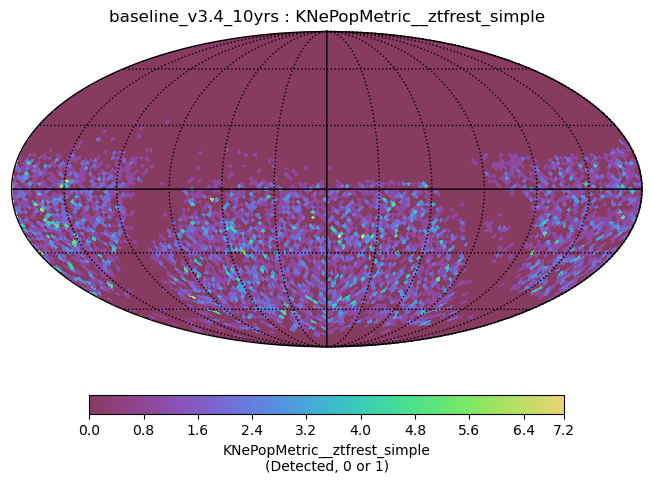}
    \caption{One of the light curve plots output by the efficiency script for the seven different metrics referred to in Section \ref{sec:metrics}. This is an example of \texttt{KNePopMetric\_ztfrest\_simple} sky map.}
    \label{fig:lightcurvesss}
\end{figure}

After the metric analysis is complete, the code uses the \texttt{plotAll} function to generate the light curve plots based on the calculated metric values as seen in Fig.~\ref{fig:lightcurvesss}. These plots visualize the spatial distribution of kilonovae detection efficiency across the celestial sphere. Finally, the code writes the collected results (efficiency values) to the output file. For the cadence specified, the script records the minimum and maximum distances, the metric name, the number of detected kilonovae, the total number of injected KNe, and the calculated efficiency. Efficiency for each cadence is calculated by dividing the number of KNe detected by the \texttt{OpSim} over the total number of kilonova injected (as seen in the second column of Table \ref{tab:table1}). 

\subsection{Time Gap (Tgap) Parameter Study}
We also evaluate the temporal characteristics of simulation cadences by examining the frequency of observation gaps between return visits to the same area in the sky using the T-gap Metric \citep{Bellm_2022}. Our use of this metric aims to uncover any potential correlation between the time intervals separating these return visits and the efficacy of detecting kilonovae. Shorter revisit intervals are valuable for capturing rapid brightness variations, aiding in the identification of fast-evolving transients, like kilonovae and early supernova light curves. Conversely, longer revisit intervals provide the opportunity to cover more of the sky, balancing transient discovery with overall survey efficiency.

Two metrics, \texttt{TgapsMetric} and \texttt{TgapsPercentMetric}, are employed to analyze time gaps between observations. The \texttt{TgapsMetric} calculates the distribution of time gaps between consecutive observations, providing insights into the frequency and duration of observation gaps. The other metric, \texttt{TgapsPercentMetric}, offers a different perspective by calculating the percentage of time spent in observation gaps relative to the total observation duration. This metric offers a measure of how efficient observing is and what amount of time gaps or time to return to a specific spot in the sky is within a specific cadence, indicating the proportion of time during which no observations are made.

These metrics are applied within a loop of cadence names and families, where the script examines if a cadence contains the substring \texttt{`Tgaps\_'}. Upon encountering such metrics, the \texttt{make\_hist} function is invoked, tailored to the specific metric, with parameters including the metric values, a slicer object, and histogram bin specifications. The function then returns histograms showing the return time between observations, as in Fig.~\ref{fig:returntime1}.

\section{Results}
\subsection{Filters}
\label{sec:filters}
We simulated event detection using the fiducial baseline strategy (\texttt{baseline\_nexp2\_v1.7.1}) of the Rubin Observatory, where observations were conducted using only the six photometric filters (u, g, r, i, z, y). These simulations allowed us to analyze how specific filter selections impact our results and quantitatively determine the most effective filter for kilonova detection. For instance, in Fig.~\ref{fig:cadencecomparison}, bars labeled with prefixes such as 'g' or 'i' indicate that the observations were carried out using the 'g' or 'i' filters, respectively. Recovery efficiencies obtained from these simulated cadences, which involved 500,000 injected light curves, are shown in Fig.~\ref{fig:cadencecomparison}. In these simulations, we employed the GW170817-like kilonova model, using a luminosity distance range of 10 -- 600,Mpc (as detailed in the first row of Table \ref{tab:table1}). 

In our criteria, we defined red
to near IR filters (i, z, y) as ``\_red'' and bluer filters (u, g, r) as ``\_blue''. In our results, all the metrics finishing with ``\_red'' perform better (have higher efficiencies) when red to near-infrared bands (i, z, y) are used in the simulation, and those finishing~with ``\_blue'' performed better when bluer bands (u, g, r) are used, consistent with expectations for these metrics. Moreover, it is important to highlight the fact that the \texttt{ztfrest\_simple} metric covered the criteria used in this study (focused on KNe discovery), whereas the \texttt{multi\_detect} metrics are normally used for simple transient discovery and, in some cases, might be sufficient to use for KNe discovery only in the case of a GW- or GRB-trigger follow-up detection. Thus, when determining how KNe detection rates vary with each filter, we more closely examine the results found with the
\texttt{ztfrest\_simple} metric.

Detection rates show that $i$-band and $r$-band dominated
surveys yield the most KNe detections. The $i$-band-dominated cadence
received the highest number of \texttt{ztfrest\_simple} detections and
consequently the highest recovery efficiency with an approximately 6\%
increase from the original \texttt{baseline} cadence, and its \texttt{multi\_detect}
metric results show a 7\% increase in detection rates from the baseline
cadence. The $r$-band-dominated cadence gave a slightly lower rate than the
$i$-band in \texttt{ztfrest\_simple} but had a similar
\textasciitilde6\% increase from the baseline cadence. The two ($r$-band and $i$-band) filters
also outperformed the rest of the filters, as well as the baseline
cadence, when both \texttt{multi\_detect} and \texttt{ztfrest\_simple}
metrics prefixed and suffixed with ``\emph{\_blue}'' for r-band
observations and ``\emph{\_red}'' for i-band observations were
considered, overall indicating that they are the optimal filters for
transient and KNe discoveries.

We initially hypothesized that red bands (i, z, y) would outperform blue bands (u, g, r) in the \texttt{ztfrest\_simple}
results, as KNe tend to fade slower in these bands. The $i$-band did have the most detections, but the $z$-band and the $y$-band did not perform as well. The z-band cadence performed worse than
the baseline cadence in both \texttt{multi\_detect} and
\texttt{ztfrest\_simple} metrics but outperformed the $g$-band cadence by
about 5\% in the \texttt{ztfrest\_simple} results. The $r$-band cadence outperformed the $z$-band consistently. The $y$-band cadence got
significantly fewer detections than the baseline cadence and all other
filters in most of the results; the \texttt{ztfrest\_simple} metric made
fewer detections than most blue-band observations and performed
\textasciitilde74\% worse than the baseline cadence. 

This is very likely a result of the interplay between KNe brightness in the redder bands and the relatively lesser depth in the reddest bands. Specifically, while kilonovae tend to be brighter in the 'i', 'z', and 'y' bands due to their red spectral energy distribution, the LSST's limiting magnitudes indicate that these bands have shallower depth compared to 'g' and 'r' bands. Over the ten-year survey, the co-added 5-sigma limiting magnitudes for 'z' and 'y' bands are expected to reach only 25.6 and 24.8, respectively, whereas the 'g' and 'r' bands reach 27.0 \citep{2009arXiv0912.0201L}, making it more challenging to detect fainter kilonovae in the reddest bands.

\subsection{v1.7 - v2.1 LSST families}

\begin{figure*}
\centering
\includegraphics[scale=0.27]{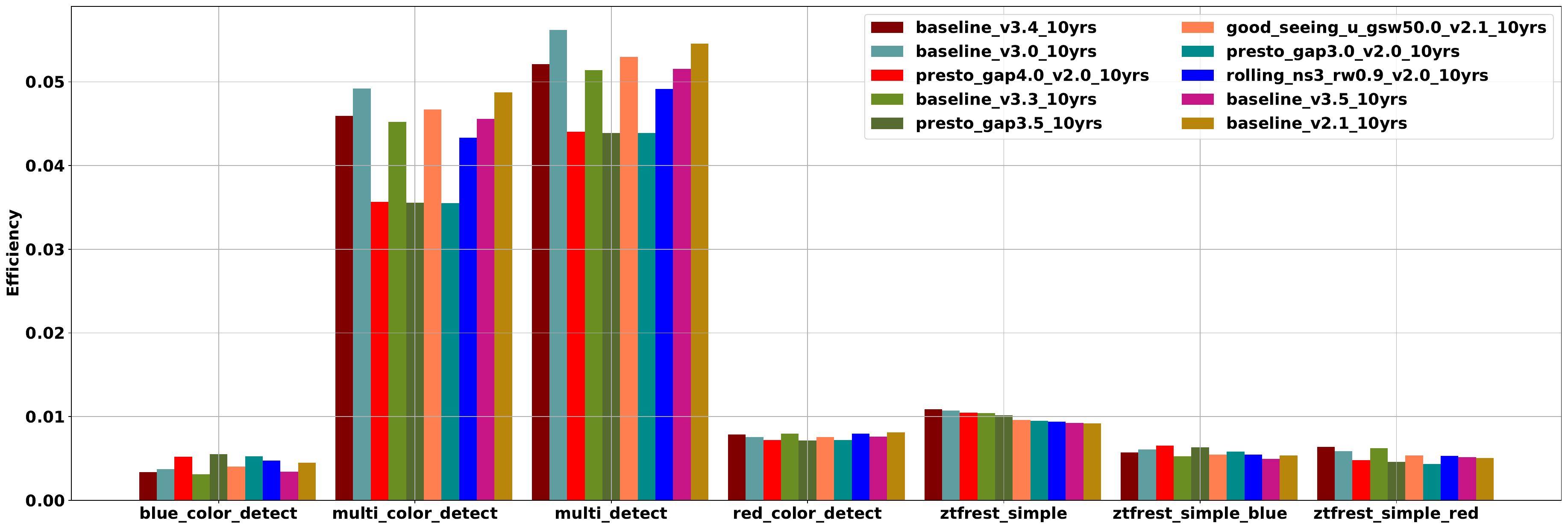}
\caption{Recovery efficiencies produced in simulations where 500,000 total light curves were injected, ranging from $10 ~Mpc - 600~ Mpc$. The data is organized by seven different metrics. The ten cadences displayed are the top 10 for the \texttt{ztfrest\_simple} metric, which imposes the most stringent criteria for kilonova detection. The cadences are arranged in descending order of efficiency within this metric.}
\centering
\label{fig:top10kilonovaerecovery}
\end{figure*}

We performed the same single-model study (see \ref{sec:filters}) on existing LSST cadence families. Fig.~\ref{fig:top10kilonovaerecovery} showcases the top 10 kilonova recovery efficiencies across 29 cadences, encompassing \texttt{baseline} through v3.6, \texttt{presto\_gap} with 3 and 4 hour gaps, \texttt{rolling} cadences, \texttt{noroll}, and \texttt{retro} cadences. In this figure, we use the single model parameters with a luminosity distance range of 10 -- 600 Mpc (first row of Table \ref{tab:table1}). These cadences were selected based on the findings in Fig.~\ref{fig:cadencecomparison}. 

Each metric in Fig.~\ref{fig:top10kilonovaerecovery} reveals a distinct pattern of efficacy. The \texttt{blue\_color\_detect} metric demonstrated the lowest performance, which aligns with expectations since kilonovae are most frequently detected in the red or near-IR wavelengths rather than in blue. However, when a rise/fall rate was applied to the blue filters, as in \texttt{ztfrest\_simple\_blue}, the number of detected kilonovae increased slightly. The cadences that performed best varied across these metrics. Notably, for metrics restricted to blue filters (u, g, r), \texttt{presto\_gap4.0\_v2.0} and \texttt{presto\_gap3.5\_v2.0} outperformed the others.

For red-specific metrics, such as \texttt{red\_color\_detect} and \texttt{ztfrest\_simple\_red}, the cadences demonstrate higher efficiency compared to their blue counterparts. Among the color-specific metrics, \texttt{red\_color\_detect} performs the best, with less variability in efficacy across different cadences. In contrast, other metrics exhibit greater differences in performance across cadences. The two \texttt{ztfrest} color metrics, \texttt{ztfrest\_simple\_red} and its blue counterpart, perform similarly, though with varying levels of success depending on the cadence. For \texttt{ztfrest\_simple\_red}, the \texttt{baseline\_v3.4\_10yrs} cadence detects the highest number of KNe, consistent with results from \texttt{ztfrest\_simple}.

All color-specific metrics did not surpass a $1\%$ efficiency rate. In contrast, metrics like \texttt{multi\_color\_detect} and \texttt{multi\_detect}, which only require two detections, show higher efficiency values across all cadences, reaching approximately \textasciitilde$4.5\%$, with the \texttt{rolling} and \texttt{presto\_gap} families performing particularly well. However, these metrics are not stringent enough for kilonova identification, as they do not distinguish fast-fading transients from other variables. Tables \ref{tab:efficient1} and \ref{tab:efficient2} summarize efficiency results across all cadences for a range of transient detection metrics, illustrating these trends. The more selective \texttt{ztfrest\_simple} metric was designed to impose stricter criteria for kilonova identification.

We find that, with the \texttt{ztfrest\_simple} metric, the \texttt{baseline\_v2.1} cadence has an approximate 2.3\% increase in number of detections from the v2.0 release (\ref{fig:evolutionbaseline}). Cadences that bridge the v1.7 to v2.0 differences, like \texttt{baseline\_retrofoot} and \texttt{retro\_baseline}, who did not perform highly, decrease in number of detections by about 1.4\% and 4.7\%, respectively.  However, despite the lack of a rolling weight in \texttt{retro\_baseline}, it is more efficient than \texttt{noroll}. This may be due to the old footprint placing more visits per pointing into the low-dust WFD (see \ref{sec:cadences}). 

Looking at the \texttt{rolling} cadence results in Fig.~\ref{fig:top10kilonovaerecovery}, the \texttt{ztfrest\_simple} metric, \texttt{rolling\_ns3\_rw0.9\_v2.0} outperforms the \texttt{baseline\_v2.1} by about 2.6\%, indicating that high rolling strength may be favorable for kilonovae detections. The \texttt{rolling\_ns2\_rw0.9\_v2.0} cadence has an identical observation schedule to the \texttt{baseline\_2.0} and thus yields similar efficiencies. This indicates a strong correlation between rolling weight, return time patterns and kilonova detection efficiency. The list of efficiency values can be seen in Tables \ref{tab:efficient1} and \ref{tab:efficient2}. 

The \texttt{noroll} cadence also makes less detections than the baseline v2.0 and v2.1 cadences with the \texttt{ztfrest\_simple} metric, with a decrease from the baseline 2.0 release of approximately 13\%, another indication that the presence of a rolling cadence improves kilonova detections. 

The final family studied in these runs and the second best cadence family shown in Fig.~\ref{fig:top10kilonovaerecovery} is \texttt{presto\_gap}. The importance of intra-night return visits is widely discussed in conversations about maximizing fast transient detections. In the \texttt{presto\_gap} cadence family, for \texttt{ztfrest\_simple}, detections improve as the time between the first pair of visits and the third one increases. This time gap parameter is studied in depth in Section \ref{sec:return}. A 1.5 hour gap, indicated by cadence \texttt{presto\_gap\_1.5} \emph{under-performs} compared to the \texttt{baseline\_v2.0}, which has a single pair of visits separated by 30 minutes, by approximately 35.1\%. A 2.0 and 2.5 hour gap under-performs by 12.5\% and 4.9\%, respectively. Subsequently, as the gaps are increased by 30 minutes from a 3 hour gap, the number of detections made in the \texttt{ztfrest\_simple} metric outperforms \texttt{baseline\_v2.0} and increases from \textasciitilde5\% to \textasciitilde16\%. Overall, we see that the \texttt{presto\_gap} cadences with the larger gap between visits make more detections of all v2.0 cadences in the \texttt{ztfrest\_simple} metric, indicating that kilonova detections are improved with three visits per night and at larger time intervals between visits.

The v1.7 – v2.1 cadence analysis highlighted strategies that improved kilonova detectability but was limited to a GW170817-like model. To generalize, we expanded to a diverse KNe population (Section \ref{sec:knepop}), varying ejecta properties and viewing angles. These findings informed the v3.0 and later cadence selections (Section \ref{sec:v3.0}), which incorporated optimizations for broader transient detectability.

\subsection{KNe Population}
\label{sec:knepop}

The observed optical counterpart of the GW170817 event allowed the research community to constrain the model parameters of its associated KNe \citep{bulla2019possis}. Even though this is, to date, the only confident KNe detection with a gravitational-wave counterpart known, some studies have attempted to predict the diversity of signals from real KNe populations \citep{Setzer:2023} and the likelihood of detection due to varying parameters \citep{pillas2025limitsejectamasssearch}. For our second set of simulations we injected three types of KNe populations using the parameters showed in the second, third and fourth rows of Table \ref{tab:table1}. This aims at considering the effect of a diversity of KNe signals in the KNe detection efficiencies for the LSST survey.

\begin{figure*}[t]
\centering
\includegraphics[scale=0.35]{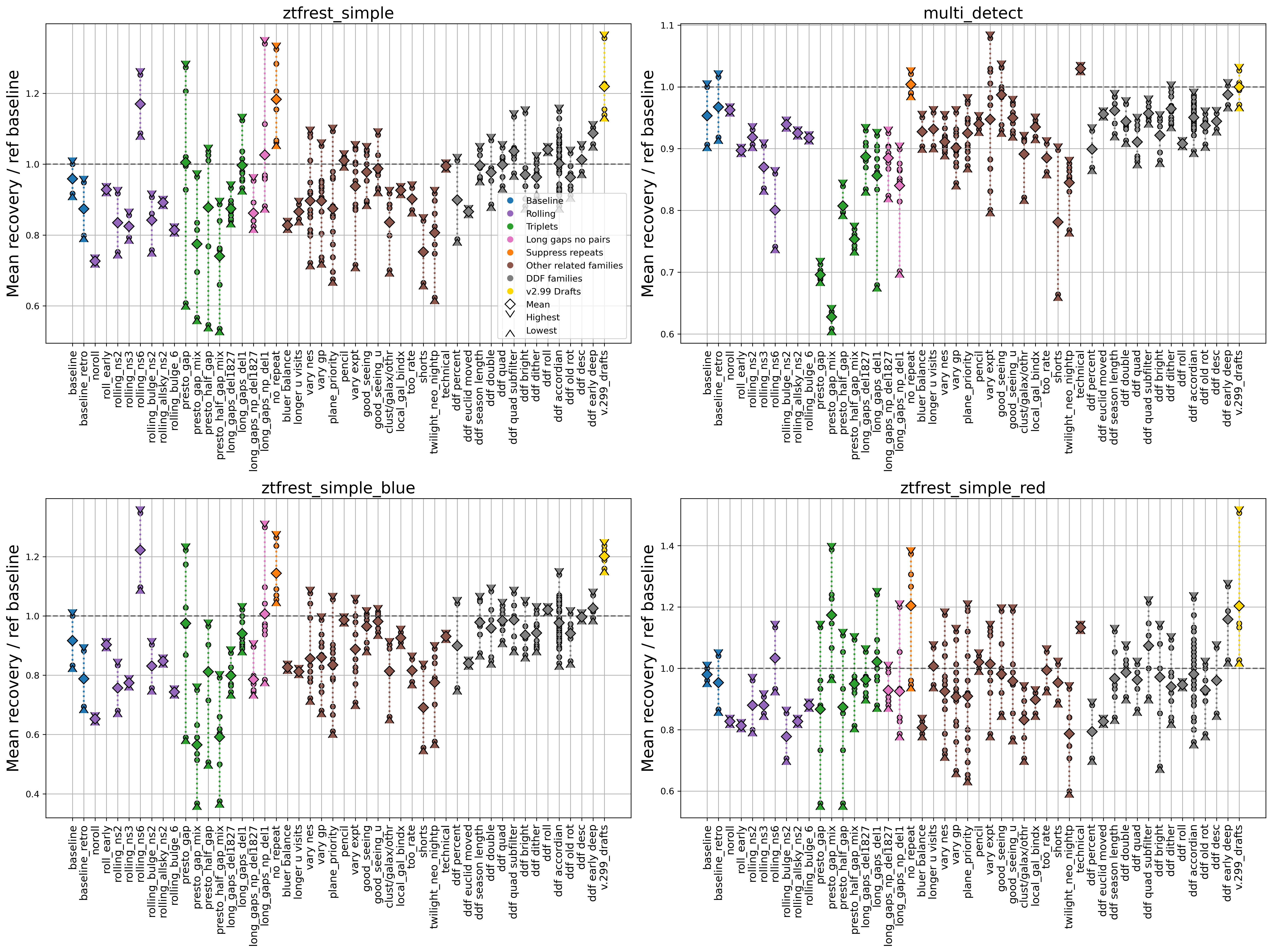}
\caption{Recovery fractions of the v2.0, v2.1 and v2.99 runs normalized using the recovery fraction of the \texttt{baseline\_v2.1} run. The runs that are variations of the same cadence are grouped vertically, and grouped by color all the cadences within the same family, except for the brown and gray colors that group more than one family. Diamonds indicate the mean of the recovery fraction within the cadence while the triangles the maximum and minimum efficiencies. All ztfrest metrics are considered (KNe identification) together with the simplest transient detection metric (\texttt{multi\_detect}). These simulations were obtained using the parameters presented in the second row of Table \ref{tab:table1}.}
\centering
\label{fig:allcadencescatter}
\end{figure*}

\begin{figure*}[t]
\centering
\includegraphics[scale=0.425]{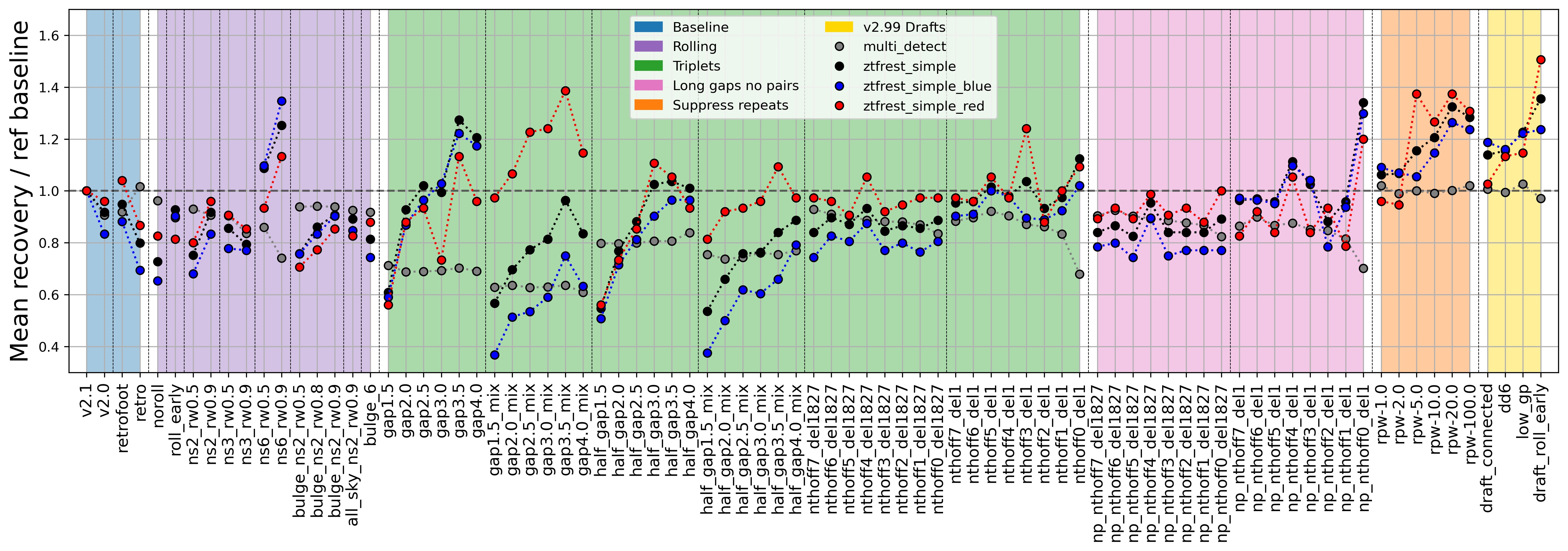}
\caption{Recovery fractions of the \texttt{Baseline}, \texttt{Rolling}, \texttt{Triplets}, \texttt{Long gaps no pairs}, \texttt{Suppress repeats} and \texttt{v2.99 Drafts} families normalized using the recovery fraction of the \texttt{baseline\_v2.1} run. The runs that are variations of the same cadence are connected with dotted lines, except when the cadence only has one run. All ztfrest metrics are considered (KNe identification) together with the simplest transient detection metric (\texttt{multi\_detect}). These simulations were obtained using the parameters presented in the second row of Table \ref{tab:table1}.}
\centering
\label{fig:allcadencemainfamilies}
\end{figure*}

In Fig.~\ref{fig:allcadencescatter}, we present normalized efficiencies of all v2.0 and v2.1 runs, in addition to the v2.99 runs, for a population of $5\times10^{5}$ KNe injected between 10 -- 1400\,Mpc, spanning optimistic, pessimistic and GW170817 ejecta masses, 11 different viewing angles $\theta$, and fixed to an angle $\phi=30$ (see second row of Table \ref{tab:table1}). From here on, our focus is on KNe serendipitous characterization (we use ztfrest-type metrics) but we include results for the \texttt{multi\_detect} metric for comparison with a transient detection metric. The first five colors indicate the cadences within the same family. The gray color indicates all DDF-type families, the brown color indicates the rest of v2.0 and v2.1 families and the yellow color indicates the v2.99 cadences. We focus our attention in the first five families since they have cadence variations traditionally optimal for transient science \citep{Gris_2023}. However the rest of families will be briefly mentioned. 

Focusing our attention on the v2.0 and v2.1 cadences, from Fig.~\ref{fig:allcadencescatter}, we see how the efficiency of the current v2.1 baseline cadence is above average for the \texttt{ztfrest\_simple}, \texttt{ztfrest\_simple\_blue} and \texttt{multi\_detect} metrics. This reflects the suggestions of the transient community for the development of the current baseline cadence. However the baseline cadence is still under average for the \texttt{ztfrest\_simple\_red} metric. This is not favorable for KNe detections because \cite{Andreoni_2021} found that red-band observations are better at finding nearby KNe. The difference between the baseline and the best performing runs is around 30\% for the \texttt{ztfrest\_simple} and \texttt{ztfrest\_simple\_blue} metric, while it is only 10\% for the \texttt{multi\_detect} metric. The 40\% difference between the baseline and the best performing run in the \texttt{ztfrest\_simple\_red} metric is the highest within all 4 metrics.

Again for the v2.0 and v2.1 cadences, for each of the \texttt{rolling}, \texttt{triplets} and \texttt{long gaps no pairs} families, there is one cadence that consistently has runs among the most efficient for each of the ztfrest-type metrics: \texttt{rolling\_ns6}, \texttt{presto\_gap} and \texttt{long\_gaps\_np\_del1}. Conversely the same set of cadences are among the worst for the \texttt{multi\_detect} metric. The \texttt{presto\_gap} cadence, depending on the specific value of its variation, can be within the best or the worst performing cadences for each of the ztfrest-type metrics, with a dispersion within 60\% and 70\% of efficiency. Most of the cadences of the \texttt{triplets} family are below the baseline efficiency for the simple and blue ztfrest metrics except for the red ztfrest metric. We notice that the only cadence that is consistently within the best performing cadences for all four metrics is the \texttt{no\_repeat} cadence from the \texttt{suppress repeats} family. 

%The gray-colored cadences (DDF-type) and brown-colored cadences have maximum efficiencies around 10\% above baseline for the \texttt{ztfrest\_simple} and the \texttt{ztfrest\_simple\_blue} metrics, while for the \texttt{ztfrest\_simple\_red} the maximum above baseline is around 20\%. 

The gray-colored cadences (DDF-type) and brown-colored cadences have maximum efficiencies around 10-20\% above baseline depending on the ztfrest metric. The draft cadences of the v3.0 simulations (in yellow) are within the most efficient cadences for all presented metrics. These for the \texttt{ztfrest\_simple} and \texttt{ztfrest\_simple\_red} metrics present high dispersion but at the same time they have a run that is the best performing of all plotted runs, while for the \texttt{ztfrest\_simple\_red} and \texttt{multi\_detect} the dispersion is smaller but despite their high efficiencies they do not have the most efficient run.

Fig.~\ref{fig:allcadencemainfamilies} shows the normalized efficiencies of individual runs for the cadences within the \texttt{baseline}, \texttt{rolling}, \texttt{triplets}, \texttt{long gaps no pairs} and \texttt{suppres repeats} families, in addition to the v2.99 runs, for the ztfrest-type metrics and the \texttt{multi\_detect} metric. We see that the most recent baseline cadence (the v2.1 baseline) is the best performing run within the \texttt{baseline} family for KNe characterization. We can see that for this type of science, the baseline cadence has been improved consistently in the direction of the new LSST baseline cadence. 

Regarding the \texttt{rolling} family, the \texttt{noroll} run exhibits the poorest performance in all ztfrest-type metrics, which supports the implementation of rolling cadences in the baseline strategy. Comparing the \texttt{noroll} run with the baseline cadences, we see that the inclusion of the 2-region rolling cadence added in the v2.0 and v2.1 baselines improve the efficiencies for KNe source characterization by about  20\% and 30\%, respectively. We see a direct correlation between the number of rolling bands and the KNe efficiencies for the \texttt{ns2} and \texttt{ns6} cadences, where we also see that the greater the rolling strength, the greater the efficiency per metric is (not valid for the \texttt{ns3}). The best performing cadence within the \texttt{rolling} family is the \texttt{ns6} cadence (as mentioned earlier), where the \texttt{ns6\_rw0.9} run can have an increase between 10\% and 30\% above the baseline, however we should note that unfortunately this particular cadence is highly susceptible to poor weather seasons.   

% It is not a surprise that the runs with rolls in the bulge are not better than the baseline cadence given the extragalactic nature of the KNe.

The best performing cadence within the \texttt{triplets} family is the \texttt{presto\_gap}, specifically for ztfrest-type metrics. A direct correlation between efficiency per metric and the separation to the third visit is observed  until a critical 3.5 hours threshold, beyond which efficiency drops. The \texttt{presto\_gap3.5\_v2.0} run shows an increment above baseline of between 15-25\% for the ztfrest metrics. Similarly, the \texttt{gap\_mix} cadence reflects this correlation, but only the \texttt{ztfrest\_simple\_red} metric improves over the baseline. This implies that, for KNe characterizations, closer filters pairs (like g+r in \texttt{presto\_gap}) are more efficient than spaced pairs (e.g., g+i in \texttt{presto\_gap\_mix}), diverging from the findings of \cite{Bianco:2019} favoring pairs with greater wavelength separation  for fast transients in general. The \texttt{presto} \texttt{half\_gap} and \texttt{half\_gap\_mix} cadences mirror the behavior of their \texttt{gap} and \texttt{gap\_mix} cadences counterpart, but with lower efficiencies, indicating that triplets all the time are preferred over triplets half the time. From the \texttt{nthoff\_del1827} and \texttt{nthoff\_del1} cadences in Fig.~\ref{fig:allcadencemainfamilies} we see that having even longer gaps at varying frequencies does not produce an improvement over the baseline for all 4 metrics (only \texttt{long\_gaps\_nigthsoff0\_delayed-1\_v2.0} run produces a small improvement). This is consistent with what was mentioned about the decrease in efficiency after a critical value for the \texttt{presto\_gap} cadence, consistent in turn with the short-lived nature of KNe.  

The behavior and efficiencies of the \texttt{long gaps no pairs} family resemble those of the \texttt{long\_gaps} runs in the \texttt{triplets} family. However when the long gap pair is taken every night and from the first year of the survey (\texttt{long\_gaps\_np\_nigthsoff0\_delayed-1\_v2.0} run) the improvement in efficiency is around 20-30\% for the ztfrest metrics, surpassing most \texttt{triplets} cadences. There is a small peak in efficiency when the long gap pair is taken every 4 nights.  

The \texttt{suppress repeats} is the only family with most of its runs above the baseline level. The peaks in efficiency, between 25 and 35\% improvement for the ztfrest-type metrics, are obtained for a suppression factor of 20. This implies that forcing the Rubin scheduler algorithm to limit its visits to the same pointings to no more than twice per night is the cadence modification with the most positive impact to KNe characterizations. 

Most of the v2.99 runs for the ztfrest-type metrics present an improvement above baseline between 10\% to 20\%, except for the \texttt{draft\_connected} with the \texttt{ztfrest\_simple\_red} for which the efficiency is barely more efficient than the v2.1 baseline. The \texttt{roll\_early\_v2.99} run goes even to higher efficiencies, for the \texttt{ztfrest\_simple} and \texttt{ztfrest\_simple\_blue} metrics the improvement in efficiency is between 20\% to 40\%, while for the \texttt{ztfrest\_simple\_red} the improvement in efficiency is around 50\%. 

From Fig.~\ref{fig:allcadencemainfamilies}, we can also see that the \texttt{ztfrest\_simple\_red} metric breaks the trends of the rest of ztfrest metrics for specific variations within a cadence (e.g. \texttt{preto\_gap3.0}, \texttt{presto\_half\_gap3.5\_mix}, \texttt{nightsoff3\_delayed-1}, and \texttt{no\_repeat\_rpw-5.0} runs). This could mean that, given the red nature of KNe, the characterization efficiency is more sensitive to changes in metrics based on red filters than to changes in metrics based on other colors. 

The families (and their cadences) that are not mainly related with transient science are presented as an appendix in Fig.~\ref{fig:allcadenceappfamilies}, in the bottom part the DDF-related families and the rest of the families in the upper part. Here we will only contribute to the idea that optimizing one science case can adversely impact other science cases. This can be evidenced with cadences within the \texttt{mini-surveys} and \texttt{micro-survey} families, as well as with the \texttt{ddf percent} cadence. An inverse correlation exist between the time allocated to mini- and micro-surveys, as well as DDF fields, and the recovery fraction of KNe. This emphasizes the notion that optimizing the cadence for a specific science case can not be carried out entirely independently of the effects on other science cases, specially if we aim to the optimize the LSST project as a whole.

\subsection{Effects of \texorpdfstring{$\phi$}{phi} and \texorpdfstring{$\theta$}{theta}}
\label{sec:Effects}

\begin{figure*}[t]
\centering
\includegraphics[scale=0.30]{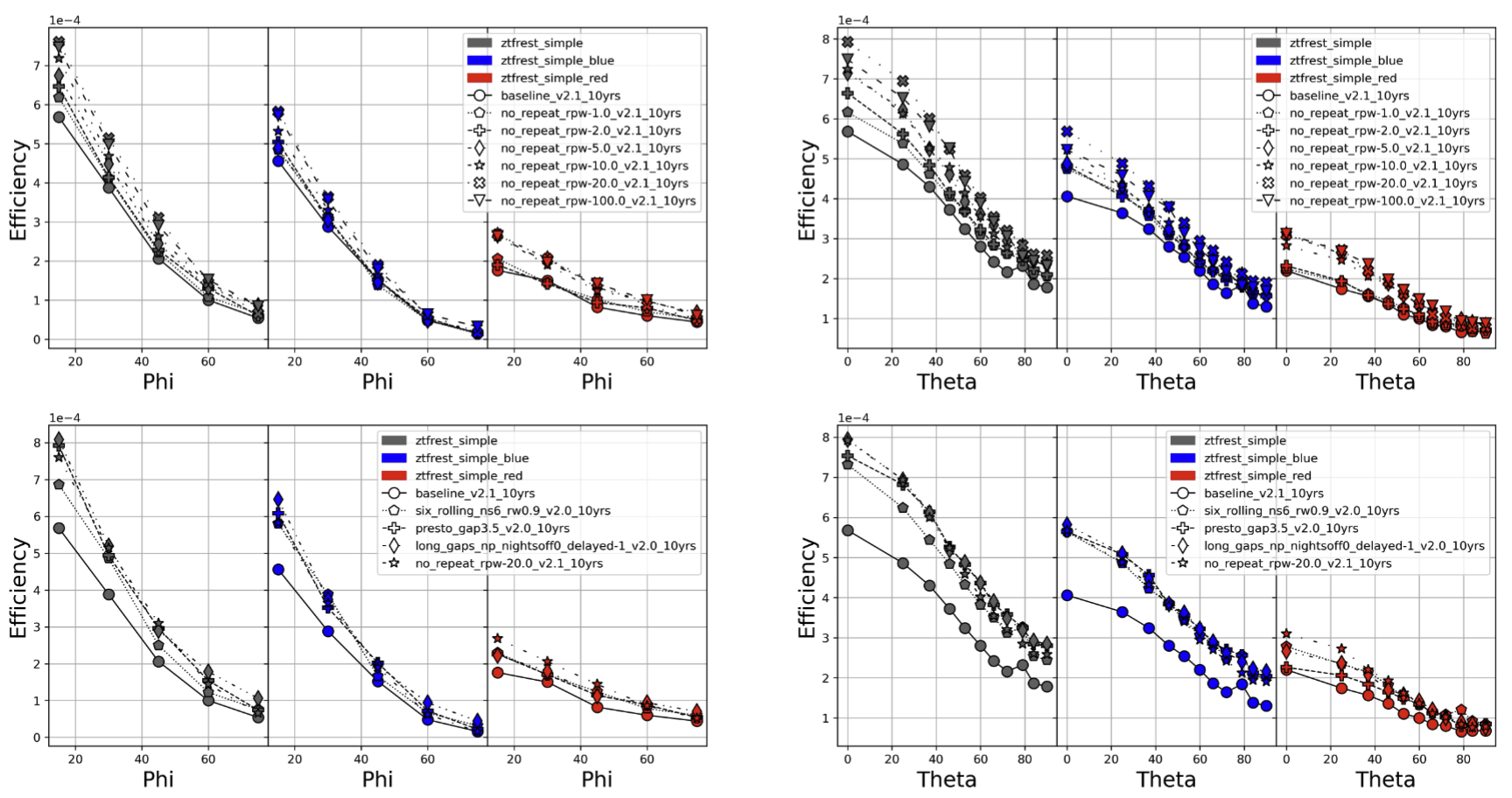}
\caption{Recovery fractions of a sample of the v2.0 and v2.1 runs without normalization (to allow comparison between metrics) as a function of the $\phi$ or $\theta$ angle. The left panels are simulations for a population of KNe fixed to each of the $\phi$ values spanning all available $\theta$ values (see fourth row of Table \ref{tab:table1}). Conversely the right panels of the figure were produced with a population of KNe fixed to each of the $\theta$ values spanning all available $\phi$ values (see third row of Table \ref{tab:table1}). The top panels present the efficiencies for the cadences within the \texttt{suppress repeats} family, while the bottom panels present the efficiencies for the best performing cadences of the \texttt{rolling}, \texttt{triplets}, \texttt{long gaps no pairs} and \texttt{suppress repeats} families. The v2.1 baseline is also included in all plots for comparison.}%
\label{fig:parameterspace}%
\centering
\end{figure*}

As mentioned, the results just presented were obtained after the injection of a population of KNe with the half-opening angle $\phi$ fixed to 30$^\circ$ ($\phi$ of the GW170817 kilonova model). We performed the same set of simulations for the other available $\phi$ angles (fourth row of Table \ref{tab:table1}). We also injected a population of KNe spanning all the available $\phi$ angles but fixed to different viewing angles $\theta$ (third row of Table \ref{tab:table1}). In Fig.~\ref{fig:parameterspace} we plot the non-normalized efficiencies for different cadences, for each of the KNe characterization metrics and for KNe populations fixed to different $\phi$ or $\theta$ values. 

The first row of Fig.~\ref{fig:parameterspace} shows the behavior of the cadences within the \texttt{suppress repeats} family for different $\phi$ and $\theta$ angles. In the upper left panel we see that the 20\% suppression factor cadence, most efficient for a $\phi$ fixed to 30° (recall Fig.~\ref{fig:allcadencemainfamilies}), is the most efficient only for smaller $\phi$ values ($\leq 45^\circ$), while the 100\% suppression factor tends to be more efficient for larger angles. The upper right panel of Fig.~\ref{fig:parameterspace} shows that the 20\% suppression factor is the most efficient for the \texttt{ztfrest\_simple} and \texttt{ztfrest\_simple\_blue} metrics independent of the viewing angle. For the \texttt{ztfrest\_simple\_red} metric, a suppression factor of 100\% is the best performing for most values of $\theta$. 

The second row of Fig.~\ref{fig:parameterspace} evaluates the dependency on $\phi$ and $\theta$ for the best-performing cadences of the transient science families. In the lower-left panel we see that for the \texttt{ztfrest\_simple} and \texttt{ztfrest\_simple\_blue} metrics, the \texttt{long gaps no pairs} run is generally the best performing cadence for the $\phi\leq$ 30$^\circ$ and $\phi\geq$ 60$^\circ$ ranges. For a $\phi$ value of 45° the most efficient cadence is either the \texttt{suppress repeats} or the \texttt{triplets} run. For the \texttt{ztfrest\_simple\_red} metric, the \texttt{suppress repeats} cadence is the best performing for $\phi\leq$ 45$^\circ$, while the \texttt{long gaps no pairs} cadence dominates for $\phi\geq$ 60° values. In the lower right panel, we see that across $\theta$ values both \texttt{triplets} and \texttt{long gaps no pairs} cadences are the most efficient in the \texttt{ztfrest\_simple} and \texttt{ztfrest\_simple\_blue} metrics. For the \texttt{ztfrest\_simple\_red}, the \texttt{suppress repeats} and the \texttt{long gaps no pairs} runs are the best performing cadences for the small and high $\theta$ ranges respectively.

We also observe trends on each side of Fig.~\ref{fig:parameterspace} that are cadence-independent. From the left plots (top and bottom) we observe that for small $\phi$ angles it is more efficient the detection of KNe with blue metrics than red metrics by a factor of $\sim$2. This could be explained because small $\phi$ angles means that the lanthanide-rich (with high opacity) component is small, making the blue component (lanthanide-free, low opacity) of the KNe the dominant. Conversely, when the $\phi$ angle is large, the red and faint component dominate, then red metrics will be slightly more efficient than blue ones, as can be seen from both left-side plots. Additionally, we note that large $\phi$ angles result in less detected KNe compared to the larger absolute number of detected KNe when $\phi$ is small. This can also be explained because at large $\phi$ the faint and red lanthanide-rich component dominates.  

From the right-side plots of Fig.~\ref{fig:parameterspace}, we also note how the KNe characterization efficiency is strongly dependent on the viewing angle $\theta$. When we observe from near the pole (small $\theta$ angles) we would observe better the blue and bright component (lanthanide-free), making the blue strategies more efficient than red ones. For large $\theta$ (i.e. the KNe seen from near the equator), the blue strategies still perform better but the difference is lower. This could be explained because the region of the solid angle not covered by the faint lanthanide-rich component is dominated by the blue and bright emission from the lanthanide-free component. And again, there is a general trend in which large $\theta$ angles imply less detected KNe. 

Finally, we note the effect of varying $\phi$ and $\theta$ angles in the dispersion between different cadences. Comparing the first row of plots with the second row of plots in Fig.~\ref{fig:parameterspace} we see a higher dispersion in efficiencies when both angles are smaller, which is related to observing a dominant blue lanthanide-free region. However this dispersion is higher for runs within the same cadence (as the runs of \texttt{suppress repeats}).

\subsection{v3.0 and Other Cadence Families}
\label{sec:v3.0}

\begin{figure*}
\centering
\includegraphics[scale=0.6]{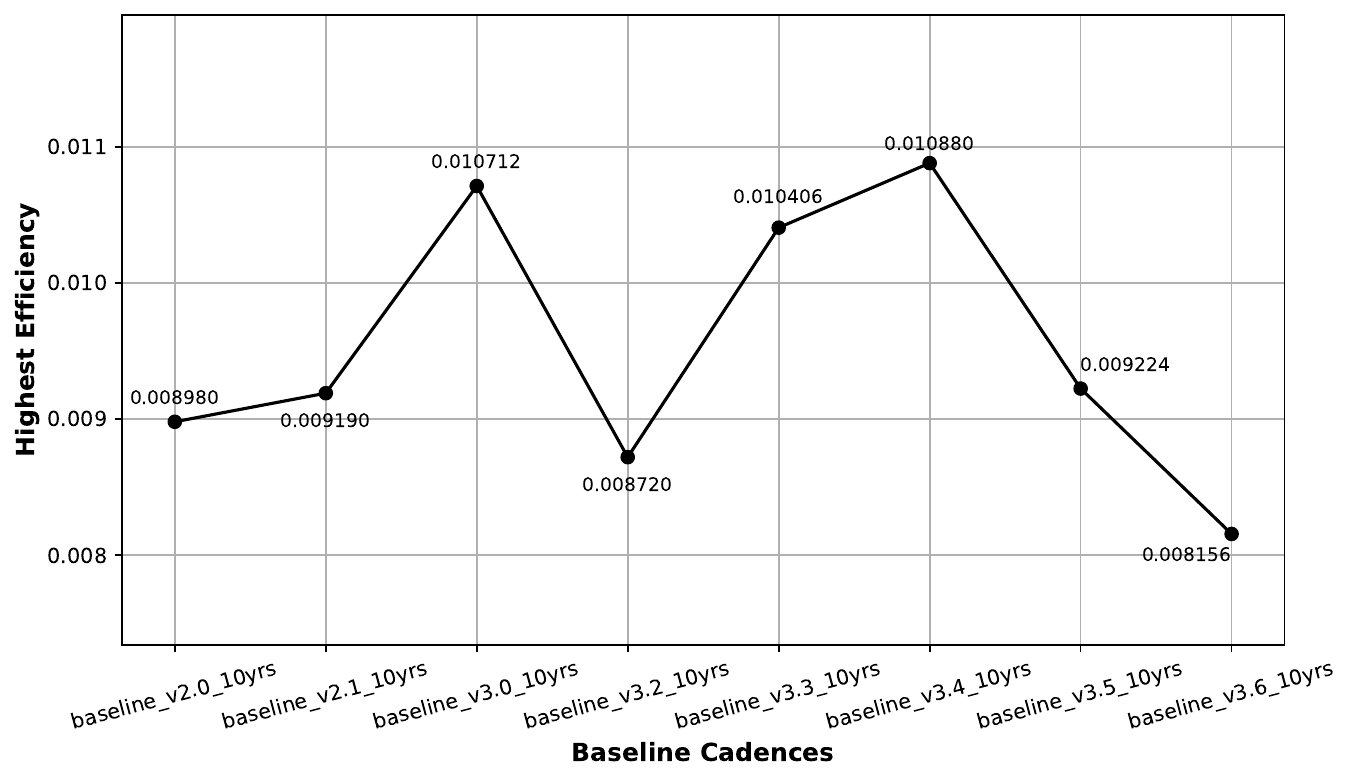}
\caption{Line plot of baseline evolution for single-model cadence efficiencies in simulations where 500,000 total light curves were injected, ranging from
$10 ~Mpc - 600~ Mpc$.}
\centering
\label{fig:evolutionbaseline}
\end{figure*}

Further exploration of the kilonova parameter space revealed a variety of performance variations across filter selections and cadences. Our findings (Section \ref{sec:knepop}) suggest that blue filters are more proficient at detecting KNe with smaller half-opening angles, while red filters prove more effective for higher angles. Consequently, fine-tuning the interval between return visits revealed a correlation with KNe detectability.

In light of these insights, our focus shifted to an evaluation of the latest baseline iterations: \texttt{baseline\_v3.0}, \texttt{baseline\_v3.2}, \texttt{baseline\_v3.3}, \texttt{baseline\_v3.4}, \texttt{baseline\_v3.5}, \texttt{baseline\_v3.6} and novel cadence families exhibiting potential as efficient kilonova detectors. The \texttt{long\_u} visits cadences extend `u' band visit times for enhanced performance, while \texttt{good\_seeing} family simulations incorporate the requirement of three high-quality images per year. Moreover, the \texttt{bluer\_indx} family, characterized by a shift towards bluer wavelengths, was considered in anticipation of varied signals emanating from the diverse simulated KNe populations. While these \texttt{long\_u} and \texttt{bluer\_indx} were not the best at detecting kilonovae, they could provide an opportunity to study which cadences may be best for early detection.

The \texttt{multi\_detect} metric, being the least stringent in terms of transients, demonstrated the most efficient outcome across the spectrum of cadences and detection criteria as can be seen in Fig.~\ref{fig:top10kilonovaerecovery}. In the \texttt{multi\_detect} metric, the \texttt{retro\_baseline\_v2.0} cadence (Table \ref{tab:efficient1} exhibited the highest efficacy at $5.9\%$ and a $9.9\%$ difference over \texttt{baseline\_v2.0}. It was followed by \texttt{noroll\_v2.0}, \texttt{baseline\_v3.0}, and \texttt{baseline\_v2.1}, respectively. We find that the novel cadence families surpass the \texttt{presto\_gap} cadences substantially and rivaling the baseline iterations in terms of detection efficacy. Notably, the \texttt{baseline\_v3.4} cadence exhibited improved efficiency compared to its v3.2 and v3.3 predecessors.

In regards to the color-specific metrics, the newly introduced cadences competed significantly with their established counterparts which provides insight into how varying cadence parameters for specific scenarios can still yield efficient kilonova detection. For \texttt{blue\_color\_detect} and \texttt{ztfrest\_simple\_blue}, the \texttt{presto\_gap} cadences, particularly those with longer intra-night gaps ($\geq 2.5$ hours), yielded the highest efficiency in detecting kilonovae while \texttt{baseline} v3.2, v3.3, v3.6 performed the worst. In red filters, like \texttt{ztfrest\_simple\_red} and \texttt{red\_color\_detect}, \texttt{baseline} v3.4 and v3.3 led in red-band kilonova detection. \texttt{Baseline} v3.2 and v3.6 performed poorly in red-specific metrics. 

However, within more stringent evaluation criteria, the \texttt{ztfrest\_simple} metric, \texttt{baseline\_v3.4} exhibited the highest degree of efficiency, with a $21.2\%$ improvement over \texttt{baseline\_v2.0}. It is closely trailed by \texttt{baseline\_v3.0} with a $19.3\%$ improvement. This was after a sharp decrease in efficacy from \texttt{baseline} v3.0 to v3.2. In v3.2 to v3.4 \texttt{baseline} improves. However, in Fig.~\ref{fig:evolutionbaseline} we see a break in trend and the newest iterations (v3.5 and v3.6) of the \texttt{baseline} cadences have a dramatic drop in efficacy. With the introduction of a realistic jerk on slew time, we believe that this contributes to a decreased efficiency in recent iterations of baseline.  

 In Fig.~\ref{fig:top10kilonovaerecovery}, some of these novel cadences were the top 10 of all cadences in terms of detection efficiency. The \texttt{presto\_gap} family and novel cadence family, \texttt{good\_seeing}, also shows remarkable efficiency within this metric. \texttt{Good\_seeing\_u\_gsw50.0}, which includes an image in the 'u' band, outperformed \texttt{presto\_gap3.0} (with a 3 hour space between triples, see \ref{sec:return}). The \texttt{presto\_gap4.0\_v2.0} cadence had a $16.6\%$ improvement, surpassing \texttt{baseline\_v3.3} and only third to \texttt{baseline\_v3.4}. The \texttt{good\_seeing\_u\_gsw50.0\_v2.1} also showed promise with a $6.9\%$ improvement. The \texttt{long\_u2\_v2.0} cadence demonstrated competitive performance akin to \texttt{baseline\_v2.1}, but was not among the top 10 most efficient. Another of the novel cadences, \texttt{bluer\_indx0\_v2.0}, displayed comparable results to \texttt{baseline\_v2.0} and \texttt{rolling\_ns3\_rw0.5\_v2.0}. Most of the \texttt{baseline} cadence family remained the most efficient cadences in detecting kilonova, only excluding v2.0, v3.2, and v3.6.

\subsection{Return Times}
\label{sec:return}

\begin{figure*}
\centering
\includegraphics[scale=0.4]{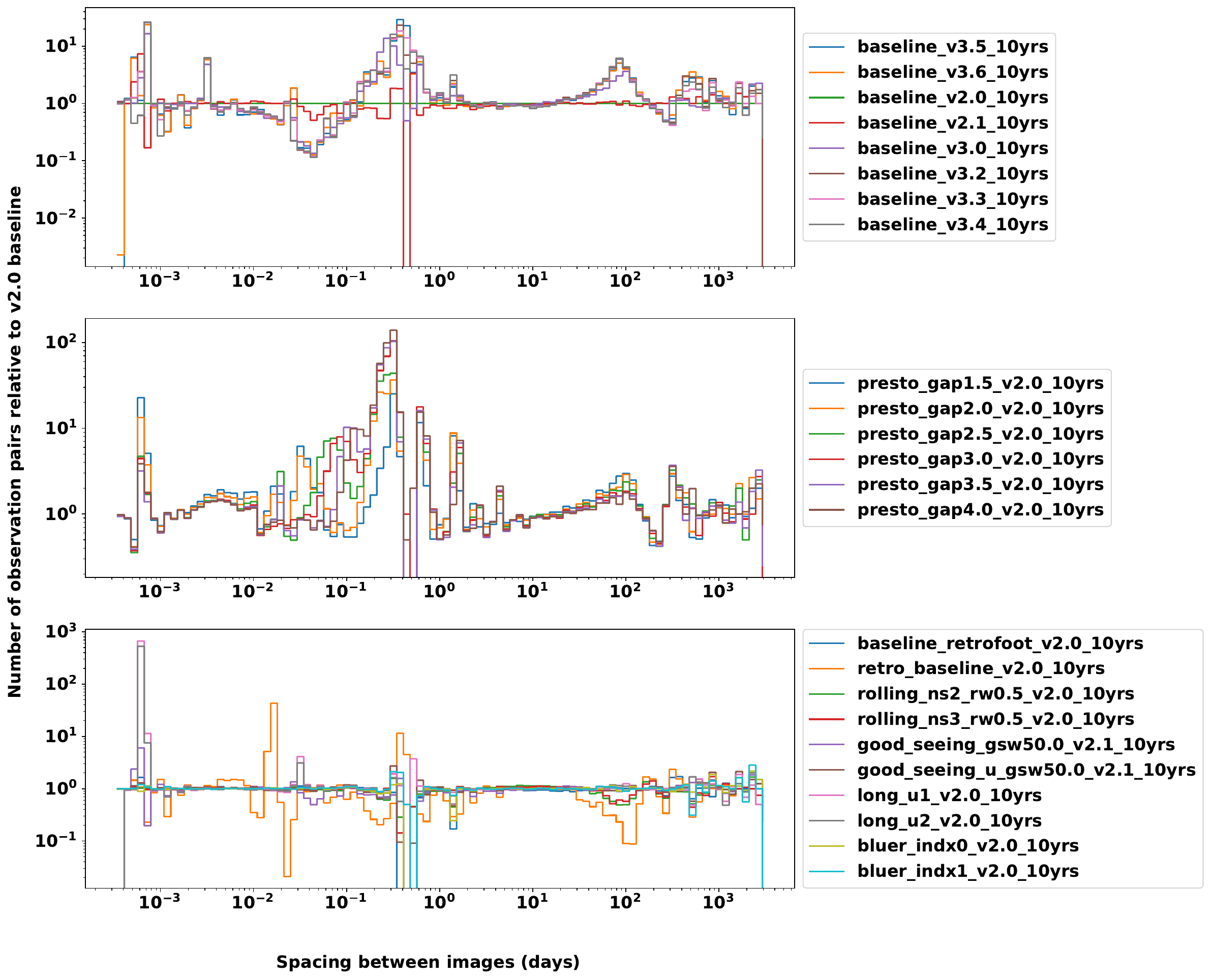}
\caption{Number of Observation Pairs Relative to \texttt{baseline\_v2.0} vs. Spacing Between Images in Days}
\label{fig:returntime1}
\centering
\end{figure*}

For transient detection, the temporal pattern of observations determines which astrophysical phenomena can be observed \citep{Bellm_2022, Kovacevic_2022}. Analyzing transients over a variety of timescales requires logarithmic distributions of visit separations to study how the number of observation pairs varies over a certain period of time.
 
 %Using the MAF \texttt{Tgap} metric: \texttt{sst.sims.maf.metrics.}
 With \texttt{TgapsMetric}, we studied the separation of visit for a range of LSST cadence simulations, as well as cadences used to perform particular analyses in this study \citep{Bellm_2022}. In order to understand the timescales in each cadence family, Fig.~\ref{fig:returntime1} presents these time gap histograms for the listed cadences relative to \texttt{baseline\_v2.0} against the spacing between images as a logarithmic plot. The logarithmic plot shows which cadence has the largest number of observation pairs within a specific time frame. Using representatives from each cadence family from Fig.~\ref{fig:returntime1} and keeping in mind the efficiencies seen in figures \ref{fig:top10kilonovaerecovery} and \ref{fig:evolutionbaseline}, we then compute the median percent of observations probing intra-night and 1 day timescales. Table \ref{tab:tablemedian} summarizes these results, which we produce by extracting the data from the x-axis (the spacing between images in days) and the y-axis (the number of observation pairs relative to v2.0 baseline). For each representative, we multiply the value from the y-axis by 100 and divide over the sum of \texttt{baseline\_v2.0} cadence. 

Fig.~\ref{fig:returntime1} divides cadences by number of returns. The upper and lower plot presents cadences with pairs of visits. The center plot presents cadences that have a third visit. These third visit cadences are part of the triplet family. The \texttt{presto\_gap} plot illustrates the distribution of observation pairs within triplet families, highlighting significant differences in coverage of intra-night and 1-day timescales compared to other cadence families. Notably, there is a high concentration of observation pairs or triplets within the $0$-$1$ hours range in table \ref{tab:tablemedian}. At $0$-$1$ hours or $\sim2 \times 10^{-2}$ days for \texttt{presto\_gap3.0\_v2.0}, the percentage of time spent in observation gaps was slightly lower than the average at $48.2\%$ whereas the other representative cadences were about $10\%$ higher. This means that triplet families return to the same area of the sky less at this time interval. This pattern aligns with LSST's nominal return time of 30 minutes per field. The distribution centered between \(10^0\) and \(10^1\) in Fig.~\ref{fig:returntime1} reflects this analysis, which indicates that the majority of cadences have pairs of visits occurring within the first hour. We note that cadences returning at $0$-$1$ hours are not optimal for detecting kilonova as we want a large enough time between observations to measure decay rate. 

For the non-triplet cadence families, time spent returning decreases dramatically at the $1$-$2$ hours mark ($\sim8.3\times10^{-2}$ days) to an average of $0.8\%$. \texttt{Baseline\_v2.0} is the highest among all baseline iterations at $1.4\%$. However, \texttt{rolling\_ns3\_rw0.9\_v2.0} led with $1.5\%$. This means baselines has an observation pair in $1$-$2$ hour time scale. For the triplets, \texttt{presto\_gap3.0\_v2.0} had a $\sim5\%$ increase of time spent compared to its triplet compatriots at the $1$-$2$ hours timescale.

For $2$-$14$ hours ($\sim8.3 \times 10^{-2}$ to $5.83 \times 10^{-1}$ days), \texttt{presto\_gap4.0\_v2.0} has the most time spent in an observation gap at $13.8\%$. The triplets are spending much more time in observation gaps compared to the non-triplet families. Those with the lowest percentage is \texttt{retro\_baseline\_v2.0} with $0.6\%$ and \texttt{noroll\_v2.0} with $0.9\%$. This makes sense as these cadences have little to no return visits in their sequence. When we look at \texttt{rolling\_ns3\_rw0.9\_v2.0}, we see that the iterations of baseline after v2.1 spend more time in observation gaps at this time scale. This coincides with the rolling weights in the cadence and the trend continues at the next timescale intervals. 

For time spent in observation gaps between 4-8 hours ($\sim1.67 \times 10^{-1}$ to $3 \times 10^{-1}$ days), a significant drop occurs in non-triplet families. This indicates that families like baseline do not spend much time in $4$-$8$ hour gaps. For the triplet cadence families, there’s an increase to $5-6\%$ of time spent at these timescales. An indication that with triplet families there is a greater variety of time between return visits.

Thus far, \texttt{presto\_gap} has been spending the most time in gaps of all the cadences in table \ref{tab:tablemedian}. When we get to the $14$-$38$ hour timescale, we see the percentages of the latest iterations of \texttt{baseline} rise above those of the triplet families. As well, \texttt{presto\_gap} does not vary from the average at the $14$-$38$ hours timescale. This is an indication that with the right time gap between pairs of visits the triplet is not necessarily a dominant factor in detecting kilonova. %The percentage of observation pairs for each representative cadence with either the $1-2$ hour or $23-25$ hour time gap is identical.
\label{sec:return2}

\begin{table*}[]
    \centering
    \begin{tabular}{|l|l|l|l|l|l|l|}
        \hline
        \textbf{OpSim Run (Hours)} &
        \textbf{E} &
        \textbf{0-1} &
        \textbf{1-2} &
        \textbf{2-14} &
        \textbf{4-8} &
        \textbf{14-38} \\ \hline
        baseline\_v2.0\_10yrs                 & 0.008980 & 57.8\% & 1.4\% & 1.1\% & 0.2\% & 9.1\%  \\ \hline
        baseline\_v2.1\_10yrs                 & 0.009190 & 60.4\% & 1.1\% & 1.0\% & 0.1\% & 8.3\%  \\ \hline
        baseline\_v3.0\_10yrs                 & 0.010712 & 54.9\% & 0.4\% & 1.8\% & 0.8\% & 11.9\% \\ \hline
        baseline\_v3.2\_10yrs                 & 0.008720 & 54.6\% & 0.5\% & 2.0\% & 0.6\% & 12.5\% \\ \hline
        baseline\_v3.3\_10yrs                 & 0.010406 & 54.5\% & 0.5\% & 2.0\% & 0.7\% & 12.7\% \\ \hline
        baseline\_v3.4\_10yrs                 & 0.010880 & 55.2\% & 0.5\% & 1.9\% & 0.6\% & 12.6\% \\ \hline
        baseline\_v3.5\_10yrs                 & 0.009224 & 56.7\% & 0.6\% & 1.9\% & 0.6\% & 10.8\% \\ \hline
        baseline\_v3.6\_10yrs                 & 0.008156 & 56.5\% & 0.6\% & 1.8\% & 0.6\% & 10.4\% \\ \hline
        presto\_gap4.0\_v2.0\_10yrs           & 0.010482 & 49.1\% & 1.3\% & 13.8\% & 6.3\% & 9.3\%  \\ \hline
        presto\_gap3.5\_v2.0\_10yrs           & 0.010150 & 48.8\% & 2.1\% & 13.7\% & 5.9\% & 9.2\%  \\ \hline
        good\_seeing\_u\_gsw50.0\_v2.1\_10yrs & 0.009596 & 57.3\% & 1.4\% & 1.1\% & 0.2\% & 9.6\%  \\ \hline
        presto\_gap3.0\_v2.0\_10yrs           & 0.009500 & 48.2\% & 6.3\% & 9.9\% & 5.1\% & 9.0\%  \\ \hline
        rolling\_ns3\_rw0.9\_v2.0\_10yrs      & 0.009422 & 58.9\% & 1.5\% & 1.3\% & 0.2\% & 9.9\%  \\ \hline
        long\_u2\_v2.0\_10yrs                 & 0.008818 & 57.1\% & 1.4\% & 1.1\% & 0.2\% & 9.1\%  \\ \hline
        bluer\_indx0\_v2.0\_10yrs             & 0.008514 & 57.9\% & 1.4\% & 1.1\% & 0.2\% & 8.9\%  \\ \hline
        retro\_baseline\_v2.0\_10yrs          & 0.008556 & 57.5\% & 1.2\% & 0.9\% & 0.1\% & 8.6\%  \\ \hline
        noroll\_v2.0\_10yrs                   & 0.007510 & 56.6\% & 1.2\% & 0.9\% & 0.1\% & 7.8\%  \\ \hline
    \end{tabular}%
    \caption{Median Percent of Observation Probing Intra-Night and 1-Day Timescales. A selection of cadence-family representatives from 0 hours to 38 hours relative to baseline\_v2.0 alongside their respective \texttt{ztfrest\_simple} efficiency (E). This table shows the percentage of time spent in observation gaps.}
    \label{tab:tablemedian}
\end{table*}

Observing the data in Table \ref{tab:tablemedian}, several noteworthy trends emerge for the v3.0 iterations of the \texttt{baseline} and other cadences. From Table \ref{tab:tablemedian}, \texttt{baseline\_v3.3} cadence achieves the highest number of time spent returning (12.7\%). \texttt{Baseline's} most recent predecessors also dominated at this timescale compared to those cadences with a third intra-night visit. Some of the most recent iterations (\texttt{baseline\_v3.4}, \texttt{baseline\_v3.2} and \texttt{baseline\_v3.0}) followed with $12.6\%$, $12.5\%$ and $11.9\%$, respectively. \texttt{Baseline} v3.5 and v3.6 spent less time in gaps than the 3 versions that came before them. 

Notably, \texttt{bluer\_indx0\_v2.0}, \texttt{long\_u2\_v2.0}, and \texttt{good\_seeing\_u\_gsw50.0\_v2.1} showcased competitive results across the board when compared to other cadences, excluding 
\texttt{baseline}. This indicates that these cadences have a balanced temporal distribution of observations which could be lucrative when observing in 'u' or 'g' filters. While not spending the most time returning, these cadences competed with known triplets like \texttt{presto\_gap} for the $14$-$38$ hour timescale.

\texttt{Presto\_gap4.0\_v2.0} and \texttt{presto\_gap3.5\_v2.0} maintained higher median percentages over almost all timescales. This indicates that the time spent returning between triplets is correlated to its efficiency as \texttt{presto\_gap4.0\_v2.0} is the cadence with the third highest ability to detect kilonova (Fig.~\ref{fig:top10kilonovaerecovery}).

To summarize the observed return times, there is more time spent in observations gaps overall in the triplet families than in other cadence strategies in an intra-night timescale. In cases where kilonova detection is important, the use of triplet cadences with higher time between triple return visits will warrant usage. However, on 1-day timescales, we see that the most recent \texttt{baseline} cadences lead with percentage of time. They also lead in detection efficiency. The difference being the inclusion of rolling weights.

\subsection{Realistic KNe Rates}
\label{sec:realistic}

In order to estimate realistic KNe rates, we performed simulations with $1,000,000$ KNe injections ($n_{\text{events}}$). These simulations are based on the binary neutron star (BNS) merger rates derived from the latest gravitational wave catalogs, including the third Gravitational-Wave Transient Catalog (GWTC-3) by the LIGO-Virgo-KAGRA collaboration \citep{theligoscientificcollaboration2022populationmergingcompactbinaries}, which summarizes the gravitational-wave detections made during the first three observing runs (O1, O2, and O3). This catalog includes data for 90 compact binary coalescence (CBC) events, such as binary neutron star (BNS), neutron star-black hole (NSBH), and binary black hole (BBH) mergers. 
The BNS merger rate was used to determine the optimistic ($295.7 \, \text{Gpc}^{-3} \, \text{yr}^{-1}$), pessimistic ($21.6 \, \text{Gpc}^{-3} \, \text{yr}^{-1}$), and median rate ($105.5 \, \text{Gpc}^{-3} \, \text{yr}^{-1}$) of kilonovae. We focused on testing the 10 most efficient cadences of the \texttt{ztfrest\_simple} metric  of the 29 initially simulated found in Fig.~\ref{fig:top10kilonovaerecovery}.

In order to define the space in which BNS mergers could occur, we define:
\[
V_{\text{sector}} = V_{\text{sphere}} \times \frac{A_\text{footprint}}{A_\text{sphere}}
\]
To account for the survey's observational limits, the survey area ($20,283$ square degrees) was compared to the total sky area ($41,252.96$ square degrees), and the volume of the sector within this survey area was calculated. This sector represents the observable volume of space where BNS mergers could be detected by the survey. In this context, $V_\text{sector}$ represents the comoving volume which is calculated using the comoving distance, or radius of the search volume (R = $1200 \, \text{Mpc}$), rather than the luminosity distance. This approach accounts for the universe's expansion and provides a more accurate estimate of the physical volume of space where kilonovae occur, independent of temporal changes. 

The expected number of kilonovae detections was calculated for a survey period of $10$ years ($T$), and the detection efficiency for each metric was extracted from simulation data. This efficiency varies based on the specific cadence and metric used, influencing the final expected detection count. The equation:
\[
N = \text{BNS\_rates} \times V_{\text{sector}} \times \text{Efficiency} \times T
\]
was used to estimate the number of kilonovae detections over the survey's duration for each metric. 

Table \ref{tab:realistictab1} and \ref{tab:realistictab2} provides a comprehensive comparison of realistic kilonova detection rates seen in Fig.~\ref{fig:MILKNE}.

For the Optimistic BNS Rate model, the \texttt{multi\_detect} metric achieves its highest detection rates with the \texttt{baseline\_v3.4} and \texttt{baseline\_v3.3} cadence, yielding $\sim$\,128.0 detections. In comparison, the \texttt{ztfrest\_simple} metric's highest detections are observed with the \texttt{baseline\_v3.3} cadence at 14 detections, the \texttt{baseline\_v3.4} cadence with 14 detections, and the \texttt{good\_seeing\_gsw50.0\_v2.1} cadence with 10 detections. The \texttt{multi\_detect} metric consistently shows higher detection rates than \texttt{ztfrest\_simple} due to the less stringent parameters set for that metric as explained in \ref{sec:metrics}. For the color-specific metrics, \texttt{baseline\_v3.3} demonstrated the highest efficiency across all categories except for \texttt{blue\_color\_detect}, where \texttt{presto\_gap3.5\_v2.0} achieved the best detection rate. This metric actually detected more KNe than the other color-specific metrics with $11.3$ detections. Overall, \texttt{Presto\_Gap} cadences with gaps $\geq 3.0$ hours performed poorly across all metrics, with the exception of \texttt{blue\_color\_detect}, where \texttt{baseline\_v3.0} had the lowest detection efficiency.

In the Median BNS Rate model, \texttt{multi\_detect} reaches its peak with the \texttt{baseline\_v3.4} cadence, achieving 45.43 detections. The \texttt{baseline\_v3.3} cadence follows with 45.29 detections. For \texttt{ztfrest\_simple}, the top cadences are \texttt{baseline\_v3.3} and \texttt{baseline\_v3.4} with 5 detections. The median rates followed the same detection pattern as the optimistic model for the color-specific metrics, with \texttt{baseline\_v3.3} leading in most categories except for \texttt{blue\_color\_detect}, where \texttt{presto\_gap3.5\_v2.0} remained the highest performer. Only a handful of events were detected for the color-specific metrics.

Under the Pessimistic BNS Rate model, the \texttt{multi\_detect} metric shows its highest detections with the \texttt{baseline\_v3.4} and \texttt{baseline\_v3.3} cadences at 9 detections. For \texttt{ztfrest\_simple}, the highest detections are seen with the \texttt{baseline\_v3.3} and \texttt{baseline\_v3.4} cadences at 1 detection and 0.99, respectively. The pessimistic rates exhibited the same detection trends as the optimistic and median models for the color-specific metrics.

In metrics designed for detecting kilonovae, such as those with \texttt{ztfrest} prefix, \texttt{baseline\_v3.3} produced the highest number of kilonovae compared to \texttt{baseline\_v3.4}. This pattern is the same for \texttt{red\_color\_detect} and \texttt{blue\_color\_detect}. However, for \texttt{multi\_color\_detect} and \texttt{multi\_detect}, \texttt{baseline\_v3.4} edged \texttt{baseline\_v3.3} out. 

Further analysis showed that cadences such as \texttt{good\_seeing\_gsw50.0\_v2.1} and \texttt{presto\_gap3.0\_v2.0} appear frequently in the top three for various metrics. This indicates that these cadences might offer a balanced performance across different detection scenarios.

Additional details, the single-model all-cadence figure, and code are accessible on GitHub\footnote{\url{https://github.com/andra104/KNe_Detectability_Study}}.

\section{Conclusion}
In this analysis, we evaluated the efficacy of both filter and cadence strategies on LSST's prospects for detecting kilonovae. This study evaluated the impact of different LSST cadence strategies on kilonova detectability, incorporating insights from simulated single-model and population studies, as well as realistic BNS merger detection rates derived from GWTC-3. Our analysis highlights key trends in observational scheduling and efficiency, emphasizing the role of both baseline and \texttt{presto\_gap} cadences in optimizing transient detection. 

Filter selection plays a crucial role in optimizing kilonova detection. We found that for filter selections, the use of the $r$-band and $i$-band generally outperforms the use of the $u$-band or $g$-band (see Fig~\ref{fig:cadencecomparison}). Although, in general, the \texttt{baseline} cadences (with mixed filter metrics) outperform these single-filter cases. Among color-specific metrics, \texttt{presto\_gap} cadences performed best in blue filters, while \texttt{baseline\_v3.3} consistently led in red-filtered metrics, including \texttt{ztfrest\_simple\_red} and \texttt{red\_color\_detect}.

When considering populations of KNe, from all the survey strategy releases (through v2.1) and for the characterization metrics, the most effective cadences are the \texttt{presto\_gap} and \texttt{no\_roll} cadences, with the latter providing the most consistent improvement over the \texttt{baseline}. This strongly supports the implementation of a scheduling system that avoids (to redistribute) the repetition of visits (more than 3) within the same night. We also explored the performance variations across the KNe parameter space, indicating potential advantages of blue filters for small half-opening angles and red filters for higher angles. These initial findings highlighted the importance of optimizing return visit intervals to maximize KNe detectability.

Simulated with single-model parameters, in the most KNe-specific detection criteria, \texttt{ztfrest\_simple}, \texttt{baseline} v3.4, v3.0, and v3.3 ranked 1st, 2nd, and 4th with \texttt{presto\_gap4.0\_v2.0} coming in 3rd. Across the GWTC-3 kilonova detection rate models (optimistic, median, and pessimistic), \texttt{baseline} v3.3 and v3.4 consistently ranked as top performers, particularly in the more selective \texttt{ztfrest\_simple} metric. While \texttt{baseline\_v3.4} outperformed \texttt{baseline\_v3.3} in \texttt{multi\_detect}, \texttt{baseline\_v3.3} excelled in more selective metrics like \texttt{ztfrest\_simple}. These results reinforce that baseline v3.4 is particularly strong in broader transient detection, whereas \texttt{baseline\_v3.3} is more optimized for kilonovae identification. However, as O4 continues and fewer BNS detections occur, these realistic kilonovae detection rates may continue to decline. 

 The distribution of return times across different cadences significantly affects kilonova detectability. The relationship between observation gaps and transient detection efficiency was further explored using time-gap metrics. The \texttt{TgapsPercentMetric} analysis (Table \ref{tab:tablemedian}) highlights that most cadences have an observation gap at the $14$–$38$ hour timescale, with baseline v3.3, v3.4, and v3.2 spending \textasciitilde$12\%$ of their time in these gaps, respectively. This reinforces the idea that longer return visits are critical for improving kilonova detection rates as these cadences consistently ranked among the most efficient across all detection metrics and models. In contrast, \texttt{presto\_gap4.0\_v2.0} spent the most time in $4$–$8$ hour observation gaps, and did relatively well in all models, but did not outperform the last iterations of \texttt{baseline}. Despite the \texttt{presto\_gap} cadence family being considered a priority contender for detecting fast transients, it did not perform the best overall nor spend the most time in observation gaps. However, larger intra-night visits ($\geq 3.0$ hours) did compete with the \texttt{baseline} family. 
 
 These results indicate that longer observation gaps in the $14$–$38$ hour range contribute to improved kilonova detections. Returning too often will eventually have diminishing returns as it affects the overall sky area covered. However, some cadences, like \texttt{baseline\_v3.2} and certain \texttt{presto\_gap} cadences, despite having long return times, performed poorly overall, suggesting that efficient scheduling, not just the length of observation gaps, plays a key role in transient detection.

For both single-model simulations and the GWTC-3 kilonovae detection rates, the cadence efficiencies followed consistent patterns, reinforcing key strategies for optimizing kilonova detection. Our findings suggest that an optimal LSST strategy for kilonova detection should adopt a hybrid approach. This can be achieved by integrating long revisit times of approximately $14$–$38$ hours, as demonstrated by \texttt{baseline} v3.3 and v3.4, with strategic intra-night revisits within the $4$–$8$ hour range, as observed in \texttt{presto\_gap} cadences. Additionally, a balanced filter selection is essential for capturing the full evolution of kilonovae, as bluer filters are more effective for early detection of rapidly evolving emission, while redder filters are crucial for tracking the later phases. These insights provide valuable guidance for future LSST cadence refinements, ultimately enhancing the observatory's ability to detect kilonovae and other fast-evolving astronomical phenomena.

\section{Acknowledgements}
C.A. and M.W.C acknowledge support from the National Science Foundation with grant numbers PHY-2308862 and PHY-2117997 and the Preparing for Astrophysics with LSST Program, funded by the Heising Simons Foundation through grant 2021-2975, and administered by Las Cumbres Observatory. NG acknowledges support from the American University of Sharjah (UAE) through the grant FRG22-C-S68. L.R.S. and L.S.M. also thank the Preparing for Astrophysics with LSST Program.

This publication was made possible through the support of Grant 62192 from the John Templeton Foundation to LSST-DA. The opinions expressed in this publication are those of the author(s) and do not necessarily reflect the views of LSST-DA or the John Templeton Foundation.

\clearpage
\clearpage

\appendix
\begin{table*}[!ht]
    \centering
    \resizebox{\textwidth}{!}{%
    \begin{tabular}{|l|l|l|l|l|}
        \hline
        \textbf{Metric} & \textbf{blue\_color\_detect} & \textbf{multi\_color\_detect} & \textbf{multi\_detect} & \textbf{red\_color\_detect} \\ \hline
        rolling\_ns3\_rw0.9\_v2.0\_10yrs & 0.004738 & 0.043344 & 0.049162 & 0.00798 \\ \hline
        rolling\_ns3\_rw0.5\_v2.0\_10yrs & 0.00394 & 0.047144 & 0.053582 & 0.007084 \\ \hline
        rolling\_ns2\_rw0.9\_v2.0\_10yrs & 0.004252 & 0.046634 & 0.052834 & 0.007698 \\ \hline
        rolling\_ns2\_rw0.5\_v2.0\_10yrs & 0.003762 & 0.048642 & 0.05541 & 0.00719 \\ \hline
        retro\_baseline\_v2.0\_10yrs & 0.004724 & 0.048438 & 0.058038 & 0.008516 \\ \hline
        presto\_gap4.0\_v2.0\_10yrs & 0.005226 & 0.035672 & 0.044068 & 0.007188 \\ \hline
        presto\_gap3.5\_v2.0\_10yrs & 0.005542 & 0.03556 & 0.0439 & 0.007162 \\ \hline
        presto\_gap3.0\_v2.0\_10yrs & 0.005242 & 0.035502 & 0.043902 & 0.007192 \\ \hline
        presto\_gap2.5\_v2.0\_10yrs & 0.005456 & 0.035326 & 0.043366 & 0.00716 \\ \hline
        presto\_gap2.0\_v2.0\_10yrs & 0.005062 & 0.035198 & 0.04358 & 0.007202 \\ \hline
        presto\_gap1.5\_v2.0\_10yrs & 0.005378 & 0.035362 & 0.044738 & 0.00724 \\ \hline
        noroll\_v2.0\_10yrs & 0.003614 & 0.050272 & 0.057212 & 0.006832 \\ \hline
        long\_u2\_v2.0\_10yrs & 0.004216 & 0.048172 & 0.05395 & 0.007606 \\ \hline
        long\_u1\_v2.0\_10yrs & 0.004168 & 0.046452 & 0.052192 & 0.007332 \\ \hline
        good\_seeing\_u\_gsw50.0\_v2.1\_10yrs & 0.004036 & 0.046688 & 0.053012 & 0.007582 \\ \hline
        good\_seeing\_u\_gsw0.0\_v2.1\_10yrs & 0.004238 & 0.04683 & 0.053306 & 0.007706 \\ \hline
        good\_seeing\_gsw50.0\_v2.1\_10yrs & 0.004312 & 0.048774 & 0.054596 & 0.007832 \\ \hline
        good\_seeing\_gsw0.0\_v2.1\_10yrs & 0.004464 & 0.048844 & 0.054852 & 0.00817 \\ \hline
        bluer\_indx1\_v2.0\_10yrs & 0.00443 & 0.045932 & 0.052246 & 0.007122 \\ \hline
        bluer\_indx0\_v2.0\_10yrs & 0.004812 & 0.046782 & 0.05323 & 0.00703 \\ \hline
        baseline\_v2.1\_10yrs & 0.00448 & 0.048774 & 0.054602 & 0.008118 \\ \hline
        baseline\_v2.0\_10yrs & 0.004252 & 0.046634 & 0.052834 & 0.007698 \\ \hline
        baseline\_retrofoot\_v2.0\_10yrs & 0.004132 & 0.04576 & 0.052036 & 0.007608 \\ \hline
        baseline\_v3.0\_10yrs & 0.003712 & 0.049212 & 0.056212 & 0.007566 \\ \hline
        baseline\_v3.2\_10yrs & 0.002996 & 0.042744 & 0.048296 & 0.006886 \\ \hline
        baseline\_v3.3\_10yrs & 0.003142 & 0.045234 & 0.051412 & 0.007956 \\ \hline
        baseline\_v3.4\_10yrs & 0.003388 & 0.045964 & 0.052102 & 0.00789 \\ \hline
        baseline\_v3.5\_10yrs & 0.003436 & 0.045604 & 0.05158 & 0.007598 \\
        \hline
        baseline\_v3.6\_10yrs & 0.003132 & 0.041942 & 0.047576 & 0.00683 \\
        \hline
    \end{tabular}%
    }
    \caption{Recovery Efficiencies (multi\_detect). 500,000 total light curves were injected into the simulations. The light curves were placed at a minimum luminosity distance of 10 Mpc and a maximum luminosity distance of 600 Mpc. A comprehensive figure can be found on the GitHub.}
    \label{tab:efficient1}
\end{table*}

\begin{table*}[!h]
    \centering
    \begin{tabular}{|l|l|l|l|}
        \hline
        \textbf{Metric} & \textbf{ztfrest\_simple} & \textbf{ztfrest\_simple\_blue} & \textbf{ztfrest\_simple\_red} \\ \hline
        rolling\_ns3\_rw0.9\_v2.0\_10yrs & 0.009422 & 0.005456 & 0.005334 \\ \hline
        rolling\_ns3\_rw0.5\_v2.0\_10yrs & 0.00833 & 0.004848 & 0.004394 \\ \hline
        rolling\_ns2\_rw0.9\_v2.0\_10yrs & 0.00898 & 0.004996 & 0.00492 \\ \hline
        rolling\_ns2\_rw0.5\_v2.0\_10yrs & 0.00825 & 0.004606 & 0.00439 \\ \hline
        retro\_baseline\_v2.0\_10yrs & 0.008556 & 0.004524 & 0.004678 \\ \hline
        presto\_gap4.0\_v2.0\_10yrs & 0.010482 & 0.006552 & 0.004784 \\ \hline
        presto\_gap3.5\_v2.0\_10yrs & 0.01015 & 0.006342 & 0.004604 \\ \hline
        presto\_gap3.0\_v2.0\_10yrs & 0.0095 & 0.005818 & 0.00435 \\ \hline
        presto\_gap2.5\_v2.0\_10yrs & 0.008612 & 0.005232 & 0.003982 \\ \hline
        presto\_gap2.0\_v2.0\_10yrs & 0.007548 & 0.004646 & 0.003344 \\ \hline
        presto\_gap1.5\_v2.0\_10yrs & 0.005826 & 0.003408 & 0.002782 \\ \hline
        noroll\_v2.0\_10yrs & 0.00751 & 0.004238 & 0.003966 \\ \hline
        long\_u2\_v2.0\_10yrs & 0.008818 & 0.004912 & 0.004924 \\ \hline
        long\_u1\_v2.0\_10yrs & 0.008322 & 0.004576 & 0.004634 \\ \hline
        good\_seeing\_u\_gsw50.0\_v2.1\_10yrs & 0.009596 & 0.005478 & 0.00535 \\ \hline
        good\_seeing\_u\_gsw0.0\_v2.1\_10yrs & 0.009006 & 0.005126 & 0.00494 \\ \hline
        good\_seeing\_gsw50.0\_v2.1\_10yrs & 0.009148 & 0.005374 & 0.005124 \\ \hline
        good\_seeing\_gsw0.0\_v2.1\_10yrs & 0.009012 & 0.005186 & 0.00499 \\ \hline
        bluer\_indx1\_v2.0\_10yrs & 0.008254 & 0.004748 & 0.00433 \\ \hline
        bluer\_indx0\_v2.0\_10yrs & 0.008514 & 0.004914 & 0.004492 \\ \hline
        baseline\_v2.1\_10yrs & 0.00919 & 0.005362 & 0.005048 \\ \hline
        baseline\_v2.0\_10yrs & 0.00898 & 0.004996 & 0.00492 \\ \hline
        baseline\_retrofoot\_v2.0\_10yrs & 0.008852 & 0.004894 & 0.004998 \\ \hline
        baseline\_v3.0\_10yrs & 0.010712 & 0.006074 & 0.005894 \\ \hline
        baseline\_v3.2\_10yrs & 0.00872 & 0.004706 & 0.004986 \\ \hline
        baseline\_v3.3\_10yrs & 0.010406 & 0.005252 & 0.00623 \\ \hline
        baseline\_v3.4\_10yrs & 0.01088 & 0.005728 & 0.00638 \\ \hline
        baseline\_v3.5\_10yrs & 0.009224 & 0.045604 & 0.00517 \\ \hline
        baseline\_v3.6\_10yrs & 0.008156 & 0.00437 & 0.004608 \\ \hline
    \end{tabular}%
    \caption{Recovery Efficiencies (ztfrest). 500,000 total light curves were injected into the simulations. The efficiencies are found by dividing the total recovered by the total injected. The light curves were placed at a minimum luminosity distance of 10 Mpc and a maximum luminosity distance of 600 Mpc. A comprehensive figure can be found on the GitHub.}
    \label{tab:efficient2}
\end{table*}

\begin{table*}[!h]
    \centering
    \resizebox{\textwidth}{!}{%
    \begin{tabular}{|c|c|c|c|c|}
    \hline
    \textbf{Cadence} & \textbf{blue\_color\_detect} & \textbf{multi\_color\_detect} & \textbf{multi\_detect} & \textbf{red\_color\_detect} \\ 
    \hline
    \multicolumn{5}{|c|}{Model: n\_events = $1e6$, Optimistic BNS Rate = 295.7, R\_Mpc = 1200} \\  
    \hline
    baseline\_v3.0 & 7.8887 & 93.6550 & 116.4874 & 6.6422  \\ 
    baseline\_v3.4 & 8.2615 & 105.3186 & 127.9691 & 8.0203 \\ 
    baseline\_v3.3 & 8.8560 & 104.8475 & 127.6938 & 8.2971 \\ 
    presto\_gap3.5\_v2.0 & 11.3044 & 60.8786 & 83.3588 & 5.9370 \\ 
    presto\_gap3.0\_v2.0 & 11.2744 & 62.1288 & 84.6062 & 6.6201 \\ 
    good\_seeing\_u\_gsw50.0\_v2.1 & 8.7655 & 89.7194 & 111.2882 & 6.0777 \\ 
    presto\_gap4.0\_v2.0 & 10.5490 & 61.6706 & 84.1268 & 6.3766 \\ 
    rolling\_ns3\_rw0.9\_v2.0 & 9.9111 & 82.7201 & 104.2066 & 6.8742 \\ 
    baseline\_v2.1 & 10.5790 & 94.3413 & 115.9502 & 6.8854 \\ 
    baseline\_v2.0 & 9.0928 & 85.9864 & 108.0682 & 6.7899 \\ 
    good\_seeing\_gsw50.0\_v2.1 & 10.2965 & 97.0070 & 116.7495 & 6.8566 \\ 
    baseline\_v3.5 & 8.8397 & 99.7418 & 120.5994 & 7.8084 \\ 
    baseline\_v3.6 & 8.9029 & 91.4282 & 111.5387 & 7.6401 \\
    \hline
    
    \multicolumn{5}{|c|}{Model: n\_events = $1e6$, Median BNS Rate = 105.5, R\_Mpc = 1200} \\  
    \hline
    baseline\_v3.0 & 2.8159 & 33.4271 & 41.6008 & 2.3729 \\ 
    baseline\_v3.4 & 2.9473 & 37.6097 & 45.4305 & 2.8497 \\ 
    baseline\_v3.3 & 3.1576 & 37.3393 & 45.2953 & 2.9586 \\ 
    presto\_gap3.5\_v2.0 & 4.0287 & 21.7090 & 29.6312 & 2.1101 \\ 
    presto\_gap3.0\_v2.0 & 4.0174 & 22.1671 & 30.0216 & 2.3729 \\ 
    good\_seeing\_u\_gsw50.0\_v2.1 & 3.2139 & 32.8864 & 40.9888 & 2.2377 \\ 
    presto\_gap4.0\_v2.0 & 3.8635 & 21.8216 & 29.7401 & 2.2678 \\ 
    rolling\_ns3\_rw0.9\_v2.0 & 3.6307 & 29.1731 & 36.9188 & 2.4367 \\ 
    baseline\_v2.1 & 3.8747 & 34.5609 & 42.3705 & 2.4405 \\ 
    baseline\_v2.0 & 3.3303 & 31.4822 & 39.7122 & 2.2190 \\ 
    good\_seeing\_gsw50.0\_v2.1 & 3.7734 & 35.5747 & 42.8135 & 2.2152 \\ 
    baseline\_v3.5 & 3.1538 & 35.5859 & 43.0275 & 2.7859 \\
    baseline\_v3.6 & 3.1764 & 32.61981 & 39.7948 & 2.7258 \\
    \hline
    \multicolumn{5}{|c|}{Model: n\_events = $1e6$, Pessimistic BNS Rate = 21.6, R\_Mpc = 1200} \\ 
    \hline
    baseline\_v3.0 & 0.5765 & 6.8438 & 8.5173 & 0.4858 \\ 
    baseline\_v3.4 & 0.6034 & 7.7002 & 9.3014 & 0.5835 \\ 
    baseline\_v3.3 & 0.6465 & 7.6448 & 9.2737 & 0.6057 \\ 
    presto\_gap3.5\_v2.0 & 0.8248 & 4.4447 & 6.0667 & 0.4320 \\ 
    presto\_gap3.0\_v2.0 & 0.8225 & 4.5385 & 6.1466 & 0.4858 \\ 
    good\_seeing\_u\_gsw50.0\_v2.1 & 0.6580 & 6.7331 & 8.3920 & 0.4582 \\ 
    presto\_gap4.0\_v2.0 & 0.7910 & 4.4677 & 6.0890 & 0.4643 \\ 
    rolling\_ns3\_rw0.9\_v2.0 & 0.7433 & 5.9729 & 7.5587 & 0.4989 \\ 
    baseline\_v2.1 & 0.7933 & 7.0760 & 8.6749 & 0.4997 \\ 
    baseline\_v2.0 & 0.6818 & 6.4456 & 8.1307 & 0.4543 \\ 
    good\_seeing\_gsw50.0\_v2.1 & 0.7726 & 7.2835 & 8.7656 & 0.4535 \\ 
    baseline\_v3.5 & 0.6457 & 7.2858 & 8.8094 & 0.5704 \\
    baseline\_v3.6 & 0.6503 & 6.6786 & 8.1476 & 0.5581 \\
    \hline
    \end{tabular}%
    }
    \caption{\textbf{GWTC-3 Rates:} The most efficient cadences from the single model simulations applied to a script to produce the most realistic BNS kilonovae detection rates based on data from GWTC-3. The \texttt{multi\_detect} and color detect metrics. See section \ref{sec:realistic}}
    \label{tab:realistictab1}
\end{table*}

\begin{table*}[!h]
    \centering
    \resizebox{\textwidth}{!}{%
    \begin{tabular}{|c|c|c|c|}
    \hline
    \textbf{Cadence} & \textbf{ztfrest\_simple} & \textbf{ztfrest\_simple\_blue} & \textbf{ztfrest\_simple\_red} \\ 
    \hline
    \multicolumn{4}{|c|}{Model: n\_events = $1e6$, Optimistic BNS Rate = 295.7, R\_Mpc = 1200} \\  
    \hline
    baseline\_v3.0 & 11.6531 & 8.0372 & 4.6218 \\ 
    baseline\_v3.4 & 13.6458 & 9.2637 & 5.4019 \\ 
    baseline\_v3.3 & 14.1061 & 9.3160 & 5.8448 \\ 
    presto\_gap3.5\_v2.0 & 9.7232 & 7.0050 & 3.4138 \\ 
    presto\_gap3.0\_v2.0 & 9.6391 & 6.6055 & 3.6065 \\ 
    good\_seeing\_u\_gsw50.0\_v2.1 & 10.2194 & 7.3152 & 4.0116 \\ 
    presto\_gap4.0\_v2.0 & 10.9320 & 8.2073 & 3.3062 \\ 
    rolling\_ns3\_rw0.9\_v2.0 & 9.7574 & 7.1697 & 3.7001 \\ 
    baseline\_v2.1 & 10.4485 & 7.6077 & 4.0322 \\ 
    baseline\_v2.0 & 9.7840 & 6.7968 & 3.7605 \\ 
    good\_seeing\_gsw50.0\_v2.1 & 10.7850 & 7.9401 & 4.2010 \\ 
    baseline\_v3.5 & 12.1967 & 8.1136 & 4.9145 \\ 
    baseline\_v3.6 & 11.3549 & 7.9768 & 4.3357 \\
    \hline
    
    \multicolumn{4}{|c|}{Model: n\_events = $1e6$, Median BNS Rate = 105.5, R\_Mpc = 1200} \\  
    \hline
    baseline\_v3.0 & 4.1638 & 2.8723 & 1.6520 \\ 
    baseline\_v3.4 & 4.8734 & 3.3078 & 1.9299 \\ 
    baseline\_v3.3 & 5.0236 & 3.3153 & 2.0800 \\ 
    presto\_gap3.5\_v2.0 & 3.4542 & 2.4893 & 1.2127 \\ 
    presto\_gap3.0\_v2.0 & 3.4242 & 2.3466 & 1.2803 \\ 
    good\_seeing\_u\_gsw50.0\_v2.1 & 3.7621 & 2.6920 & 1.4756 \\ 
    presto\_gap4.0\_v2.0 & 3.8860 & 2.9173 & 1.1752 \\ 
    rolling\_ns3\_rw0.9\_v2.0 & 3.4692 & 2.5494 & 1.3141 \\ 
    baseline\_v2.1 & 3.7133 & 2.6995 & 1.4305 \\ 
    baseline\_v2.0 & 3.4767 & 2.4142 & 1.3366 \\ 
    good\_seeing\_gsw50.0\_v2.1 & 3.8034 & 2.8009 & 1.4831 \\ 
    baseline\_v3.5 & 4.3516 & 2.8948 & 1.7534 \\
    baseline\_v3.6 & 4.0512 & 2.8459 & 1.5469 \\
    \hline
    \multicolumn{4}{|c|}{Model: n\_events = $1e6$, Pessimistic BNS Rate = 21.6, R\_Mpc = 1200} \\ 
    \hline
    baseline\_v3.0 & 0.8525 & 0.5881 & 0.3382 \\ 
    baseline\_v3.4 & 0.9978 & 0.6772 & 0.3951 \\ 
    baseline\_v3.3 & 1.0285 & 0.6788 & 0.4259 \\ 
    presto\_gap3.5\_v2.0 & 0.7072 & 0.5097 & 0.2483 \\ 
    presto\_gap3.0\_v2.0 & 0.7011 & 0.4804 & 0.2621 \\ 
    good\_seeing\_u\_gsw50.0\_v2.1 & 0.7702 & 0.5512 & 0.3021 \\ 
    presto\_gap4.0\_v2.0 & 0.7956 & 0.5973 & 0.2406 \\ 
    rolling\_ns3\_rw0.9\_v2.0 & 0.7103 & 0.5220 & 0.2690 \\ 
    baseline\_v2.1 & 0.7603 & 0.5527 & 0.2929 \\ 
    baseline\_v2.0 & 0.7118 & 0.4943 & 0.2737 \\ 
    good\_seeing\_gsw50.0\_v2.1 & 0.7787 & 0.5735 & 0.3036 \\ 
    baseline\_v3.5 & 0.8909 & 0.5927 & 0.3589 \\
    baseline\_v3.6 & 0.8294 & 0.5827 & 0.3167 \\
    \hline
    \end{tabular}%
    }
    \caption{\textbf{GWTC-3 Rates:} The most efficient cadences from the single model simulations adapted to a script to produce the most realistic BNS kilonovae detection rates based on data from GWTC-3. The \texttt{ztfrest\_simple} metrics. See section \ref{sec:realistic}}
    \label{tab:realistictab2}
\end{table*}

\begin{figure*}[!h]
\centering
\includegraphics[scale=0.35]{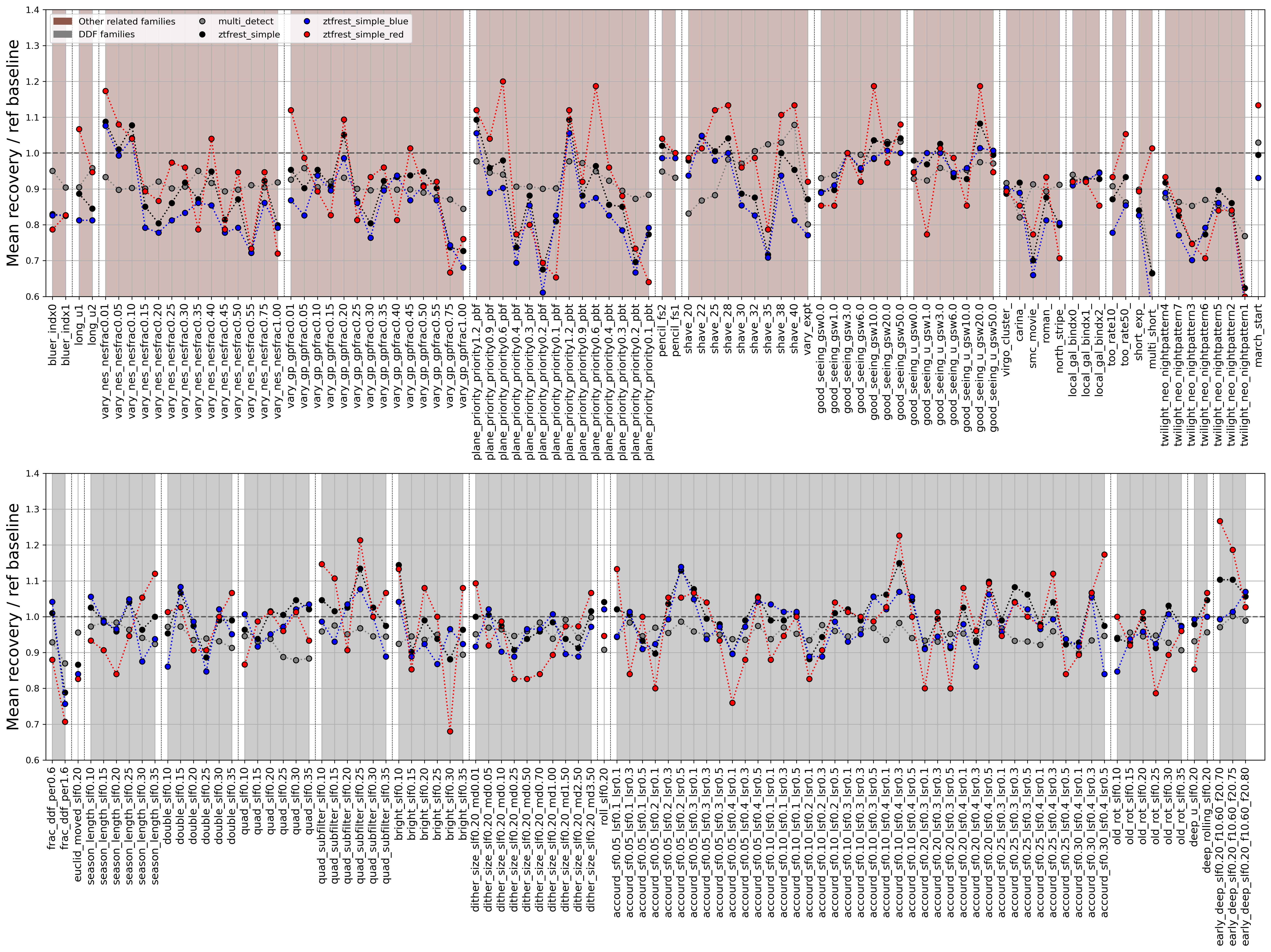}
\caption{Same description as Fig.~\ref{fig:allcadencemainfamilies}, but for the DDF families (bottom) and for all the rest of v2.0 and v2.1 families (top).}
\centering
\label{fig:allcadenceappfamilies}
\end{figure*}

\begin{figure*}[!h]
\centering
\includegraphics[scale=0.5]{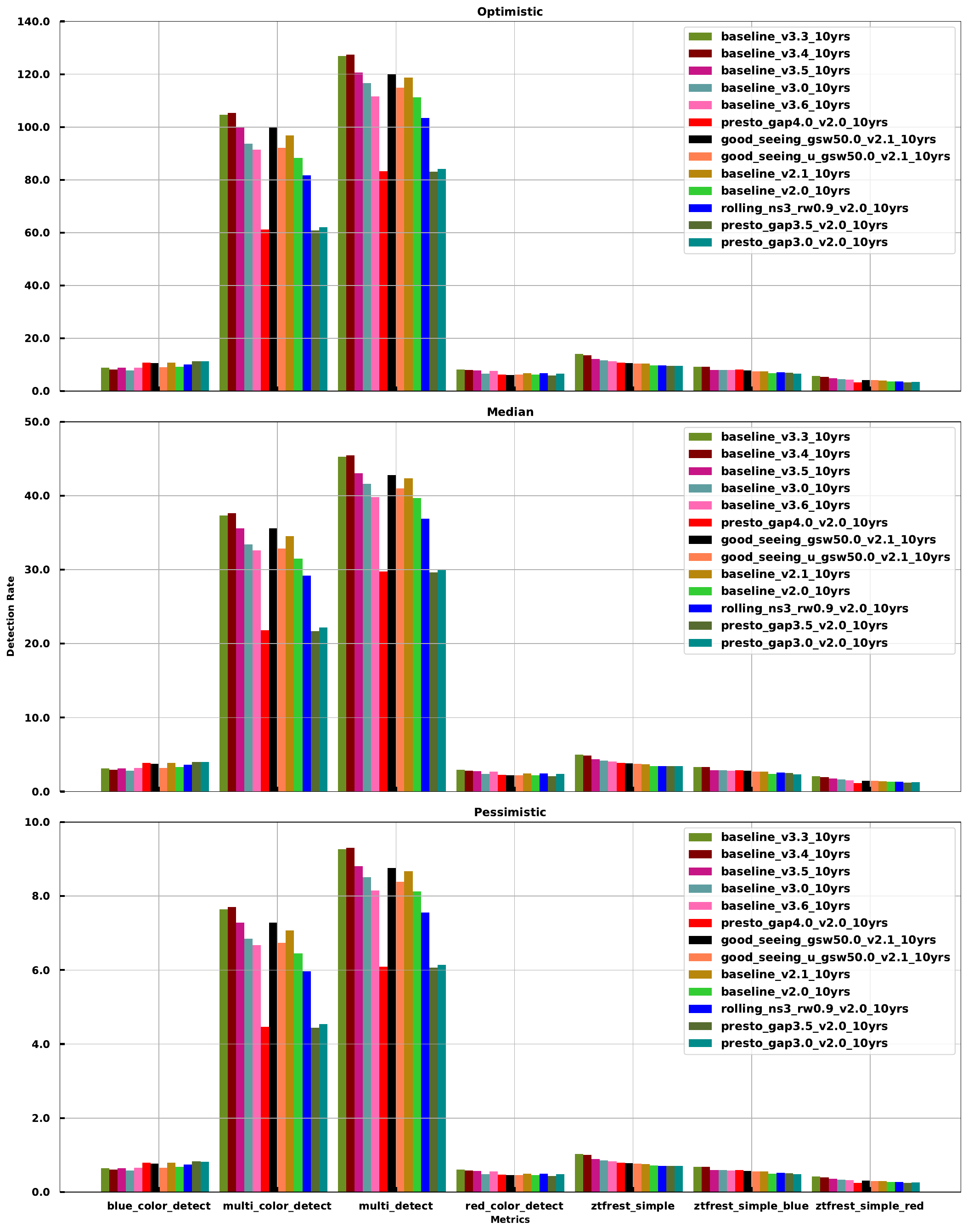}
\caption{Top 10 cadences including all \texttt{Baseline} cadences with GWTC-3 parameters for all metrics. See \ref{sec:realistic}.}
\centering
\label{fig:MILKNE}
\end{figure*}

%\begin{figure*}[!h]
%\centering
%\includegraphics[scale=0.30]{plots/fig4_3.6+newnewnew+2024_vertical.pdf}
%\caption{All cadences simulated using single model parameters for all metrics. Expanded Fig.~\ref{fig:top10kilonovaerecovery}.}
%\centering
%\label{fig:allcadencebig}
%\end{figure*}

\clearpage
\bibliographystyle{aasjournal}
\bibliography{references}

\begin{thebibliography}{}
\expandafter\ifx\csname natexlab\endcsname\relax\def\natexlab#1{#1}\fi
\providecommand{\url}[1]{\href{#1}{#1}}
\providecommand{\dodoi}[1]{doi:~\href{http://doi.org/#1}{\nolinkurl{#1}}}
\providecommand{\doeprint}[1]{\href{http://ascl.net/#1}{\nolinkurl{http://ascl.net/#1}}}
\providecommand{\doarXiv}[1]{\href{https://arxiv.org/abs/#1}{\nolinkurl{https://arxiv.org/abs/#1}}}

\bibitem[{Abbott {et~al.}(2016)Abbott, Abbott, Abbott, Abernathy, Acernese, \&
  et~al.}]{Abbott_2016}
Abbott, B.~P., Abbott, R., Abbott, T.~D., {et~al.} 2016, The Astrophysical
  Journal Letters, 826, L13, \dodoi{10.3847/2041-8205/826/1/L13}

\bibitem[{Abbott {et~al.}(2017{\natexlab{a}})Abbott, Abbott, Abbott, Acernese,
  Ackley, Adams, Adams, Addesso, Adhikari, Adya, Affeldt, Afrough, Agarwal,
  Agathos, Agatsuma, Aggarwal, Aguiar, Aiello, Ain, Ajith, Allen, Allen,
  Allocca, Altin, Amato, Ananyeva, Anderson, Anderson, Antier, Appert, Arai,
  Araya, Areeda, Arnaud, Arun, Ascenzi, Ashton, Ast, Aston, Astone, Aufmuth,
  Aulbert, AultONeal, Avila-Alvarez, Babak, Bacon, Bader, Bae, Baker,
  Baldaccini, Ballardin, Ballmer, Banagiri, Barayoga, Barclay, Barish, Barker,
  Barone, Barr, Barsotti, Barsuglia, Barta, Bartlett, Bartos, Bassiri, Basti,
  Batch, Baune, Bawaj, Bazzan, B\'ecsy, Beer, Bejger, Belahcene, Bell, Berger,
  Bergmann, Berry, Bersanetti, Bertolini, Betzwieser, Bhagwat, Bhandare,
  Bilenko, Billingsley, Billman, Birch, Birney, Birnholtz, Biscans, Bisht,
  Bitossi, Biwer, Bizouard, Blackburn, Blackman, Blair, Blair, Blair, Bloemen,
  Bock, Bode, Boer, Bogaert, Bohe, Bondu, Bonnand, Boom, Bork, Boschi, Bose,
  Bouffanais, Bozzi, Bradaschia, Brady, Braginsky, Branchesi, Brau, Briant,
  Brillet, Brinkmann, Brisson, Brockill, Broida, Brooks, Brown, Brown, Brown,
  Brunett, Buchanan, Buikema, Bulik, Bulten, Buonanno, Buskulic, Buy, Byer,
  Cabero, Cadonati, Cagnoli, Cahillane, Calder\'on~Bustillo, Callister,
  Calloni, Camp, Canepa, Canizares, Cannon, Cao, Cao, Capano, Capocasa,
  Carbognani, Caride, Carney, Casanueva~Diaz, Casentini, Caudill, Cavagli\`a,
  Cavalier, Cavalieri, Cella, Cepeda, Cerboni~Baiardi, Cerretani, Cesarini,
  Chamberlin, Chan, Chao, Charlton, Chassande-Mottin, Chatterjee,
  Chatziioannou, Cheeseboro, Chen, Chen, Cheng, Chincarini, Chiummo, Chmiel,
  Cho, Cho, Chow, Christensen, Chu, Chua, Chua, Chung, Chung, Ciani, Ciolfi,
  Cirelli, Cirone, Clara, Clark, Cleva, Cocchieri, Coccia, Cohadon, Colla,
  Collette, Cominsky, Constancio, Conti, Cooper, Corban, Corbitt, Corley,
  Cornish, Corsi, Cortese, Costa, Coughlin, Coughlin, Coulon, Countryman,
  Couvares, Covas, Cowan, Coward, Cowart, Coyne, Coyne, Creighton, Creighton,
  Cripe, Crowder, Cullen, Cumming, Cunningham, Cuoco, Dal~Canton, Danilishin,
  D'Antonio, Danzmann, Dasgupta, Da~Silva~Costa, Dattilo, Dave, Davier, Davis,
  Daw, Day, De, DeBra, Deelman, Degallaix, De~Laurentis, Del\'eglise,
  Del~Pozzo, Denker, Dent, Dergachev, De~Rosa, DeRosa, DeSalvo, Devenson,
  Devine, Dhurandhar, D\'{\i}az, Di~Fiore, Di~Giovanni, Di~Girolamo, Di~Lieto,
  Di~Pace, Di~Palma, Di~Renzo, Doctor, Dolique, Donovan, Dooley, Doravari,
  Dorrington, Douglas, Dovale~\'Alvarez, Downes, Drago, Drever, Driggers, Du,
  Ducrot, Duncan, Dwyer, Edo, Edwards, Effler, Eggenstein, Ehrens, Eichholz,
  Eikenberry, Eisenstein, Essick, Etienne, Etzel, Evans, Evans, Factourovich,
  Fafone, Fair, Fairhurst, Fan, Farinon, Farr, Farr, Fauchon-Jones, Favata,
  Fays, Fehrmann, Feicht, Fejer, Fernandez-Galiana, Ferrante, Ferreira,
  Ferrini, Fidecaro, Fiori, Fiorucci, Fisher, Flaminio, Fletcher, Fong,
  Forsyth, Forsyth, Fournier, Frasca, Frasconi, Frei, Freise, Frey, Frey,
  Fries, Fritschel, Frolov, Fulda, Fyffe, Gabbard, Gabel, Gadre, Gaebel, Gair,
  Gammaitoni, Ganija, Gaonkar, Garufi, Gaudio, Gaur, Gayathri, Gehrels, Gemme,
  Genin, Gennai, George, George, Gergely, Germain, Ghonge, Ghosh, Ghosh, Ghosh,
  Giaime, Giardina, Giazotto, Gill, Glover, Goetz, Goetz, Gomes, Gonz\'alez,
  Gonzalez~Castro, Gopakumar, Gorodetsky, Gossan, Gosselin, Gouaty, Grado,
  Graef, Granata, Grant, Gras, Gray, Greco, Green, Groot, Grote, Grunewald,
  Gruning, Guidi, Guo, Gupta, Gupta, Gushwa, Gustafson, Gustafson, Hall, Hall,
  Hammond, Haney, Hanke, Hanks, Hanna, Hannam, Hannuksela, Hanson, Hardwick,
  Harms, Harry, Harry, Hart, Haster, Haughian, Healy, Heidmann, Heintze,
  Heitmann, Hello, Hemming, Hendry, Heng, Hennig, Henry, Heptonstall, Heurs,
  Hild, Hoak, Hofman, Holt, Holz, Hopkins, Horst, Hough, Houston, Howell, Hu,
  Huerta, Huet, Hughey, Husa, Huttner, Huynh-Dinh, Indik, Ingram, Inta, Intini,
  Isa, Isac, Isi, Iyer, Izumi, Jacqmin, Jani, Jaranowski, Jawahar,
  Jim\'enez-Forteza, Johnson, Johnson-McDaniel, Jones, Jones, Jonker, Ju,
  Junker, Kalaghatgi, Kalogera, Kandhasamy, Kang, Kanner, Karki, Karvinen,
  Kasprzack, Katolik, Katsavounidis, Katzman, Kaufer, Kawabe, K\'ef\'elian,
  Keitel, Kemball, Kennedy, Kent, Key, Khalili, Khan, Khan, Khan, Khazanov,
  Kijbunchoo, Kim, Kim, Kim, Kim, Kim, Kimbrell, King, King, Kirchhoff, Kissel,
  Kleybolte, Klimenko, Koch, Koehlenbeck, Koley, Kondrashov, Kontos, Korobko,
  Korth, Kowalska, Kozak, Kr\"amer, Kringel, Krishnan, Kr\'olak, Kuehn, Kumar,
  Kumar, Kumar, Kuo, Kutynia, Kwang, Lackey, Lai, Landry, Lang, Lange, Lantz,
  Lanza, Lartaux-Vollard, Lasky, Laxen, Lazzarini, Lazzaro, Leaci, Leavey, Lee,
  Lee, Lee, Lee, Lee, Lehmann, Lenon, Leonardi, Leroy, Letendre, Levin, Li,
  Libson, Littenberg, Liu, Lo, Lockerbie, London, Lord, Lorenzini, Loriette,
  Lormand, Losurdo, Lough, Lovelace, L\"uck, Lumaca, Lundgren, Lynch, Ma,
  Macfoy, Machenschalk, MacInnis, Macleod, Maga\~na Hernandez, Maga\~na
  Sandoval, Maga\~na Zertuche, Magee, Majorana, Maksimovic, Man, Mandic,
  Mangano, Mansell, Manske, Mantovani, Marchesoni, Marion, M\'arka, M\'arka,
  Markakis, Markosyan, Maros, Martelli, Martellini, Martin, Martynov, Marx,
  Mason, Masserot, Massinger, Masso-Reid, Mastrogiovanni, Matas, Matichard,
  Matone, Mavalvala, Mayani, Mazumder, McCarthy, McClelland, McCormick,
  McCuller, McGuire, McIntyre, McIver, McManus, McRae, McWilliams, Meacher,
  Meadors, Meidam, Mejuto-Villa, Melatos, Mendell, Mercer, Merilh, Merzougui,
  Meshkov, Messenger, Messick, Metzdorff, Meyers, Mezzani, Miao, Michel,
  Middleton, Mikhailov, Milano, Miller, Miller, Miller, Miller, Millhouse,
  Minazzoli, Minenkov, Ming, Mishra, Mitra, Mitrofanov, Mitselmakher,
  Mittleman, Moggi, Mohan, Mohapatra, Montani, Moore, Moore, Moraru, Moreno,
  Morriss, Mours, Mow-Lowry, Mueller, Muir, Mukherjee, Mukherjee, Mukherjee,
  Mukund, Mullavey, Munch, Muniz, Murray, Napier, Nardecchia, Naticchioni,
  Nayak, Nelemans, Nelson, Neri, Nery, Neunzert, Newport, Newton, Ng, Nguyen,
  Nichols, Nielsen, Nissanke, Nitz, Noack, Nocera, Nolting, Normandin, Nuttall,
  Oberling, Ochsner, Oelker, Ogin, Oh, Oh, Ohme, Oliver, Oppermann, Oram,
  O'Reilly, Ormiston, Ortega, O'Shaughnessy, Ottaway, Overmier, Owen, Pace,
  Page, Page, Pai, Pai, Palamos, Palashov, Palomba, Pal-Singh, Pan, Pang, Pang,
  Pankow, Pannarale, Pant, Paoletti, Paoli, Papa, Paris, Parker, Pascucci,
  Pasqualetti, Passaquieti, Passuello, Patricelli, Pearlstone, Pedraza,
  Pedurand, Pekowsky, Pele, Penn, Perez, Perreca, Perri, Pfeiffer, Phelps,
  Piccinni, Pichot, Piergiovanni, Pierro, Pillant, Pinard, Pinto, Pitkin,
  Poggiani, Popolizio, Porter, Post, Powell, Prasad, Pratt, Predoi, Prestegard,
  Prijatelj, Principe, Privitera, Prodi, Prokhorov, Puncken, Punturo, Puppo,
  P\"urrer, Qi, Qin, Qiu, Quetschke, Quintero, Quitzow-James, Raab, Rabeling,
  Radkins, Raffai, Raja, Rajan, Rakhmanov, Ramirez, Rapagnani, Raymond,
  Razzano, Read, Regimbau, Rei, Reid, Reitze, Rew, Reyes, Ricci, Ricker,
  Rieger, Riles, Rizzo, Robertson, Robie, Robinet, Rocchi, Rolland, Rollins,
  Roma, Romano, Romano, Romel, Romie, Rosi\ifmmode~\acute{n}\else
  \'{n}\fi{}ska, Ross, Rowan, R\"udiger, Ruggi, Ryan, Rynge, Sachdev, Sadecki,
  Sadeghian, Sakellariadou, Salconi, Saleem, Salemi, Samajdar, Sammut, Sampson,
  Sanchez, Sandberg, Sandeen, Sanders, Sassolas, Sathyaprakash, Saulson,
  Sauter, Savage, Sawadsky, Schale, Scheuer, Schmidt, Schmidt, Schmidt,
  Schnabel, Schofield, Sch\"onbeck, Schreiber, Schuette, Schulte, Schutz,
  Schwalbe, Scott, Scott, Seidel, Sellers, Sengupta, Sentenac, Sequino,
  Sergeev, Shaddock, Shaffer, Shah, Shahriar, Shao, Shapiro, Shawhan, Sheperd,
  Shoemaker, Shoemaker, Siellez, Siemens, Sieniawska, Sigg, Silva, Singer,
  Singer, Singh, Singh, Singhal, Sintes, Slagmolen, Smith, Smith, Smith, Son,
  Sonnenberg, Sorazu, Sorrentino, Souradeep, Spencer, Srivastava, Staley,
  Steinke, Steinlechner, Steinlechner, Steinmeyer, Stephens, Stevenson, Stone,
  Strain, Stratta, Strigin, Sturani, Stuver, Summerscales, Sun, Sunil, Sutton,
  Swinkels, Szczepa\ifmmode~\acute{n}\else \'{n}\fi{}czyk, Tacca, Talukder,
  Tanner, T\'apai, Taracchini, Taylor, Taylor, Theeg, Thomas, Thomas, Thomas,
  Thorne, Thorne, Thrane, Tiwari, Tiwari, Tokmakov, Toland, Tonelli, Tornasi,
  Torrie, T\"oyr\"a, Travasso, Traylor, Trifir\`o, Trinastic, Tringali, Trozzo,
  Tsang, Tse, Tso, Tuyenbayev, Ueno, Ugolini, Unnikrishnan, Urban, Usman, Vahi,
  Vahlbruch, Vajente, Valdes, Vallisneri, van Bakel, van Beuzekom, van~den
  Brand, Van Den~Broeck, Vander-Hyde, van~der Schaaf, van Heijningen, van
  Veggel, Vardaro, Varma, Vass, Vas\'uth, Vecchio, Vedovato, Veitch, Veitch,
  Venkateswara, Venugopalan, Verkindt, Vetrano, Vicer\'e, Viets, Vinciguerra,
  Vine, Vinet, Vitale, Vo, Vocca, Vorvick, Voss, Vousden, Vyatchanin, Wade,
  Wade, Wade, Wald, Walet, Walker, Wallace, Walsh, Wang, Wang, Wang, Wang,
  Wang, Wang, Ward, Warner, Was, Watchi, Weaver, Wei, Weinert, Weinstein,
  Weiss, Wen, Wessel, We\ss{}els, Westphal, Wette, Whelan, Whiting, Whittle,
  Williams, Williams, Williamson, Willis, Willke, Wimmer, Winkler, Wipf,
  Wittel, Woan, Woehler, Wofford, Wong, Worden, Wright, Wu, Wu, Yam, Yamamoto,
  Yancey, Yap, Yu, Yu, Yvert, Zadro\ifmmode~\dot{z}\else \.{z}\fi{}ny, Zanolin,
  Zelenova, Zendri, Zevin, Zhang, Zhang, Zhang, Zhang, Zhao, Zhou, Zhou, Zhu,
  Zimmerman, Zucker, \& Zweizig}]{PhysRevLett.118.221101}
---. 2017{\natexlab{a}}, Phys. Rev. Lett., 118, 221101,
  \dodoi{10.1103/PhysRevLett.118.221101}

\bibitem[{Abbott {et~al.}(2017{\natexlab{b}})Abbott, Abbott, Abbott, Acernese,
  Ackley, Adams, Adams, Addesso, Adhikari, Adya, Affeldt, Afrough, Agarwal,
  Agathos, Agatsuma, Aggarwal, Aguiar, Aiello, Ain, Ajith, Allen, Allen,
  Allocca, Aloy, Altin, Amato, Ananyeva, Anderson, Anderson, Angelova, Antier,
  Appert, Arai, Araya, Areeda, Arnaud, Arun, Ascenzi, Ashton, Ast, Aston,
  Astone, Atallah, Aufmuth, Aulbert, AultONeal, Austin, Avila-Alvarez, Babak,
  Bacon, Bader, Bae, Baker, Baldaccini, Ballardin, Ballmer, Banagiri, Barayoga,
  Barclay, Barish, Barker, Barkett, Barone, Barr, Barsotti, Barsuglia, Barta,
  Bartlett, Bartos, Bassiri, Basti, Batch, Bawaj, Bayley, Bazzan, Bécsy, Beer,
  Bejger, Belahcene, Bell, Berger, Bergmann, Bero, Berry, Bersanetti,
  Bertolini, Betzwieser, Bhagwat, Bhandare, Bilenko, Billingsley, Billman,
  Birch, Birney, Birnholtz, Biscans, Biscoveanu, Bisht, Bitossi, Biwer,
  Bizouard, Blackburn, Blackman, Blair, Blair, Blair, Bloemen, Bock, Bode,
  Boer, Bogaert, Bohe, Bondu, Bonilla, Bonnand, Boom, Bork, Boschi, Bose,
  Bossie, Bouffanais, Bozzi, Bradaschia, Brady, Branchesi, Brau, Briant,
  Brillet, Brinkmann, Brisson, Brockill, Broida, Brooks, Brown, Brown, Brunett,
  Buchanan, Buikema, Bulik, Bulten, Buonanno, Buskulic, Buy, Byer, Cabero,
  Cadonati, Cagnoli, Cahillane, Calderón~Bustillo, Callister, Calloni, Camp,
  Canepa, Canizares, Cannon, Cao, Cao, Capano, Capocasa, Carbognani, Caride,
  Carney, Diaz, Casentini, Caudill, Cavaglià, Cavalier, Cavalieri, Cella,
  Cepeda, Cerdá-Durán, Cerretani, Cesarini, Chamberlin, Chan, Chao, Charlton,
  Chase, Chassande-Mottin, Chatterjee, Chatziioannou, Cheeseboro, Chen, Chen,
  Chen, Cheng, Chia, Chincarini, Chiummo, Chmiel, Cho, Cho, Chow, Christensen,
  Chu, Chua, Chua, Chung, Chung, Ciani, Ciolfi, Cirelli, Cirone, Clara, Clark,
  Clearwater, Cleva, Cocchieri, Coccia, Cohadon, Cohen, Colla, Collette,
  Cominsky, Constancio~Jr., Conti, Cooper, Corban, Corbitt, Cordero-Carrión,
  Corley, Cornish, Corsi, Cortese, Costa, Coughlin, Coughlin, Coulon,
  Countryman, Couvares, Covas, Cowan, Coward, Cowart, Coyne, Coyne, Creighton,
  Creighton, Cripe, Crowder, Cullen, Cumming, Cunningham, Cuoco, Canton,
  Dálya, Danilishin, D’Antonio, Danzmann, Dasgupta, Costa, Dattilo, Dave,
  Davier, Davis, Daw, Day, De, DeBra, Degallaix, Laurentis, Deléglise, Pozzo,
  Demos, Denker, Dent, Pietri, Dergachev, Rosa, DeRosa, Rossi, DeSalvo, Varona,
  Devenson, Dhurandhar, Díaz, Fiore, Giovanni, Girolamo, Lieto, Pace, Palma,
  Renzo, Doctor, Dolique, Donovan, Dooley, Doravari, Dorrington, Douglas,
  Dovale~Álvarez, Downes, Drago, Dreissigacker, Driggers, Du, Ducrot, Dupej,
  Dwyer, Edo, Edwards, Effler, Eggenstein, Ehrens, Eichholz, Eikenberry,
  Eisenstein, Essick, Estevez, Etienne, Etzel, Evans, Evans, Factourovich,
  Fafone, Fair, Fairhurst, Fan, Farinon, Farr, Farr, Fauchon-Jones, Favata,
  Fays, Fee, Fehrmann, Feicht, Fejer, Fernandez-Galiana, Ferrante, Ferreira,
  Ferrini, Fidecaro, Finstad, Fiori, Fiorucci, Fishbach, Fisher, Fitz-Axen,
  Flaminio, Fletcher, Fong, Font, Forsyth, Forsyth, Fournier, Frasca, Frasconi,
  Frei, Freise, Frey, Frey, Fries, Fritschel, Frolov, Fulda, Fyffe, Gabbard,
  Gadre, Gaebel, Gair, Gammaitoni, Ganija, Gaonkar, Garcia-Quiros, Garufi,
  Gateley, Gaudio, Gaur, Gayathri, Gehrels, Gemme, Genin, Gennai, George,
  George, Gergely, Germain, Ghonge, Ghosh, Ghosh, Ghosh, Giaime, Giardina,
  Giazotto, Gill, Glover, Goetz, Goetz, Gomes, Goncharov, González, Castro,
  Gopakumar, Gorodetsky, Gossan, Gosselin, Gouaty, Grado, Graef, Granata,
  Grant, Gras, Gray, Greco, Green, Gretarsson, Groot, Grote, Grunewald,
  Gruning, Guidi, Guo, Gupta, Gupta, Gushwa, Gustafson, Gustafson, Halim, Hall,
  Hall, Hamilton, Hammond, Haney, Hanke, Hanks, Hanna, Hannam, Hannuksela,
  Hanson, Hardwick, Harms, Harry, Harry, Hart, Haster, Haughian, Healy,
  Heidmann, Heintze, Heitmann, Hello, Hemming, Hendry, Heng, Hennig,
  Heptonstall, Heurs, Hild, Hinderer, Hoak, Hofman, Holt, Holz, Hopkins, Horst,
  Hough, Houston, Howell, Hreibi, Hu, Huerta, Huet, Hughey, Husa, Huttner,
  Huynh-Dinh, Indik, Inta, Intini, Isa, Isac, Isi, Iyer, Izumi, Jacqmin, Jani,
  Jaranowski, Jawahar, Jiménez-Forteza, Johnson, Johnson-McDaniel, Jones,
  Jones, Jonker, Ju, Junker, Kalaghatgi, Kalogera, Kamai, Kandhasamy, Kang,
  Kanner, Kapadia, Karki, Karvinen, Kasprzack, Kastaun, Katolik, Katsavounidis,
  Katzman, Kaufer, Kawabe, Kéfélian, Keitel, Kemball, Kennedy, Kent, Key,
  Khalili, Khan, Khan, Khan, Khazanov, Kijbunchoo, Kim, Kim, Kim, Kim, Kim,
  Kim, Kimbrell, King, King, Kinley-Hanlon, Kirchhoff, Kissel, Kleybolte,
  Klimenko, Knowles, Koch, Koehlenbeck, Koley, Kondrashov, Kontos, Korobko,
  Korth, Kowalska, Kozak, Krämer, Kringel, Krishnan, Królak, Kuehn, Kumar,
  Kumar, Kumar, Kuo, Kutynia, Kwang, Lackey, Lai, Landry, Lang, Lange, Lantz,
  Lanza, Lartaux-Vollard, Lasky, Laxen, Lazzarini, Lazzaro, Leaci, Leavey, Lee,
  Lee, Lee, Lee, Lee, Lehmann, Lenon, Leonardi, Leroy, Letendre, Levin, Li,
  Linker, Littenberg, Liu, Lo, Lockerbie, London, Lord, Lorenzini, Loriette,
  Lormand, Losurdo, Lough, Lousto, Lovelace, Lück, Lumaca, Lundgren, Lynch,
  Ma, Macas, Macfoy, Machenschalk, MacInnis, Macleod, Magaña~Hernandez,
  Magaña-Sandoval, Magaña~Zertuche, Magee, Majorana, Maksimovic, Man, Mandic,
  Mangano, Mansell, Manske, Mantovani, Marchesoni, Marion, Márka, Márka,
  Markakis, Markosyan, Markowitz, Maros, Marquina, Martelli, Martellini,
  Martin, Martin, Martynov, Mason, Massera, Masserot, Massinger, Masso-Reid,
  Mastrogiovanni, Matas, Matichard, Matone, Mavalvala, Mazumder, McCarthy,
  McClelland, McCormick, McCuller, McGuire, McIntyre, McIver, McManus, McNeill,
  McRae, McWilliams, Meacher, Meadors, Mehmet, Meidam, Mejuto-Villa, Melatos,
  Mendell, Mercer, Merilh, Merzougui, Meshkov, Messenger, Messick, Metzdorff,
  Meyers, Miao, Michel, Middleton, Mikhailov, Milano, Miller, Miller, Miller,
  Millhouse, Milovich-Goff, Minazzoli, Minenkov, Ming, Mishra, Mitra,
  Mitrofanov, Mitselmakher, Mittleman, Moffa, Moggi, Mogushi, Mohan, Mohapatra,
  Montani, Moore, Moraru, Moreno, Morriss, Mours, Mow-Lowry, Mueller, Muir,
  Mukherjee, Mukherjee, Mukherjee, Mukund, Mullavey, Munch, Muñiz, Muratore,
  Murray, Napier, Nardecchia, Naticchioni, Nayak, Neilson, Nelemans, Nelson,
  Nery, Neunzert, Nevin, Newport, Newton, Ng, Nguyen, Nichols, Nielsen,
  Nissanke, Nitz, Noack, Nocera, Nolting, North, Nuttall, Oberling, O’Dea,
  Ogin, Oh, Oh, Ohme, Okada, Oliver, Oppermann, Oram, O’Reilly, Ormiston,
  Ortega, O’Shaughnessy, Ossokine, Ottaway, Overmier, Owen, Pace, Page, Page,
  Pai, Pai, Palamos, Palashov, Palomba, Pal-Singh, Pan, Pan, Pang, Pang,
  Pankow, Pannarale, Pant, Paoletti, Paoli, Papa, Parida, Parker, Pascucci,
  Pasqualetti, Passaquieti, Passuello, Patil, Patricelli, Pearlstone, Pedraza,
  Pedurand, Pekowsky, Pele, Penn, Perez, Perreca, Perri, Pfeiffer, Phelps,
  Piccinni, Pichot, Piergiovanni, Pierro, Pillant, Pinard, Pinto, Pirello,
  Pitkin, Poe, Poggiani, Popolizio, Porter, Post, Powell, Prasad, Pratt,
  Pratten, Predoi, Prestegard, Prijatelj, Principe, Privitera, Prodi,
  Prokhorov, Puncken, Punturo, Puppo, Pürrer, Qi, Quetschke, Quintero,
  Quitzow-James, Raab, Rabeling, Radkins, Raffai, Raja, Rajan, Rajbhandari,
  Rakhmanov, Ramirez, Ramos-Buades, Rapagnani, Raymond, Razzano, Read,
  Regimbau, Rei, Reid, Reitze, Ren, Reyes, Ricci, Ricker, Rieger, Riles, Rizzo,
  Robertson, Robie, Robinet, Rocchi, Rolland, Rollins, Roma, Romano, Romel,
  Romie, Rosińska, Ross, Rowan, Rüdiger, Ruggi, Rutins, Ryan, Sachdev,
  Sadecki, Sadeghian, Sakellariadou, Salconi, Saleem, Salemi, Samajdar, Sammut,
  Sampson, Sanchez, Sanchez, Sanchis-Gual, Sandberg, Sanders, Sassolas,
  Sathyaprakash, Saulson, Sauter, Savage, Sawadsky, Schale, Scheel, Scheuer,
  Schmidt, Schmidt, Schnabel, Schofield, Schönbeck, Schreiber, Schuette,
  Schulte, Schutz, Schwalbe, Scott, Scott, Seidel, Sellers, Sengupta, Sentenac,
  Sequino, Sergeev, Shaddock, Shaffer, Shah, Shahriar, Shaner, Shao, Shapiro,
  Shawhan, Sheperd, Shoemaker, Shoemaker, Siellez, Siemens, Sieniawska, Sigg,
  Silva, Singer, Singh, Singhal, Sintes, Slagmolen, Smith, Smith, Smith,
  Somala, Son, Sonnenberg, Sorazu, Sorrentino, Souradeep, Spencer, Srivastava,
  Staats, Staley, Steinke, Steinlechner, Steinlechner, Steinmeyer, Stevenson,
  Stone, Stops, Strain, Stratta, Strigin, Strunk, Sturani, Stuver,
  Summerscales, Sun, Sunil, Suresh, Sutton, Swinkels, Szczepańczyk, Tacca,
  Tait, Talbot, Talukder, Tanner, Tápai, Taracchini, Tasson, Taylor, Taylor,
  Tewari, Theeg, Thies, Thomas, Thomas, Thomas, Thorne, Thorne, Thrane, Tiwari,
  Tiwari, Tokmakov, Toland, Tonelli, Tornasi, Torres-Forné, Torrie, Töyrä,
  Travasso, Traylor, Trinastic, Tringali, Trozzo, Tsang, Tse, Tso, Tsukada,
  Tsuna, Tuyenbayev, Ueno, Ugolini, Unnikrishnan, Urban, Usman, Vahlbruch,
  Vajente, Valdes, Bakel, Beuzekom, Brand, Broeck, Vander-Hyde, Schaaf,
  Heijningen, Veggel, Vardaro, Varma, Vass, Vasúth, Vecchio, Vedovato, Veitch,
  Veitch, Venkateswara, Venugopalan, Verkindt, Vetrano, Viceré, Viets,
  Vinciguerra, Vine, Vinet, Vitale, Vo, Vocca, Vorvick, Vyatchanin, Wade, Wade,
  Wade, Walet, Walker, Wallace, Walsh, Wang, Wang, Wang, Wang, Wang, Ward,
  Warner, Was, Watchi, Weaver, Wei, Weinert, Weinstein, Weiss, Wen, Wessel,
  Weßels, Westerweck, Westphal, Wette, Whelan, Whitcomb, Whiting, Whittle,
  Wilken, Williams, Williams, Williamson, Willis, Willke, Wimmer, Winkler,
  Wipf, Wittel, Woan, Woehler, Wofford, Wong, Worden, Wright, Wu, Wysocki,
  Xiao, Yamamoto, Yancey, Yang, Yap, Yazback, Yu, Yu, Yvert, Zadrożny,
  Zanolin, Zelenova, Zendri, Zevin, Zhang, Zhang, Zhang, Zhang, Zhao, Zhou,
  Zhou, Zhu, Zhu, Zimmerman, Zucker, Zweizig, Collaboration, Collaboration),
  Burns, Veres, Kocevski, Racusin, Goldstein, Connaughton, Briggs, Blackburn,
  Hamburg, Hui, Kienlin, McEnery, Preece, Wilson-Hodge, Bissaldi, Cleveland,
  Gibby, Giles, Kippen, McBreen, Meegan, Paciesas, Poolakkil, Roberts, Stanbro,
  ray Burst~Monitor), Savchenko, Ferrigno, Kuulkers, Bazzano, Bozzo, Brandt,
  Chenevez, Courvoisier, Diehl, Domingo, Hanlon, Jourdain, Laurent, Lebrun,
  Lutovinov, Mereghetti, Natalucci, Rodi, Roques, Sunyaev, Ubertini, \&
  (INTEGRAL)}]{Abbott_2017}
---. 2017{\natexlab{b}}, The Astrophysical Journal Letters, 848, L13,
  \dodoi{10.3847/2041-8213/aa920c}

\bibitem[{Abbott {et~al.}(2023)Abbott, Abbott, Acernese, Ackley, Adams,
  Adhikari, Adhikari, Adya, Affeldt, Agarwal, Agathos, Agatsuma, Aggarwal,
  Aguiar, Aiello, Ain, Ajith, Akutsu, de~Alarc\'on, Akcay, Albanesi, Allocca,
  Altin, Amato, Anand, Anand, Ananyeva, Anderson, Anderson, Ando, Andrade,
  Andres, Andri\ifmmode~\acute{c}\else \'{c}\fi{}, Angelova, Ansoldi, Antelis,
  Antier, Antonini, Appert, Arai, Arai, Arai, Araki, Araya, Araya, Areeda,
  Ar\`ene, Aritomi, Arnaud, Arogeti, Aronson, Arun, Asada, Asali, Ashton, Aso,
  Assiduo, Aston, Astone, Aubin, Austin, Babak, Badaracco, Bader, Badger, Bae,
  Bae, Baer, Bagnasco, Bai, Baiotti, Baird, Bajpai, Ball, Ballardin, Ballmer,
  Balsamo, Baltus, Banagiri, Bankar, Barayoga, Barbieri, Barish, Barker,
  Barneo, Barone, Barr, Barsotti, Barsuglia, Barta, Bartlett, Barton, Bartos,
  Bassiri, Basti, Bawaj, Bayley, Baylor, Bazzan, B\'ecsy, Bedakihale, Bejger,
  Belahcene, Benedetto, Beniwal, Bennett, Bentley, BenYaala, Bergamin, Berger,
  Bernuzzi, Berry, Bersanetti, Bertolini, Betzwieser, Beveridge, Bhandare,
  Bhardwaj, Bhattacharjee, Bhaumik, Bilenko, Billingsley, Bini, Birney,
  Birnholtz, Biscans, Bischi, Biscoveanu, Bisht, Biswas, Bitossi, Bizouard,
  Blackburn, Blair, Blair, Blair, Bobba, Bode, Boer, Bogaert, Boldrini,
  Bonavena, Bondu, Bonilla, Bonnand, Booker, Boom, Bork, Boschi, Bose, Bose,
  Bossilkov, Boudart, Bouffanais, Bozzi, Bradaschia, Brady, Bramley, Branch,
  Branchesi, Brandt, Brau, Breschi, Briant, Briggs, Brillet, Brinkmann,
  Brockill, Brooks, Brooks, Brown, Brunett, Bruno, Bruntz, Bryant, Bulik,
  Bulten, Buonanno, Buscicchio, Buskulic, Buy, Byer, Cadonati, Cagnoli,
  Cahillane, Bustillo, Callaghan, Callister, Calloni, Cameron, Camp, Canepa,
  Canevarolo, Cannavacciuolo, Cannon, Cao, Cao, Capocasa, Capote, Carapella,
  Carbognani, Carlin, Carney, Carpinelli, Carrillo, Carullo, Carver, Diaz,
  Casentini, Castaldi, Caudill, Cavagli\`a, Cavalier, Cavalieri, Ceasar, Cella,
  Cerd\'a-Dur\'an, Cesarini, Chaibi, Chakravarti, Subrahmanya, Champion, Chan,
  Chan, Chan, Chan, Chan, Chandra, Chanial, Chao, Chapman-Bird, Charlton,
  Chase, Chassande-Mottin, Chatterjee, Chatterjee, Chatterjee, Chaturvedi,
  Chaty, Chatziioannou, Chen, Chen, Chen, Chen, Chen, Chen, Chen, Chen, Cheng,
  Cheong, Cheung, Chia, Chiadini, Chiang, Chiarini, Chierici, Chincarini,
  Chiofalo, Chiummo, Cho, Cho, Choudhary, Choudhary, Christensen, Chu, Chu,
  Chu, Chua, Chung, Ciani, Ciecielag, Cie\ifmmode~\acute{s}\else \'{s}\fi{}lar,
  Cifaldi, Ciobanu, Ciolfi, Cipriano, Cirone, Clara, Clark, Clark, Clarke,
  Clearwater, Clesse, Cleva, Coccia, Codazzo, Cohadon, Cohen, Cohen, Colleoni,
  Collette, Colombo, Colpi, Compton, Constancio, Conti, Cooper, Corban,
  Corbitt, Cordero-Carri\'on, Corezzi, Corley, Cornish, Corre, Corsi, Cortese,
  Costa, Cotesta, Coughlin, Coulon, Countryman, Cousins, Couvares, Coward,
  Cowart, Coyne, Coyne, Creighton, Creighton, Criswell, Croquette, Crowder,
  Cudell, Cullen, Cumming, Cummings, Cunningham, Cuoco, Cury\l{}o, Dabadie,
  Canton, Dall'Osso, D\'alya, Dana, DaneshgaranBajastani, D'Angelo, Danila,
  Danilishin, D'Antonio, Danzmann, Darsow-Fromm, Dasgupta, Datrier, Datta,
  Dattilo, Dave, Davier, Davies, Davis, Davis, Daw, Dean, DeBra, Deenadayalan,
  Degallaix, De~Laurentis, Del\'eglise, Del~Favero, De~Lillo, De~Lillo,
  Del~Pozzo, DeMarchi, De~Matteis, D'Emilio, Demos, Dent, Depasse, De~Pietri,
  De~Rosa, De~Rossi, DeSalvo, De~Simone, Dhurandhar, D\'{\i}az, Diaz-Ortiz,
  Didio, Dietrich, Di~Fiore, Di~Fronzo, Di~Giorgio, Di~Giovanni, Di~Giovanni,
  Di~Girolamo, Di~Lieto, Ding, Di~Pace, Di~Palma, Di~Renzo, Divakarla,
  Dmitriev, Doctor, D'Onofrio, Donovan, Dooley, Doravari, Dorrington, Drago,
  Driggers, Drori, Ducoin, Dupej, Durante, D'Urso, Duverne, Dwyer, Eassa,
  Easter, Ebersold, Eckhardt, Eddolls, Edelman, Edo, Edy, Effler, Eguchi,
  Eichholz, Eikenberry, Eisenmann, Eisenstein, Ejlli, Engelby, Enomoto, Errico,
  Essick, Estell\'es, Estevez, Etienne, Etzel, Evans, Evans, Ewing, Fafone,
  Fair, Fairhurst, Farah, Farinon, Farr, Farr, Farrow, Fauchon-Jones, Favaro,
  Favata, Fays, Fazio, Feicht, Fejer, Fenyvesi, Ferguson, Fernandez-Galiana,
  Ferrante, Ferreira, Fidecaro, Figura, Fiori, Fishbach, Fisher, Fittipaldi,
  Fiumara, Flaminio, Floden, Fong, Font, Fornal, Forsyth, Franke, Frasca,
  Frasconi, Frederick, Freed, Frei, Freise, Frey, Fritschel, Frolov, Fronz\'e,
  Fujii, Fujikawa, Fukunaga, Fukushima, Fulda, Fyffe, Gabbard, Gadre, Gair,
  Gais, Galaudage, Gamba, Ganapathy, Ganguly, Gao, Gaonkar, Garaventa,
  Garc\'{\i}a, Garc\'{\i}a-N\'u\~nez, Garc\'{\i}a-Quir\'os, Garufi, Gateley,
  Gaudio, Gayathri, Ge, Gemme, Gennai, George, George, Gerberding, Gergely,
  Gewecke, Ghonge, Ghosh, Ghosh, Ghosh, Ghosh, Giacomazzo, Giacoppo, Giaime,
  Giardina, Gibson, Gier, Giesler, Giri, Gissi, Glanzer, Gleckl, Godwin,
  Golomb, Goetz, Goetz, Gohlke, Goncharov, Gonz\'alez, Gopakumar, Gosselin,
  Gouaty, Gould, Grace, Grado, Granata, Granata, Grant, Gras, Grassia, Gray,
  Gray, Greco, Green, Green, Gretarsson, Gretarsson, Griffith, Griffiths,
  Griggs, Grignani, Grimaldi, Grimm, Grote, Grunewald, Gruning, Guerra, Guidi,
  Guimaraes, Guix\'e, Gulati, Guo, Guo, Gupta, Gupta, Gupta, Gustafson,
  Gustafson, Guzman, Ha, Haegel, Hagiwara, Haino, Halim, Hall, Hamilton,
  Hammond, Han, Haney, Hanks, Hanna, Hannam, Hannuksela, Hansen, Hansen,
  Hanson, Harder, Hardwick, Haris, Harms, Harry, Harry, Hartwig, Hasegawa,
  Haskell, Hasskew, Haster, Hattori, Haughian, Hayakawa, Hayama, Hayes, Healy,
  Heidmann, Heidt, Heintze, Heinze, Heinzel, Heitmann, Hellman, Hello,
  Helmling-Cornell, Hemming, Hendry, Heng, Hennes, Hennig, Hennig, Hernandez,
  Vivanco, Heurs, Hild, Hill, Himemoto, Hines, Hiranuma, Hirata, Hirose,
  Hochheim, Hofman, Hohmann, Holcomb, Holland, Hollows, Holmes, Holt, Holz,
  Hong, Hopkins, Hough, Hourihane, Howell, Hoy, Hoyland, Hreibi, Hsieh, Hsu,
  Huang, Huang, Huang, Huang, Huang, Huang, H\"ubner, Huddart, Hughey, Hui,
  Hui, Husa, Huttner, Huxford, Huynh-Dinh, Ide, Idzkowski, Iess, Ikenoue, Imam,
  Inayoshi, Ingram, Inoue, Ioka, Isi, Isleif, Ito, Itoh, Iyer, Izumi,
  JaberianHamedan, Jacqmin, Jadhav, Jadhav, James, Jan, Jani, Janquart,
  Janssens, Janthalur, Jaranowski, Jariwala, Jaume, Jenkins, Jenner, Jeon,
  Jeunon, Jia, Jin, Johns, Jones, Jones, Jones, Jones, Jones, Jonker, Ju, Jung,
  Jung, Junker, Juste, Kaihotsu, Kajita, Kakizaki, Kalaghatgi, Kalogera, Kamai,
  Kamiizumi, Kanda, Kandhasamy, Kang, Kanner, Kao, Kapadia, Kapasi, Karat,
  Karathanasis, Karki, Kashyap, Kasprzack, Kastaun, Katsanevas, Katsavounidis,
  Katzman, Kaur, Kawabe, Kawaguchi, Kawai, Kawasaki, K\'ef\'elian, Keitel, Key,
  Khadka, Khalili, Khan, Khazanov, Khetan, Khursheed, Kijbunchoo, Kim, Kim,
  Kim, Kim, Kim, Kim, Kimball, Kimura, Kinley-Hanlon, Kirchhoff, Kissel, Kita,
  Kitazawa, Kleybolte, Klimenko, Knee, Knowles, Knyazev, Koch, Koekoek, Kojima,
  Kokeyama, Koley, Kolitsidou, Kolstein, Komori, Kondrashov, Kong, Kontos,
  Koper, Korobko, Kotake, Kovalam, Kozak, Kozakai, Kozu, Kringel, Krishnendu,
  Kr\'olak, Kuehn, Kuei, Kuijer, Kulkarni, Kumar, Kumar, Kumar, Kumar, Kume,
  Kuns, Kuo, Kuo, Kuromiya, Kuroyanagi, Kusayanagi, Kuwahara, Kwak, Lagabbe,
  Laghi, Lalande, Lam, Lamberts, Landry, Landry, Lane, Lang, Lange, Lantz,
  La~Rosa, Lartaux-Vollard, Lasky, Laxen, Lazzarini, Lazzaro, Leaci, Leavey,
  Lecoeuche, Lee, Lee, Lee, Lee, Lee, Lee, Lehmann, Lema\^{\i}tre, Leonardi,
  Leroy, Letendre, Levesque, Levin, Leviton, Leyde, Li, Li, Li, Li, Li, Li,
  Lin, Lin, Lin, Lin, Lin, Linde, Linker, Linley, Littenberg, Liu, Liu, Liu,
  Liu, Llamas, Llorens-Monteagudo, Lo, Lockwood, Loh, London, Longo, Lopez,
  Portilla, Lorenzini, Loriette, Lormand, Losurdo, Lott, Lough, Lousto,
  Lovelace, Lucaccioni, L\"uck, Lumaca, Lundgren, Luo, Lynam, Macas, MacInnis,
  Macleod, MacMillan, Macquet, Hernandez, Magazz\`u, Magee, Maggiore, Magnozzi,
  Mahesh, Majorana, Makarem, Maksimovic, Maliakal, Malik, Man, Mandic, Mangano,
  Mango, Mansell, Manske, Mantovani, Mapelli, Marchesoni, Marchio, Marion,
  Mark, M\'arka, M\'arka, Markakis, Markosyan, Markowitz, Maros, Marquina,
  Marsat, Martelli, Martin, Martin, Martinez, Martinez, Martinez, Martinovic,
  Martynov, Marx, Masalehdan, Mason, Massera, Masserot, Massinger, Masso-Reid,
  Mastrogiovanni, Matas, Mateu-Lucena, Matichard, Matiushechkina, Mavalvala,
  McCann, McCarthy, McClelland, McClincy, McCormick, McCuller, McGhee, McGuire,
  McIsaac, McIver, McRae, McWilliams, Meacher, Mehmet, Mehta, Meijer, Melatos,
  Melchor, Mendell, Menendez-Vazquez, Menoni, Mercer, Mereni, Merfeld, Merilh,
  Merritt, Merzougui, Meshkov, Messenger, Messick, Meyers, Meylahn, Mhaske,
  Miani, Miao, Michaloliakos, Michel, Michimura, Middleton, Milano, Miller,
  Miller, Miller, Miller, Millhouse, Mills, Milotti, Minazzoli, Minenkov, Mio,
  Mir, Miravet-Ten\'es, Mishra, Mishra, Mistry, Mitra, Mitrofanov,
  Mitselmakher, Mittleman, Miyakawa, Miyamoto, Miyazaki, Miyo, Miyoki, Mo,
  Modafferi, Moguel, Mogushi, Mohapatra, Mohite, Molina, Molina-Ruiz, Mondin,
  Montani, Moore, Moraru, Morawski, More, Moreno, Moreno, Mori, Morisaki,
  Moriwaki, Morr\'as, Mours, Mow-Lowry, Mozzon, Muciaccia, Mukherjee,
  Mukherjee, Mukherjee, Mukherjee, Mukherjee, Mukund, Mullavey, Munch, Mu\~niz,
  Murray, Musenich, Muusse, Nadji, Nagano, Nagano, Nagar, Nakamura, Nakano,
  Nakano, Nakashima, Nakayama, Napolano, Nardecchia, Narikawa, Naticchioni,
  Nayak, Nayak, Negishi, Neil, Neilson, Nelemans, Nelson, Nery, Neubauer,
  Neunzert, Ng, Ng, Nguyen, Nguyen, Nguyen, Quynh, Ni, Nichols, Nishizawa,
  Nissanke, Nitoglia, Nocera, Norman, North, Nozaki, Siles, Nuttall, Oberling,
  O'Brien, Obuchi, O'Dell, Oelker, Ogaki, Oganesyan, Oh, Oh, Oh, Ohashi,
  Ohishi, Ohkawa, Ohme, Ohta, Okada, Okutani, Okutomi, Olivetto, Oohara, Ooi,
  Oram, O'Reilly, Ormiston, Ormsby, Ortega, O'Shaughnessy, O'Shea, Oshino,
  Ossokine, Osthelder, Otabe, Ottaway, Overmier, Pace, Pagano, Page,
  Pagliaroli, Pai, Pai, Palamos, Palashov, Palomba, Pan, Pan, Panda, Pang,
  Pang, Pankow, Pannarale, Pant, Panther, Paoletti, Paoli, Paolone, Parisi,
  Park, Park, Parker, Pascucci, Pasqualetti, Passaquieti, Passuello, Patel,
  Pathak, Patricelli, Patron, Paul, Payne, Pedraza, Pegoraro, Pele, Arellano,
  Penn, Perego, Pereira, Pereira, Perez, P\'erigois, Perkins, Perreca,
  Perri\`es, Petermann, Petterson, Pfeiffer, Pham, Phukon, Piccinni, Pichot,
  Piendibene, Piergiovanni, Pierini, Pierro, Pillant, Pillas, Pilo, Pinard,
  Pinto, Pinto, Piotrzkowski, Piotrzkowski, Pirello, Pitkin, Placidi, Planas,
  Plastino, Pluchar, Poggiani, Polini, Pong, Ponrathnam, Popolizio, Porter,
  Poulton, Powell, Pracchia, Pradier, Prajapati, Prasai, Prasanna, Pratten,
  Principe, Prodi, Prokhorov, Prosposito, Prudenzi, Puecher, Punturo, Puosi,
  Puppo, P\"urrer, Qi, Quetschke, Quitzow-James, Raab, Raaijmakers, Radkins,
  Radulesco, Raffai, Rail, Raja, Rajan, Ramirez, Ramirez, Ramos-Buades, Rana,
  Rapagnani, Rapol, Ray, Raymond, Raza, Razzano, Read, Rees, Regimbau, Rei,
  Reid, Reid, Reitze, Relton, Renzini, Rettegno, Reza, Rezac, Ricci, Richards,
  Richardson, Richardson, Riemenschneider, Riles, Rinaldi, Rink, Rizzo,
  Robertson, Robie, Robinet, Rocchi, Rodriguez, Rolland, Rollins, Romanelli,
  Romano, Romel, Romero-Rodr\'{\i}guez, Romero-Shaw, Romie, Ronchini, Rosa,
  Rose, Rosi\ifmmode~\acute{n}\else \'{n}\fi{}ska, Ross, Rowan, Rowlinson, Roy,
  Roy, Roy, Rozza, Ruggi, Ryan, Sachdev, Sadecki, Sadiq, Sago, Saito, Saito,
  Sakai, Sakai, Sakellariadou, Sakuno, Salafia, Salconi, Saleem, Salemi,
  Samajdar, Sanchez, Sanchez, Sanchez, Sanchis-Gual, Sanders, Sanuy, Saravanan,
  Sarin, Sassolas, Satari, Sathyaprakash, Sato, Sato, Sauter, Savage, Sawada,
  Sawant, Sawant, Sayah, Schaetzl, Scheel, Scheuer, Schiworski, Schmidt,
  Schmidt, Schnabel, Schneewind, Schofield, Sch\"onbeck, Schulte, Schutz,
  Schwartz, Scott, Scott, Seglar-Arroyo, Sekiguchi, Sekiguchi, Sellers,
  Sengupta, Sentenac, Seo, Sequino, Sergeev, Setyawati, Shaffer, Shahriar,
  Shams, Shao, Sharma, Sharma, Shawhan, Shcheblanov, Shibagaki, Shikauchi,
  Shimizu, Shimoda, Shimode, Shinkai, Shishido, Shoda, Shoemaker, Shoemaker,
  ShyamSundar, Sieniawska, Sigg, Singer, Singh, Singh, Singha, Sintes, Sipala,
  Skliris, Slagmolen, Slaven-Blair, Smetana, Smith, Smith, Soldateschi, Somala,
  Somiya, Son, Soni, Soni, Sordini, Sorrentino, Sorrentino, Sotani, Soulard,
  Souradeep, Sowell, Spagnuolo, Spencer, Spera, Srinivasan, Srivastava,
  Srivastava, Staats, Stachie, Steer, Steinhoff, Steinlechner, Steinlechner,
  Stevenson, Stops, Stover, Strain, Strang, Stratta, Strunk, Sturani, Stuver,
  Sudhagar, Sudhir, Sugimoto, Suh, Sullivan, Summerscales, Sun, Sun, Sunil,
  Sur, Suresh, Sutton, Suzuki, Suzuki, Swinkels, Szczepa\ifmmode~\acute{n}\else
  \'{n}\fi{}czyk, Szewczyk, Tacca, Tagoshi, Tait, Takahashi, Takahashi,
  Takamori, Takano, Takeda, Takeda, Talbot, Talbot, Tanaka, Tanaka, Tanaka,
  Tanaka, Tanaka, Tanasijczuk, Tanioka, Tanner, Tao, Tao, Mart\'{\i}n, Taranto,
  Tasson, Telada, Tenorio, Terhune, Terkowski, Thirugnanasambandam, Thomas,
  Thomas, Thomas, Thompson, Thondapu, Thorne, Thrane, Tiwari, Tiwari, Tiwari,
  Toivonen, Toland, Tolley, Tomaru, Tomigami, Tomura, Tonelli, Torres-Forn\'e,
  Torrie, e~Melo, T\"oyr\"a, Trapananti, Travasso, Traylor, Trevor, Tringali,
  Tripathee, Troiano, Trovato, Trozzo, Trudeau, Tsai, Tsai, Tsang, Tsang, Tsao,
  Tse, Tso, Tsubono, Tsuchida, Tsukada, Tsuna, Tsutsui, Tsuzuki, Turbang,
  Turconi, Tuyenbayev, Ubhi, Uchikata, Uchiyama, Udall, Ueda, Uehara, Ueno,
  Ueshima, Unnikrishnan, Uraguchi, Urban, Ushiba, Utina, Vahlbruch, Vajente,
  Vajpeyi, Valdes, Valentini, Valsan, van Bakel, van Beuzekom, van~den Brand,
  Van Den~Broeck, Vander-Hyde, van~der Schaaf, van Heijningen, Vanosky, van
  Putten, van Remortel, Vardaro, Vargas, Varma, Vas\'uth, Vecchio, Vedovato,
  Veitch, Veitch, Venneberg, Venugopalan, Verkindt, Verma, Verma, Veske,
  Vetrano, Vicer\'e, Vidyant, Viets, Vijaykumar, Villa-Ortega, Vinet, Virtuoso,
  Vitale, Vo, Vocca, von Reis, von Wrangel, Vorvick, Vyatchanin, Wade, Wade,
  Wagner, Walet, Walker, Wallace, Wallace, Walsh, Wang, Wang, Wang, Ward,
  Warner, Was, Washimi, Washington, Watchi, Weaver, Webster, Weinert,
  Weinstein, Weiss, Weller, Wellmann, Wen, We\ss{}els, Wette, Whelan, White,
  Whiting, Whittle, Wilken, Williams, Williams, Williamson, Willis, Willke,
  Wilson, Winkler, Wipf, Wlodarczyk, Woan, Woehler, Wofford, Wong, Wu, Wu, Wu,
  Wu, Wysocki, Xiao, Xu, Yamada, Yamamoto, Yamamoto, Yamamoto, Yamamoto,
  Yamashita, Yamazaki, Yang, Yang, Yang, Yang, Yang, Yap, Yeeles, Yelikar,
  Ying, Yokogawa, Yokoyama, Yokozawa, Yoo, Yoshioka, Yu, Yu, Yuzurihara,
  Zadro\ifmmode~\dot{z}\else \.{z}\fi{}ny, Zanolin, Zeidler, Zelenova, Zendri,
  Zevin, Zhan, Zhang, Zhang, Zhang, Zhang, Zhang, Zhao, Zhao, Zhao, Zhao,
  Zheng, Zhou, Zhou, Zhu, Zhu, Zimmerman, Zlochower, Zucker, \&
  Zweizig}]{theligoscientificcollaboration2022populationmergingcompactbinaries}
Abbott, R., Abbott, T.~D., Acernese, F., {et~al.} 2023, Population of Merging
  Compact Binaries Inferred Using Gravitational Waves through GWTC-3,  American
  Physical Society, \dodoi{10.1103/PhysRevX.13.011048}

\bibitem[{Alves {et~al.}(2023)Alves, Peiris, Lochner, McEwen, Kessler, \&
  Collaboration}]{Alves_2023}
Alves, C.~S., Peiris, H.~V., Lochner, M., {et~al.} 2023, The Astrophysical
  Journal Supplement Series, 265, 43, \dodoi{10.3847/1538-4365/acbb09}

\bibitem[{Andreoni {et~al.}(2021{\natexlab{a}})Andreoni, Coughlin, Almualla,
  Bellm, Bianco, Bulla, Cucchiara, Dietrich, Goobar, Kool, Li, Ragosta,
  Sagu{\'{e} }s-Carracedo, \& Singer}]{Andreoni_2021}
Andreoni, I., Coughlin, M.~W., Almualla, M., {et~al.} 2021{\natexlab{a}}, The
  Astrophysical Journal Supplement Series, 258, 5,
  \dodoi{10.3847/1538-4365/ac3bae}

\bibitem[{Andreoni {et~al.}(2021{\natexlab{b}})Andreoni, Coughlin, Kool,
  Kasliwal, Kumar, Bhalerao, Carracedo, Ho, Pang, Saraogi, Sharma, Shenoy,
  Burns, Ahumada, Anand, Singer, Perley, De, Fremling, Bellm, Bulla,
  Crellin-Quick, Dietrich, Drake, Duev, Goobar, Graham, Kaplan, Kulkarni,
  Laher, Mahabal, Shupe, Sollerman, Walters, \& Yao}]{Andreoni_Coughlin_2021}
Andreoni, I., Coughlin, M.~W., Kool, E.~C., {et~al.} 2021{\natexlab{b}}, The
  Astrophysical Journal, 918, 63, \dodoi{10.3847/1538-4357/ac0bc7}

\bibitem[{Andreoni {et~al.}(2022)Andreoni, Margutti, Salafia, Parazin, Villar,
  Coughlin, Yoachim, Mortensen, Brethauer, Smartt, Kasliwal, Alexander, Anand,
  Berger, Bernardini, Bianco, Blanchard, Bloom, Brocato, Bulla, Cartier, Cenko,
  Chornock, Copperwheat, Corsi, D’Ammando, D’Avanzo, Hélène~Datrier,
  Foley, Ghirlanda, Goobar, Grindlay, Hajela, Holz, Karambelkar, Kool, Lamb,
  Laskar, Levan, Maguire, May, Melandri, Milisavljevic, Miller, Nicholl,
  Nissanke, Palmese, Piranomonte, Rest, Sagués-Carracedo, Siellez, Singer,
  Smith, Steeghs, \& Tanvir}]{Andreoni_2022}
Andreoni, I., Margutti, R., Salafia, O.~S., {et~al.} 2022, The Astrophysical
  Journal Supplement Series, 260, 18, \dodoi{10.3847/1538-4365/ac617c}

\bibitem[{Barnes(2020)}]{Barnes_2020}
Barnes, J. 2020, Frontiers in Physics, 8, \dodoi{10.3389/fphy.2020.00355}

\bibitem[{Bartos \& Kowalski(2017)}]{Bartos&Kowalski_2017}
Bartos, I., \& Kowalski, M. 2017, Multimessenger Astronomy, 2399-2891 (IOP
  Publishing), \dodoi{10.1088/978-0-7503-1369-8}

\bibitem[{Bellm {et~al.}(2022)Bellm, Burke, Coughlin, Andreoni, Raiteri, \&
  Bonito}]{Bellm_2022}
Bellm, E.~C., Burke, C.~J., Coughlin, M.~W., {et~al.} 2022, The Astrophysical
  Journal Supplement Series, 258, 13, \dodoi{10.3847/1538-4365/ac4602}

\bibitem[{{Bianco} {et~al.}(2019){Bianco}, {Drout}, {Graham}, {Pritchard},
  {Biswas}, {Narayan}, {Andreoni}, {Cowperthwaite}, {Ribeiro}, {LSST
  Transient}, \& {Variable Stars Collaboration}}]{Bianco:2019}
{Bianco}, F.~B., {Drout}, M.~R., {Graham}, M.~L., {et~al.} 2019, Publications
  of the Astronomical Society of the Pacific, 131, 068002,
  \dodoi{10.1088/1538-3873/ab121a}

\bibitem[{Bianco {et~al.}(2021)Bianco, Ivezi{\'{c}}, Jones, Graham, Marshall,
  Saha, Strauss, Yoachim, Ribeiro, Anguita, Bauer, Bauer, Bellm, Blum, Brandt,
  Brough, Catelan, Clarkson, Connolly, Gawiser, Gizis, Hlo{\v{z}}ek, Kaviraj,
  Liu, Lochner, Mahabal, Mandelbaum, McGehee, Jr., Olsen, Peiris, Rhodes,
  Richards, Ridgway, Schwamb, Scolnic, Shemmer, Slater, Slosar, Smartt,
  Strader, Street, Trilling, Verma, Vivas, Wechsler, \& Willman}]{Bianco_2021}
Bianco, F.~B., Ivezi{\'{c}}, {\v{Z} }., Jones, R.~L., {et~al.} 2021, The
  Astrophysical Journal Supplement Series, 258, 1,
  \dodoi{10.3847/1538-4365/ac3e72}

\bibitem[{Bonito {et~al.}(2023)Bonito, Venuti, Ustamujic, Yoachim, Street,
  Prisinzano, Hartigan, Guarcello, Stassun, Giannini, Feigelson, o~Garatti,
  Orlando, Clarkson, McGehee, Bellm, \& Gizis}]{Bonito_2023}
Bonito, R., Venuti, L., Ustamujic, S., {et~al.} 2023, The Astrophysical Journal
  Supplement Series, 265, 27, \dodoi{10.3847/1538-4365/acb684}

\bibitem[{Bricman {et~al.}(2023)Bricman, van Velzen, Nicholl, \&
  Gomboc}]{Bucar_Bricman_2023}
Bricman, K.~B., van Velzen, S., Nicholl, M., \& Gomboc, A. 2023, The
  Astrophysical Journal Supplement Series, 268, 13,
  \dodoi{10.3847/1538-4365/ace1e7}

\bibitem[{{Bulla}(2019)}]{bulla2019possis}
{Bulla}, M. 2019, \mnras, 489, 5037, \dodoi{10.1093/mnras/stz2495}

\bibitem[{Chaudhary {et~al.}(2024)Chaudhary, Toivonen, Waratkar, Mo,
  Chatterjee, Antier, Brockill, Coughlin, Essick, Ghosh, Morisaki, Baral,
  Baylor, Adhikari, Brady, Cabourn~Davies, Dal~Canton, Cavaglia, Creighton,
  Choudhary, Chu, Clearwater, Davis, Dent, Drago, Ewing, Godwin, Guo, Hanna,
  Huxford, Harry, Katsavounidis, Kovalam, Li, Magee, Marx, Meacher, Messick,
  Morice-Atkinson, Pace, De~Pietri, Piotrzkowski, Roy, Sachdev, Singer, Singh,
  Szczepanczyk, Tang, Trevor, Tsukada, Villa-Ortega, Wen, \&
  Wysocki}]{Chaudhary_2024}
Chaudhary, S.~S., Toivonen, A., Waratkar, G., {et~al.} 2024, Proceedings of the
  National Academy of Sciences, 121, \dodoi{10.1073/pnas.2316474121}

\bibitem[{Clarke {et~al.}(2024)Clarke, Davenport, Gizis, Graham, Li, Fortino,
  Honaker, Sullivan, Alsayyad, Bosch, Knop, \& Bianco}]{Clarke_2024}
Clarke, R.~W., Davenport, J. R.~A., Gizis, J., {et~al.} 2024, The Astrophysical
  Journal Supplement Series, 272, 41, \dodoi{10.3847/1538-4365/ad4110}

\bibitem[{Coughlin {et~al.}(2020)Coughlin, Dietrich, Heinzel, Khetan, Antier,
  Bulla, Christensen, Coulter, \& Foley}]{Coughlin_2020}
Coughlin, M.~W., Dietrich, T., Heinzel, J., {et~al.} 2020, Phys. Rev. Res., 2,
  022006, \dodoi{10.1103/PhysRevResearch.2.022006}

\bibitem[{Coulter {et~al.}(2017)Coulter, Foley, Kilpatrick, Drout, Piro,
  Shappee, Siebert, Simon, Ulloa, Kasen, Madore, Murguia-Berthier, Pan,
  Prochaska, Ramirez-Ruiz, Rest, \& Rojas-Bravo}]{Coulter_2017}
Coulter, D.~A., Foley, R.~J., Kilpatrick, C.~D., {et~al.} 2017, Science, 358,
  1556, \dodoi{10.1126/science.aap9811}

\bibitem[{Criscienzo {et~al.}(2023)Criscienzo, Leccia, Braga, Musella, Bono,
  Dall’Ora, Fiorentino, Marconi, Molinaro, Ripepi, Carrell, Choi, Savarese,
  \& Schreiber}]{Di_Criscienzo_2023}
Criscienzo, M.~D., Leccia, S., Braga, V., {et~al.} 2023, The Astrophysical
  Journal Supplement Series, 265, 41, \dodoi{10.3847/1538-4365/acb825}

\bibitem[{Delgado {et~al.}(2014)Delgado, Saha, Chandrasekharan, Cook, Petry, \&
  Ridgway}]{Delgado_2014}
Delgado, F., Saha, A., Chandrasekharan, S., {et~al.} 2014, in Modeling, Systems
  Engineering, and Project Management for Astronomy VI, ed. G.~Z. Angeli \&
  P.~Dierickx, Vol. 9150, International Society for Optics and Photonics
  (SPIE), 422 -- 445, \dodoi{10.1117/12.2056898}

\bibitem[{Dietrich {et~al.}(2020)Dietrich, Coughlin, Pang, Bulla, Heinzel,
  Issa, Tews, \& Antier}]{Dietrich_2020}
Dietrich, T., Coughlin, M.~W., Pang, P. T.~H., {et~al.} 2020, Science, 370,
  1450, \dodoi{10.1126/science.abb4317}

\bibitem[{Feigelson {et~al.}(2023)Feigelson, Bianco, \&
  Bonito}]{Feigelson_2023}
Feigelson, E.~D., Bianco, F.~B., \& Bonito, R. 2023, The Astrophysical Journal
  Supplement Series, 268, 11, \dodoi{10.3847/1538-4365/ace616}

\bibitem[{Gizis {et~al.}(2022)Gizis, Yoachim, Jones, Hilligoss, \&
  Ji}]{Gizis_2022}
Gizis, J.~E., Yoachim, P., Jones, R.~L., Hilligoss, D., \& Ji, J. 2022, The
  Astrophysical Journal Supplement Series, 263, 23,
  \dodoi{10.3847/1538-4365/ac961f}

\bibitem[{Gris {et~al.}(2023)Gris, Regnault, Awan, Hook, Jha, Lochner, Sanchez,
  Scolnic, Sullivan, Yoachim, \& Collaboration}]{Gris_2023}
Gris, P., Regnault, N., Awan, H., {et~al.} 2023, The Astrophysical Journal
  Supplement Series, 264, 22, \dodoi{10.3847/1538-4365/ac9e58}

\bibitem[{Hernitschek \& Stassun(2021)}]{Hernitschek_2022}
Hernitschek, N., \& Stassun, K.~G. 2021, The Astrophysical Journal Supplement
  Series, 258, 4, \dodoi{10.3847/1538-4365/ac3baf}

\bibitem[{Ivezic {et~al.}(2019)Ivezic, Kahn, Tyson, Abel, Acosta, Allsman,
  Alonso, AlSayyad, Anderson, Andrew, Angel, Angeli, Ansari, Antilogus, Araujo,
  Armstrong, Arndt, Astier, Aubourg, \& Zhan}]{Ivezic_2019}
Ivezic, Z., Kahn, S., Tyson, J., {et~al.} 2019, The Astrophysical Journal, 873,
  111, \dodoi{10.3847/1538-4357/ab042c}

\bibitem[{Jones {et~al.}(2020)Jones, Yoachim, Ivezic, Neilsen, \&
  Ribeiro}]{jones_r_lynne_2020_4048838}
Jones, R.~L., Yoachim, P., Ivezic, Z., Neilsen, E.~H., \& Ribeiro, T. 2020,
  \dodoi{10.5281/zenodo.4048838}

\bibitem[{Jones {et~al.}(2014)Jones, Yoachim, Chandrasekharan, Connolly, Cook,
  Željko Ivezic, Krughoff, Petry, \& Ridgway}]{Jones_2014}
Jones, R.~L., Yoachim, P., Chandrasekharan, S., {et~al.} 2014, in Observatory
  Operations: Strategies, Processes, and Systems V, ed. A.~B. Peck, C.~R. Benn,
  \& R.~L. Seaman, Vol. 9149, International Society for Optics and Photonics
  (SPIE), 91490B, \dodoi{10.1117/12.2056835}

\bibitem[{Kova{\v c}evi{\'c} {et~al.}(2022)Kova{\v c}evi{\'c}, Radović, Ilić,
  Popović, Assef, Sánchez-Sáez, Nikutta, Raiteri, Yoon, Homayouni, Li,
  Caplar, Czerny, Panda, Ricci, Jankov, Landt, Wolf, Kovačević-Dojčinović,
  Lakićević, Đorđe V.~Savić, Vince, Simić, Čvorović Hajdinjak, \&
  Marčeta-Mandić}]{Kovacevic_2022}
Kova{\v c}evi{\'c}, A.~B., Radović, V., Ilić, D., {et~al.} 2022, The
  Astrophysical Journal Supplement Series, 262, 49,
  \dodoi{10.3847/1538-4365/ac88ce}

\bibitem[{Li \& Paczy{\'{n} }ski(1998)}]{Li_1998}
Li, L.-X., \& Paczy{\'{n} }ski, B. 1998, The Astrophysical Journal, 507, L59,
  \dodoi{10.1086/311680}

\bibitem[{Li {et~al.}(2021)Li, Ragosta, Clarkson, \& Bianco}]{Li_2022}
Li, X., Ragosta, F., Clarkson, W.~I., \& Bianco, F.~B. 2021, The Astrophysical
  Journal Supplement Series, 258, 2, \dodoi{10.3847/1538-4365/ac3bca}

\bibitem[{Lochner {et~al.}(2022)Lochner, Scolnic, Almoubayyed, Anguita, Awan,
  Gawiser, Gontcho, Graham, Gris, Huber, Jha, Jones, Kim, Mandelbaum, Marshall,
  Petrushevska, Regnault, Setzer, Suyu, Yoachim, Biswas, Blaineau, Hook,
  Moniez, Neilsen, Peiris, Rothchild, \& Stubbs}]{Lochner_2022}
Lochner, M., Scolnic, D., Almoubayyed, H., {et~al.} 2022, The Astrophysical
  Journal Supplement Series, 259, 58, \dodoi{10.3847/1538-4365/ac5033}

\bibitem[{{LSST Science Collaboration} {et~al.}(2009){LSST Science
  Collaboration}, {Abell}, {Allison}, {Anderson}, {Andrew}, {Angel}, {Armus},
  {Arnett}, {Asztalos}, {Axelrod}, \& et~al.}]{2009arXiv0912.0201L}
{LSST Science Collaboration}, {Abell}, P.~A., {Allison}, J., {et~al.} 2009,
  ArXiv e-prints.
\newblock \doarXiv{0912.0201}

\bibitem[{Metzger(2017)}]{Metzger_2017}
Metzger, B.~D. 2017, Welcome to the Multi-Messenger Era! Lessons from a Neutron
  Star Merger and the Landscape Ahead,  arXiv,
  \dodoi{10.48550/ARXIV.1710.05931}

\bibitem[{Pian(2021)}]{10.3389/fspas.2020.609460}
Pian, E. 2021, Frontiers in Astronomy and Space Sciences, 7,
  \dodoi{10.3389/fspas.2020.609460}

\bibitem[{Pillas {et~al.}(2025)Pillas, Antier, Ackley, Ahumada, Akl,
  de~Almeida, Anand, Andrade, Andreoni, Bostroem, Bulla, Burns, Cabrera, Chang,
  Choi, O'Connor, Coughlin, Corradi, Gibbs, Dietrich, Dornic, Ducoin, Duverne,
  Dyer, Eggenstein, Freeberg, Fausnaugh, Fong, Foucart, Frostig, Guessoum,
  Gupta, Hello, Hosseinzadeh, Hu, Hussenot-Desenonges, Im, Jayaraman, Jeong,
  Karambelkar, Karpov, Kasliwal, Kilpatrick, Kim, Kochiashvili, Kunnumkai,
  Lamoureux, Lee, Lourie, Lyman, Magnani, Masek, Mo, Molham, Navarete, O'Neill,
  Nicholl, Nitz, Noysena, Paek, Palmese, Poggiani, Pradier, Pyshna, Rajabov,
  Rastinejad, Sand, Shawhan, Shrestha, Simcoe, Smartt, Steeghs, Stein,
  Stevance, Sun, Takey, Toivonen, Turpin, Ulaczyk, Wold, \&
  Wouters}]{pillas2025limitsejectamasssearch}
Pillas, M., Antier, S., Ackley, K., {et~al.} 2025, Limits on the Ejecta Mass
  During the Search for Kilonovae Associated with Neutron Star-Black Hole
  Mergers: A case study of S230518h, GW230529, S230627c and the
  Low-Significance Candidate S240422ed.
\newblock \doarXiv{2503.15422}

\bibitem[{Prisinzano {et~al.}(2023)Prisinzano, Bonito, Mazzi, Damiani,
  Ustamujic, Yoachim, Street, Guarcello, Venuti, Clarkson, Jones, \&
  Girardi}]{Prisinzano_2023}
Prisinzano, L., Bonito, R., Mazzi, A., {et~al.} 2023, The Astrophysical Journal
  Supplement Series, 265, 39, \dodoi{10.3847/1538-4365/acbd3b}

\bibitem[{Ragosta {et~al.}(2024)Ragosta, Ahumada, Piranomonte, Andreoni,
  Melandri, Colombo, \& Coughlin}]{Ragosta_2024}
Ragosta, F., Ahumada, T., Piranomonte, S., {et~al.} 2024, The Astrophysical
  Journal, 966, 214, \dodoi{10.3847/1538-4357/ad35c1}

\bibitem[{Raiteri {et~al.}(2021)Raiteri, Carnerero, Balmaverde, Bellm,
  Clarkson, D’Ammando, Paolillo, Richards, Villata, Yoachim, \&
  Yoon}]{Raiteri_2022}
Raiteri, C.~M., Carnerero, M.~I., Balmaverde, B., {et~al.} 2021, The
  Astrophysical Journal Supplement Series, 258, 3,
  \dodoi{10.3847/1538-4365/ac3bb0}

\bibitem[{Schwamb {et~al.}(2023)Schwamb, Jones, Yoachim, Volk, Dorsey, Opitom,
  Greenstreet, Lister, Snodgrass, Bolin, Inno, Bannister, Eggl, Solontoi,
  Kelley, Jurić, Lin, Ragozzine, Bernardinelli, Chesley, Daylan, Ďurech,
  Fraser, Granvik, Knight, Lisse, Malhotra, Oldroyd, Thirouin, \&
  Ye}]{Schwamb_2023}
Schwamb, M.~E., Jones, R.~L., Yoachim, P., {et~al.} 2023, The Astrophysical
  Journal Supplement Series, 266, 22, \dodoi{10.3847/1538-4365/acc173}

\bibitem[{{Setzer} {et~al.}(2023){Setzer}, {Peiris}, {Korobkin}, \&
  {Rosswog}}]{Setzer:2023}
{Setzer}, C.~N., {Peiris}, H.~V., {Korobkin}, O., \& {Rosswog}, S. 2023,
  Monthly Notices of the Royal Astronomical Society, 520, 2829,
  \dodoi{10.1093/mnras/stad257}

\bibitem[{Smartt {et~al.}(2017)Smartt, Chen, Jerkstrand, Coughlin, Kankare,
  Sim, Fraser, Inserra, Maguire, Chambers, Huber, Krühler, Leloudas, Magee,
  Shingles, Smith, Young, Tonry, Kotak, Gal-Yam, Lyman, Homan, Agliozzo,
  Anderson, Angus, Ashall, Barbarino, Bauer, Berton, Botticella, Bulla, Bulger,
  Cannizzaro, Cano, Cartier, Cikota, Clark, De~Cia, Della~Valle, Denneau,
  Dennefeld, Dessart, Dimitriadis, Elias-Rosa, Firth, Flewelling, Flörs,
  Franckowiak, Frohmaier, Galbany, González-Gaitán, Greiner, Gromadzki,
  Guelbenzu, Gutiérrez, Hamanowicz, Hanlon, Harmanen, Heintz, Heinze,
  Hernandez, Hodgkin, Hook, Izzo, James, Jonker, Kerzendorf, Klose,
  Kostrzewa-Rutkowska, Kowalski, Kromer, Kuncarayakti, Lawrence, Lowe, Magnier,
  Manulis, Martin-Carrillo, Mattila, McBrien, Müller, Nordin, O’Neill,
  Onori, Palmerio, Pastorello, Patat, Pignata, Podsiadlowski, Pumo, Prentice,
  Rau, Razza, Rest, Reynolds, Roy, Ruiter, Rybicki, Salmon, Schady, Schultz,
  Schweyer, Seitenzahl, Smith, Sollerman, Stalder, Stubbs, Sullivan, Szegedi,
  Taddia, Taubenberger, Terreran, van Soelen, Vos, Wainscoat, Walton, Waters,
  Weiland, Willman, Wiseman, Wright, Wyrzykowski, \& Yaron}]{Smartt_2017}
Smartt, S.~J., Chen, T.-W., Jerkstrand, A., {et~al.} 2017, Nature, 551,
  75–79, \dodoi{10.1038/nature24303}

\bibitem[{Street {et~al.}(2023)Street, Li, Khakpash, Bellm, Girardi, Jones,
  Abrams, Tsapras, Hundertmark, Bachelet, Gandhi, Szkody, Clarkson, Szabó,
  Prisinzano, Bonito, Buckley, Marais, \& Stefano}]{Street_2023}
Street, R.~A., Li, X., Khakpash, S., {et~al.} 2023, The Astrophysical Journal
  Supplement Series, 267, 15, \dodoi{10.3847/1538-4365/acd6f4}

\bibitem[{Tanvir {et~al.}(2013)Tanvir, Levan, Fruchter, Hjorth, Hounsell,
  Wiersema, \& Tunnicliffe}]{Tanvir_2013}
Tanvir, N.~R., Levan, A.~J., Fruchter, A.~S., {et~al.} 2013, Nature, 500, 547,
  \dodoi{10.1038/nature12505}

\bibitem[{Tanvir {et~al.}(2017)Tanvir, Levan, González-Fernández, Korobkin,
  Mandel, Rosswog, Hjorth, D’Avanzo, Fruchter, Fryer, Kangas, Milvang-Jensen,
  Rosetti, Steeghs, Wollaeger, Cano, Copperwheat, Covino, D’Elia,
  de~Ugarte~Postigo, Evans, Even, Fairhurst, Jaimes, Fontes, Fujii, Fynbo,
  Gompertz, Greiner, Hodosan, Irwin, Jakobsson, Jørgensen, Kann, Lyman,
  Malesani, McMahon, Melandri, O’Brien, Osborne, Palazzi, Perley, Pian,
  Piranomonte, Rabus, Rol, Rowlinson, Schulze, Sutton, Thöne, Ulaczyk, Watson,
  Wiersema, \& Wijers}]{Tanvir_2017}
Tanvir, N.~R., Levan, A.~J., González-Fernández, C., {et~al.} 2017, The
  Astrophysical Journal Letters, 848, L27, \dodoi{10.3847/2041-8213/aa90b6}

\bibitem[{Team(2022)}]{lsst_pstn053}
Team, L. P.~S. 2022, PSTN-053: Survey Cadence Optimization Committee’s Phase
  1 Recommendations.
\newblock \url{https://pstn-053.lsst.io/}

\bibitem[{Team(2023)}]{lsst_pstn055}
---. 2023, PSTN-055: Survey Cadence Optimization Committee’s Phase 2
  Recommendations.
\newblock \url{https://pstn-055.lsst.io/}

\bibitem[{Team(2024)}]{lsst_pstn056}
---. 2024, PSTN-056: Survey Cadence Optimization Committee’s Phase 3
  Recommendations.
\newblock \url{https://pstn-056.lsst.io/}

\bibitem[{Tio {et~al.}(2022)Tio, Pastorelli, Mazzi, Trabucchi, Costa, Jacques,
  Pieres, Girardi, Chen, Olsen, Juric, Željko Ivezić, Yoachim, Clarkson,
  Marigo, Rodrigues, Zaggia, Barbieri, Momany, Bressan, Nikutta, \&
  da~Costa}]{Dal_Tio_2022}
Tio, P.~D., Pastorelli, G., Mazzi, A., {et~al.} 2022, The Astrophysical Journal
  Supplement Series, 262, 22, \dodoi{10.3847/1538-4365/ac7be6}

\bibitem[{Toivonen {et~al.}(2024)Toivonen, Mansingh, Griffin, Kazemi, Kerkow,
  Mahanty, Markus, Tsukamoto, Chaudhary, Antier, Coughlin, Chatterjee, Essick,
  Ghosh, Dietrich, \& Landry}]{toivonen2024expectkilonovalightcurve}
Toivonen, A., Mansingh, G., Griffin, H., {et~al.} 2024, What to expect:
  kilonova light curve predictions via equation of state marginalization.
\newblock \doarXiv{2410.10702}

\end{thebibliography}

\end{document}